\documentclass[british,nofootinbib, 11pt,a4paper]{revtex4-1}
\usepackage[T1]{fontenc}
\usepackage[latin9]{inputenc}
\usepackage{mathrsfs}
\usepackage{mathtools}
\usepackage{amsmath}
\usepackage{amsthm}
\usepackage{amssymb}
\usepackage{esint}

\makeatletter
\theoremstyle{remark}
\newtheorem*{acknowledgement*}{\protect\acknowledgementname}
\theoremstyle{remark}
\newtheorem*{notation*}{\protect\notationname}

\usepackage{tikz-cd} % For elegant diagrams

\makeatother

\usepackage{babel}
\providecommand{\acknowledgementname}{Acknowledgement}
\providecommand{\notationname}{Notation}

\begin{document}

\title{Comments on Minitwistors and the Celestial Supersphere}

\author{Igor Mol}

\affiliation{State University of Campinas (Unicamp)}
\email{igormol@ime.unicamp.br}

\begin{abstract}
Continuing our program of deriving aspects of celestial holography
from string theory, we extend the Roiban-Spradlin-Volovich-Witten
(RSVW) formalism to celestial amplitudes. We reformulate the tree-level
maximally-helicity-violating (MHV) celestial leaf amplitudes for gluons
in $\mathcal{N}=4$ supersymmetric Yang-Mills (SYM) theory and for
gravitons in $\mathcal{N}=8$ Supergravity in terms of \emph{minitwistor
wavefunctions}. These are defined as representatives of cohomology
classes on the minitwistor space $\mathbf{MT}$, associated to the
three-dimensional Euclidean anti-de Sitter space. In this framework,
celestial leaf amplitudes are expressed as integrals over the moduli
space of minitwistor lines. We construct a minitwistor generating
functional for MHV leaf amplitudes using the Quillen determinant line
bundle, extending the approach originally developed by Boels, Mason
and Skinner. Building on this formalism, we propose supersymmetric
celestial conformal field theories (CFTs) as $\sigma$-models, where
the worldsheet is given by the celestial supersphere $\mathbf{CP}^{1|2}$,
and the target space is the minitwistor superspace $\mathbf{MT}^{2|\mathcal{N}}$.
We demonstrate that the semiclassical effective action of these $\sigma$-models
reproduces the MHV gluonic and gravitational leaf amplitudes in $\mathcal{N}=4$
SYM theory and $\mathcal{N}=8$ Supergravity. This construction provides
a concrete realisation of the supersymmetric celestial CFT framework
recently introduced by Tropper (2024).
\end{abstract}
\maketitle
\tableofcontents{}

\newpage{}

\section{Introduction}

\citet{Tropper:2024evi} proposed a general framework for constructing
celestial CFTs that are expected to serve as holographic duals to
spacetime \emph{supersymmetric} field theories. In this paper, we
provide a concrete realisation of Tropper's proposal by building supersymmetric
celestial CFTs on the celestial \emph{supersphere}. These theories
are formulated as sigma models, where the target space is the supersymmetric
extension of minitwistor space $\mathbf{MT}$, associated with three-dimensional
Euclidean anti-de Sitter space, $H_{3}^{+}$.

The models we construct are defined by an action functional that ensures
a well-posed variational principle, allowing the use of the path-integral
formalism to quantise these theories. We demonstrate that, in the
semiclassical limit, these minitwistor celestial CFTs reproduce the
tree-level maximally-helicity-violating (MHV) celestial leaf amplitudes
for gluons in $\mathcal{N}=4$ supersymmetric Yang-Mills (SYM) theory
and for gravitons in $\mathcal{N}=8$ Supergravity. 

This work builds on the observation of \citet{bu2023celestial}, who
demonstrated that the Mellin transform defines a mapping between cohomology
classes in projective twistor space $\mathbf{PT}$ and minitwistor
space $\mathbf{MT}$. By applying both the Mellin and Penrose transforms,
one establishes a correspondence where twistor wavefunctions are mapped
to bulk-to-boundary propagators on $H_{3}^{+}$. Building on this,
we introduce \emph{minitwistor wavefunctions}, which represent cohomology
classes on $\mathbf{MT}$ and, through the Penrose transform, generate
solutions to the covariant wave equation on $H_{3}^{+}$. These wavefunctions
satisfy an integral relation, which we refer to as the \emph{celestial
Roiban-Spradlin-Volovich-Witten }(RSVW) \emph{identity.}

We also propose a new interpretation of celestial leaf amplitudes.
The standard approach involves foliating Klein space $\mathbf{R}^{\left(2,2\right)}$
into hyperbolic leaves and expressing physical scattering amplitudes
as integrals over these leaves. In our reformulation, we introduce
a different basis by expressing celestial amplitudes in terms of minitwistor
wavefunctions. Using the celestial RSVW identity, we show that the
leaf amplitudes for gluons in $\mathcal{N}=4$ SYM theory and for
gravitons in $\mathcal{N}=8$ Supergravity can be written as integrals
over the moduli space of minitwistor lines\footnote{See Appendix \ref{sec:Minitwistor-Geometry} for a definition.}.
Moreover, we prove that these leaf amplitudes vanish if the gluon
or graviton insertion points do not lie on a common minitwistor line.

It is also shown that minitwistor wavefunctions satisfy another integral
relation, which we call the \emph{celestial Boels-Mason-Skinner} \emph{identity}.
This property is derived from an $n$-fold application of the Penrose
transform. Using this result, we construct a generating functional
for MHV celestial amplitudes in terms of the Quillen determinant line
bundle, which establishes a direct link between the celestial RSVW
formalism and the semiclassical effective action of the minitwistor
celestial CFTs developed in this work.

\paragraph*{Organisation.}

In Section \ref{subsec:Minitwistor-Wavefunctions}, we begin with
a brief review of the Penrose integral-geometric transform on minitwistor
space $\mathbf{MT}$, followed by the introduction of minitwistor
wavefunctions. We then derive both the celestial RSVW identity and
the Boels-Mason-Skinner identity. In Section \ref{sec:-Supersymmetric-Yang-Mills},
we reformulate the tree-level MHV gluonic celestial leaf amplitudes
in $\mathcal{N}=4$ SYM theory using the celestial RSVW formalism
and derive the corresponding minitwistor generating functional. Section
\ref{sec:-Supergravity} extends these results to $\mathcal{N}=8$
Supergravity. In Section \ref{sec:Minitwistor-String-Theories}, we
construct the minitwistor celestial CFTs in detail and show that the
minitwistor generating functional arises as the semiclassical effective
action obtained through the path-integral quantisation of our models.
Finally, in Section \ref{sec:Discussion}, we outline future directions
of our research program aimed at deriving aspects of celestial holography
from string theory.
\begin{acknowledgement*}
I thank Roland Bittleston for his helpful insights on minitwistor
geometry. I also thank V. P. Nair for bringing his work with C. Kim
to my attention\footnote{See \citet{kim1997recursion}.}.
\end{acknowledgement*}

\section{Minitwistor Wavefunctions\label{subsec:Minitwistor-Wavefunctions}}

\subsection{Physical Motivation}

The main goal of this paper is to formulate a sigma model on the celestial
\emph{supersphere}, whose target space is the non-singular quadric
$\mathbf{MT}$, representing the \emph{minitwistor space} associated
with the three-dimensional hyperboloid $H_{3}^{+}$. This model is
designed to reproduce semiclassically the tree-level MHV celestial
leaf amplitudes for gluons in $\mathcal{N}=4$ SYM theory and for
gravitons in $\mathcal{N}=8$ Supergravity.

A key part of this construction is the introduction of \emph{minitwistor
wavefunctions}, which are precisely defined in Subsection \ref{subsec:Minitwistor-Wavefunctions-1}.
These wavefunctions serve as the basic building blocks for the vertex
operators in the minitwistor sigma model.

Based on the recent work by \citet{bu2023celestial}, the \emph{celestial
leaf amplitudes }may be derived through an alternative, yet mathematically
equivalent, formalism. The original aproach of \citet{casali2022celestial}
and \citet{melton2024celestial} involves foliating Klein space $\mathbf{R}^{\left(2.2\right)}$
by hyperbolic leaves and expressing physical scattering amplitudes
as integrals over these leaves. In the alternative approach, one begins
with twistor wavefunctions and subsequently performs a Mellin transform
with respect to one of the spinor variables. This is followed by the
application of the Penrose transform from minitwistor space $\mathbf{MT}$
to the hyperboloid $H_{3}^{+}$, thus recovering the bulk-to-boundary
propagator $K_{\Delta}\left(X;z,\bar{z}\right)$ for the covariant
Laplacian $\square_{H_{3}^{+}}$.

The minitwistor wavefunctions satisfy two important properties: the
\emph{celestial Boels-Mason-Skinner} (BMS) and \emph{Roiban-Spradlin-Volovich-Witten}
(RSVW) \emph{integral identities}. The BMS identities (discussed in
Subsection \ref{subsec:BMS}) will be employed in the construction
of a generating functional for the celestial leaf amplitudes. On the
other hand, the RSVW identity (derived in Subsection \ref{RSVW})
will be used in reformulating these amplitudes as a Fourier transform
in minitwistor space. 

In the next subsection, we review the extension of the Penrose integral-geometric
transform to the minitwistor space $\mathbf{MT}$.

\subsection{Review: Minitwistor Penrose Transform and $AdS_{3}$ Wave Equations\label{subsec:Minitwistor-Penrose-Transform}}

The formalisation of the Penrose transform is naturally expressed
through the language of integral geometry\footnote{A general review of the subject is provided in \citet{guillemin1987perspectives}.
For a rigorous textbook treatment, we direct the reader to \citet{helgason2011integral,helgason2022groups,helgason2024geometric}.
Additionally, the text by \citet{quinto2013geometric} provides an
accessible account well-suited for those interested in physical applications.}. The basic structure in this branch of geometric analysis is a \emph{double
fibration},\begin{equation}  \label{diagram}  \begin{tikzcd}[row sep=large, column sep=large] & \mathscr{Z} \arrow[dl, "\textstyle q_{1}"' ] \arrow[dr, "\textstyle q_{2}"] & \\ \mathscr{X} & & \mathscr{Y} \end{tikzcd}  \end{equation} where
$\mathscr{X}$, $\mathscr{Y}$ and $\mathscr{Z}$ denote smooth manifolds,
and $q_{1},q_{2}$ are fibre maps. The diagram (\ref{diagram}) qualifies
as a double fibration if the product map:\begin{equation} \label{eq:-41} \begin{tikzcd}[column sep=large] q_{1} \times q_{2} : \mathscr{Z} \arrow[r] & \mathscr{X} \times \mathscr{Y} \end{tikzcd}\end{equation} is
an embedding of $\mathscr{Z}$ as a submanifold of the product space
$\mathscr{X}\times\mathscr{Y}$. 

Under the assumption that $q_{1}\times q_{2}$ is indeed an embedding,
it follows directly that, for any point $y\in\mathscr{Y}$, the fibre
$F\left(y\right)\coloneqq q_{2}^{-1}\left(y\right)$ is smoothly embedded
as a submanifold of $\mathscr{X}$. Therefore, the double fibration
(\ref{diagram}) induces a family of submanifolds $\{F\left(y\right)\}_{y\in\mathscr{Y}}$,
with each submanifold $F\left(y\right)$ smoothly embedded into $\mathscr{X}$
and parametrised by points in $\mathscr{Y}$.

The key idea of integral geometry is that geometric objects on $\mathscr{X}$
can be transported to $\mathscr{Y}$ via the intermediate space $\mathscr{Z}$.
For example, given a differential form (or holomorphic section) $a$
defined on $\mathscr{X}$, the pullback $q_{1}^{*}\left(a\right)$
yields a corresponding form on $\mathscr{Z}$. The \emph{integral-geometric
transform} is then constructed by integrating the pullback over the
fibres of the map $q_{2}$, resulting in a function or section on
$\mathscr{Y}$. Formally, the integral-geometric transform is defined
as:
\begin{equation}
a\mapsto\widetilde{a}\left(y\right)\coloneqq\underset{q_{2}^{-1}\left(y\right)}{\int}\,q_{1}^{*}\left(a\right),\label{eq:-43}
\end{equation}
where $\widetilde{a}$ denotes the resulting form on $\mathscr{Y}$.

An important feature of this integral transform is that, whenever
the pullback $q_{1}^{*}\left(a\right)$ remains constant along the
fibres of $q_{1}$, the transformed form $\widetilde{a}$ must satisfy
certain differential equations on $\mathscr{Y}$. This observation
formalises a result originally noted by \citet{bateman1918solution}
and rediscovered by \citet{penrose1969solutions}.

\subsubsection{Minitwistor Geometry}

In Appendix \ref{sec:Minitwistor-Geometry}, we provide a concise
discussion of how the hyperbolic geometry of the three-dimensional
real manifold $H_{3}^{+}$ can be realised within the projective geometry
of $\mathbf{CP}^{3}$, subject to appropriate reality conditions.
Additionally, we give a brief but mathematically rigorous introduction
to the minitwistor space\footnote{See also \citet{jones1984minitwistors,jones1985minitwistor}, \citet{hitchin1982monopoles,hitchin1982twistor}
and \citet{honda2011minitwistor}.} $\mathbf{MT}$ associated to the hyperboloid $H_{3}^{+}$. Now, we
proceed to specialise the abstract framework of integral geometry,
as delineated above, to the geometry of minitwistor space.

Let $\mathbf{H}_{3}$ denote the complexification of the real hyperboloid
$H_{3}^{+}$, and introduce the complexified projective spinor bundle
$\mathbf{PS}_{3}\coloneqq\mathbf{CP}^{3}\times\mathbf{CP}^{1}$ as
the trivial bundle whose base space is $\mathbf{CP}^{3}$ and the
typical fibre consists of the projectivised space of (undotted) two-component
spinors. The relevant double fibration in this geometric setup is
given by the diagram:\begin{equation}\begin{tikzcd}[row sep=large, column sep=large] & \mathbf{PS}_{3} \arrow[dl, "\displaystyle\widetilde{q}_{1}"'] \arrow[dr, "\displaystyle\widetilde{q}_{2}"] & \\ \mathbf{MT} & & \mathbf{H}_{3} \end{tikzcd} \end{equation}The
first bundle map $\widetilde{q}_{1}$ is defined via the minitwistor
incidence relation:
\begin{equation}
\widetilde{q}_{1}:\,\left(X_{A\dot{A}}\,,\,\lambda^{A}\right)\,\mapsto\,\left(\lambda^{A},\,\lambda^{A}X_{A\dot{A}}\right),
\end{equation}
where $X_{A\dot{A}}$ represents homogeneous coordinates on $\mathbf{CP}^{3}$,
and $\lambda^{A}$ denotes homogeneous spinor coordinates in the fibre
$\mathbf{CP}^{1}$. The second bundle map $\widetilde{q}_{2}$ is
the trivial surjection $\widetilde{q}_{2}:(X_{A\dot{A}},\lambda^{A})\mapsto X_{A\dot{A}}$.

Having established the double fibration structure, it remains to specify
the module of sections of the holomorphic vector bundle on $\mathbf{MT}$
upon which the integral-geometric transform defined in Eq. (\ref{eq:-43})
operates. Therefore, let $\mathscr{C}_{p,q}^{\infty}\left(\mathbf{MT}\right)$
denote the space of $\mathscr{C}^{\infty}$ complex-valued functions
$h$ defined on $\left(\mathbf{C}^{*}\right)^{2}\times\left(\mathbf{C}^{*}\right)^{2}$
that satisfy the following homogeneity property:
\begin{equation}
h\left(a\cdot\lambda^{A},b\cdot\mu_{\dot{A}}\right)=a^{p}b^{q}\,\,\,h\left(\lambda^{A},\mu_{\dot{A}}\right),
\end{equation}
for every pair of nonzero complex scalars $a$ and $b$. In Appendix
\ref{subsec:The-Holomorphic-Vector}, we prove that the function space
$\mathscr{C}_{p,q}^{\infty}\left(\mathbf{MT}\right)$ is canonically
identified with the module $\Gamma^{\infty}(\mathcal{O}(p,q))$ of
smooth sections on the holomorphic vector bundle $\mathcal{O}\left(p,q\right)\longrightarrow\mathbf{MT}$.

\subsubsection{Scalar Representatives}

In four-dimensional physics, twistor theory\footnote{For a modern introduction, see \citet{adamo2017lectures,atiyah2017twistor}
and \citet[Appendix A]{witten2004perturbative}.} characterises solutions to (linearised) massless field equations
on spacetime through equivalence classes in the cohomology group of
projective twistor space $\mathbf{PT}$, associated with the module
of sections of the holomorphic vector bundle $\mathcal{O}\left(p\right)\longrightarrow\mathbf{PT}$.
We now extend this framework to the setting of $AdS_{3}$.\footnote{A similar discussion for the $AdS_{5}$ case can be found in \citet{adamo2016twistor}.}
To explain this construction, we restrict our attention to the case
of scalar representatives:
\begin{equation}
[\mathsf{f}]\in H^{0,1}\left(\mathbf{MT},\mathcal{O}\left(-2,-\Delta\right)\right).
\end{equation}

Let $X^{A\dot{A}}\in\mathbf{H}_{3}$ be a point on the complexified
hyperboloid embedded in projective space, $\mathbf{H}_{3}\subset\mathbf{CP}^{3}$,
where $X^{A\dot{A}}$ denotes homogeneous coordinates. Define the
rational curve $\mathcal{L}\left(X\right)$ as the line of incidence
in minitwistor space:
\begin{equation}
\mathcal{L}\left(X\right)=\left\{ \left(\lambda^{A},\mu_{\dot{A}}\right)\in\mathbf{MT}\,\big|\,\mu_{\dot{A}}=\lambda^{A}X_{A\dot{A}}\right\} .
\end{equation}
Employing the sheaf-theoretic notation of \citet{forster1981compact},
let $\rho_{\mathcal{L}\left(X\right)}$ denote the restriction homomorphism
to the incidence line $\mathcal{L}\left(X\right)$, defined by:
\begin{equation}
\rho_{\mathcal{L}\left(X\right)}(g)\left(\lambda^{A}\right)\coloneqq g\left(\lambda^{A},\lambda^{A}X_{A\dot{A}}\right),\,\,\,\forall\,[g]\in H^{0,1}\left(\mathbf{MT},\mathcal{O}\left(-2,-\Delta\right)\right).
\end{equation}

Let us restrict our attention to a $\left(0,1\right)$-form on $\mathbf{MT}$
valued in $\mathcal{O}\left(-2,-\Delta\right)$. The Penrose transform
can be introduced as a mapping from representatives of the cohomology
class $H^{0,1}\left(\mathbf{MT},\mathcal{O}\left(-2,-\Delta\right)\right)$,
given by the Penrose integral:
\begin{equation}
\mathsf{f}\mapsto\underset{\mathcal{L}\left(X\right)}{\int}\left\langle \lambda d\lambda\right\rangle \,\,\,\rho_{\mathcal{L}\left(X\right)}\big(\mathsf{f}\big)\left(\lambda\right).
\end{equation}
Thus, define:
\begin{equation}
F_{\Delta}\left(X\right)\coloneqq\left|X\right|^{\Delta}\underset{\mathcal{L}\left(X\right)}{\int}\left\langle \lambda d\lambda\right\rangle \,\,\,\rho_{\mathcal{L}\left(X\right)}\big(\mathsf{f}\big)\left(\lambda\right).
\end{equation}
Accordingly, $F_{\Delta}\left(X\right)$ is invariant under the flow
of the Euler vector field $\Upsilon$, such that $\text{£}_{\Upsilon}F_{\Delta}=0$.
Consequently, $F_{\Delta}\left(X\right)$ is homogeneous of degree
zero and, therefore, yields a well-defined \emph{function} on $H_{3}^{+}$.

To verify the consistency of this construction, note that:
\begin{equation}
\frac{\partial}{\partial X}\cdot\frac{\partial}{\partial X}\left(\frac{F_{\Delta}\left(X\right)}{\left|X\right|^{\Delta}}\right)=0\iff\overline{\partial}\,\big|_{\mathcal{L}\left(X\right)}\,\mathsf{f}=0.
\end{equation}
Since every $\mathsf{f}$ that is $\overline{\partial}$-exact integrates
to zero, it follows that $F_{\Delta}\left(X\right)$ is determined
by the cohomology class $[\mathsf{f}]$ in $H^{0,1}\left(\mathbf{MT},\mathcal{O}\left(-2,-\Delta\right)\right)$.

Furthermore, from the invariance of $F_{\Delta}\left(X\right)$ under
the flow of $\Upsilon$,
\begin{equation}
X^{A\dot{A}}\frac{\partial}{\partial X^{A\dot{A}}}F_{\Delta}\left(X\right)=0,
\end{equation}
we deduce that:
\begin{equation}
\varepsilon^{AB}\varepsilon^{\dot{A}\dot{B}}\frac{\partial}{\partial X^{A\dot{A}}}\frac{\partial}{\partial X^{B\dot{B}}}\left(\frac{F_{\Delta}\left(X\right)}{\left|X\right|^{\Delta}}\right)=0,
\end{equation}
which is equivalent to:
\begin{equation}
\square_{\mathbf{H}_{3}}\,F_{\Delta}\left(X\right)=\Delta\left(\Delta+2\right)F_{\Delta}\left(X\right).
\end{equation}

We conclude that the Penrose transform establishes a correspondence
between cohomology classes of the non-singular quadric $\mathbf{MT}$
associated to the holomorphic vector bundle $\mathcal{O}\left(-2,-\Delta\right)\longrightarrow\mathbf{MT}$,
and conformal primaries of the minisuperspace limit of the $H_{3}^{+}$-WZNW
model.

\subsection{Minitwistor Wavefunctions\label{subsec:Minitwistor-Wavefunctions-1}}

We are now prepared to introduce the notion of a minitwistor wavefunction.
This object shows an interesting connection between the Mellin and
Penrose transforms. It will be demonstrated that the Mellin transform
induces a mapping between cohomology classes on projective twistor
space $\mathbf{PT}$, and corresponding cohomology classes on the
nonsingular quadric $\mathbf{MT}$. By subsequently applying the Penrose
transform, it will become clear that twistor wavefunctions are mapped
to bulk-to-boundary Green's functions\footnote{See \citet{teschner1997conformal,teschner1999mini,teschner1999structure,teschner2000operator,ribault2005h3+}.}
for the covariant Laplacian $\square_{H_{3}^{+}}$ on the hyperboloid
$H_{3}^{+}$\emph{.}

\subsubsection{Definitions}
\begin{notation*}
From now on, we shall adopt a simplified notational convention for
spinor functions. The explicit display of abstract spinor indices
in the arguments of such functions will be reserved exclusively for
their initial definition. Thereafter, the spinor type (dotted \emph{vs.}
undotted) comprising the domain of each function will remain implicit.
\end{notation*}

\paragraph*{Twistor Scalar Wavefunction.}

We consider the twistor scalar wavefunction:
\begin{equation}
\mathsf{f}_{w}\in\Omega^{0,1}\left(\mathbf{PT},\mathcal{O}\left(-w\right)\right),\label{eq:-41}
\end{equation}
which admits the following integral representation:
\begin{equation}
\mathsf{f}_{w}\left(\lambda^{A},\mu_{\dot{A}};\text{m}\right)\,\coloneqq\,\underset{\mathbf{C}^{*}}{\int}\,\frac{dt}{t}\,t^{w}\,\,\,\overline{\delta}^{2}\big(z^{A}-t\lambda^{A}\big)\,\exp\left(i\,t\left[\mu\bar{z}\right]\right).\label{eq:-13}
\end{equation}
In the above expression, the spinor delta function is defined as:
\begin{equation}
\overline{\delta}^{2}\left(\lambda^{A}\right)\,\coloneqq\,\frac{1}{\left(2\pi i\right)^{2}}\,\bigwedge_{A\in\{1,2\}}\,\overline{\partial}\,\left(\frac{1}{\lambda^{A}}\right).
\end{equation}
The ordered pair $\text{m}\coloneqq\left(z^{A},\bar{z}_{\dot{A}}\right)$
denotes the collection of quantum numbers characterising the state
of the particle associated with $\mathsf{f}_{w}$. For the massless
gauge bosons participating in the scattering processes to be considered
in the subsequent sections, the spinors are chosen be to normalised
as $z^{A}=\left(z,1\right)^{T}$ and $\bar{z}_{\dot{A}}=\left(1,-\bar{z}\right)$,
where $\left(z,\bar{z}\right)$ parametrises the insertion points
on the celestial sphere. In what follows, we shall omit the explicit
dependence on $z^{A}$ and $\bar{z}_{\dot{A}}$ from the arguments
of any wavefunctions to streamline the notation.

The structure of the affine integral (\ref{eq:-13}) implies that
the twistor scalar wavefunction satisfies the homogeneity property:
\begin{equation}
\mathsf{f}_{w}\left(a\lambda,a\mu\right)=a^{-w}\mathsf{f}_{w}\left(\lambda,\mu\right),
\end{equation}
 for all $a\in\mathbf{C}^{*}$, as expected from the corresponding
cohomology class (\ref{eq:-41}).

Performing the affine integral in Eq. (\ref{eq:-13}) yields the explicit
form of the twistor scalar wavefunction:
\begin{equation}
\mathsf{f}_{w}\left(\lambda,\mu\right)\,=\,\overline{\delta}\left(\left\langle \lambda z\right\rangle \right)\,\left(\frac{\left\langle \lambda\iota\right\rangle }{\left\langle z\iota\right\rangle }\right)^{1-w}\exp\left(i\frac{\left\langle z\iota\right\rangle }{\left\langle \lambda\iota\right\rangle }\left[\mu\bar{z}\right]\right).\label{eq:-42}
\end{equation}

We now define the \emph{minitwistor wavefunction} with \emph{celestial
conformal weight }$\Delta$ as the Mellin transform of $\mathsf{f}_{w}$
with respect to the dotted spinor $\mu_{\dot{A}}$:
\begin{equation}
\widehat{\mathsf{f}}_{\Delta,w}\left(\lambda^{A},\mu_{\dot{A}}\right)\,\coloneqq\,\underset{\mathbf{R}_{+}^{\times}}{\int}\,\frac{ds}{s}\,s^{\Delta}\,\,\,\mathsf{f}_{w}\left(\lambda^{A},s\mu_{\dot{A}}\right),\label{eq:-44}
\end{equation}
where $\mathbf{R}_{+}^{\times}\coloneqq\left(\mathbf{R}_{+},\times\right)$
denotes the multiplicative group of positive real numbers, and $\frac{ds}{s}$
is the associated Haar measure.

Substituting Eq. (\ref{eq:-13}) into Definition (\ref{eq:-44}),
we obtain the complete integral representation of the minitwistor
wavefunction:
\begin{equation}
\widehat{\mathsf{f}}_{\Delta,w}\left(\lambda,\mu\right)\,=\,\underset{\mathbf{R}_{+}^{\times}}{\int}\,\frac{ds}{s}\,s^{\Delta}\,\underset{\mathbf{C}^{*}}{\int}\,\frac{dt}{t}\,t^{w}\,\,\,\overline{\delta}^{2}\big(z^{A}-t\lambda^{A}\big)\,\exp\left(i\,st\left[\mu\bar{z}\right]\right).\label{eq:-16}
\end{equation}
The structure of the Mellin and affine integrals in the above expression
reveals that the minitwistor wavefunction exhibits the homogeneity
property:
\begin{equation}
\widehat{\mathsf{f}}_{\Delta,w}\left(a\lambda,b\mu\right)\,=\,a^{\Delta-w}b^{-\Delta}\,\,\,\widehat{\mathsf{f}}_{\Delta,w}\left(\lambda,\mu\right),
\end{equation}
for all $a,b\in\mathbf{C}^{*}$.

Consequently,
\begin{equation}
\hat{\mathsf{f}}_{\Delta,w}\in\Omega^{0,1}\left(\mathbf{MT},\mathcal{O}\left(\Delta-w,-\Delta\right)\right).
\end{equation}
Finally, performing the integrals in the expression (\ref{eq:-16})
for $\mathsf{f}_{\Delta,w}$, we derive the explicit form of the minitwistor
scalar wavefunction:
\begin{equation}
\widehat{\mathsf{f}}_{\Delta,w}\left(\lambda,\mu\right)\,=\,\overline{\delta}\left(\left\langle \lambda z\right\rangle \right)\,\left(\frac{\left\langle \lambda\iota\right\rangle }{\left\langle z\iota\right\rangle }\right)^{1+\Delta-w}\frac{\mathcal{C}\left(\Delta\right)}{\left[\mu\bar{z}\right]^{\Delta}}.
\end{equation}

\subsubsection{Celestial Boels-Mason-Skinner Integral Identity\label{subsec:BMS}}

In this subsection, we derive an integral identity that will serve
as a key step in constructing the generating functional for celestial
leaf amplitudes, analogous to the twistor-space generating functional
introduced by \citet{boels2007twistor}. This identity will henceforth
be referred to as the \emph{celestial Boels-Mason-Skinner }(BMS) \emph{identity}\footnote{There should be no confusion with the BMS group, as each applies to
a different physical context.}\emph{.}

We begin by considering a distinguished representative of minitwistor
scalar wavefunctions, 
\begin{equation}
\mathcal{F}_{\Delta}\in\Omega^{0,1}\left(\mathbf{MT},\mathcal{O}\left(0,-\Delta\right)\right),
\end{equation}
defined by the expression:
\begin{equation}
\mathcal{F}_{\Delta}\left(\lambda^{A},\mu_{\dot{A}}\right)\,\coloneqq\,\widehat{\mathsf{f}}_{\Delta,\Delta}\left(\lambda^{A},\mu_{\dot{A}}\right),\label{eq:-22}
\end{equation}
and taking the explicit form:
\begin{equation}
\mathcal{F}_{\Delta}\left(\lambda,\mu\right)\,=\,\overline{\delta}\left(\left\langle \lambda z\right\rangle \right)\,\frac{\left\langle \lambda\iota\right\rangle }{\left\langle z\iota\right\rangle }\,\frac{\mathcal{C}\left(\Delta\right)}{\left[\mu\bar{z}\right]^{\Delta}}.
\end{equation}

The Penrose transform of $\mathcal{F}_{\Delta}$ is computed by restricting
this wavefunction to a specific holomorphic curve in minitwistor space.
Let the minitwistor line corresponding to the ``spacetime'' point
$X_{A\dot{A}}$ be defined by the locus of incidence:
\begin{equation}
\mathcal{L}\left(X\right)\,\coloneqq\,\big\{\,\big(\lambda^{A},\mu_{\dot{A}}\big)\,\in\mathbf{MT}\,\big|\,\mu_{\dot{A}}=\lambda^{A}X_{A\dot{A}}\big\}.
\end{equation}
Following the sheaf-theoretic notation of \citet{forster1981compact},
the restriction homomorphism to the conic $\mathcal{L}\left(X\right)$
is denoted by $\rho_{X}$, which acts on cohomology representatives:
\[
[g]\in H^{0,1}\left(\mathbf{MT},\mathcal{O}\left(0,-\Delta\right)\right)
\]
according to:
\begin{equation}
\rho_{X}\left(g\right)\left(\lambda\right)\,\coloneqq\,g\left(\lambda^{A},\lambda^{A}X_{A\dot{A}}\right).
\end{equation}

Let $\pi:\mathcal{L}\left(X\right)\longrightarrow\mathbf{CP}^{1}$
denote the canonical projection of the conic onto the projective line.
We introduce a trivialisation of the fibration $\pi$ by a choice
of homogeneous coordinates $\lambda^{A}$ on $\mathbf{CP}^{1}$. The
natural orientation of the fibre is induced by the volume form $D\lambda\coloneqq\varepsilon_{AB}\lambda^{A}d\lambda^{B}$.
In addition, let $a^{A}$ and $b^{A}$ be a pair of constant two-component
spinors.

Consequently, the Penrose transform of: 
\begin{equation}
\frac{\mathcal{F}_{\Delta}\left(\lambda,\mu;z,\bar{z}\right)}{\left\langle a\lambda\right\rangle \left\langle \lambda b\right\rangle },
\end{equation}
is given by the integral-geometric transform:
\begin{equation}
\frac{\mathcal{F}_{\Delta}\left(\lambda,\mu;z,\bar{z}\right)}{\left\langle a\lambda\right\rangle \left\langle \lambda b\right\rangle }\,\mapsto\,\underset{\mathcal{L}\left(X\right)}{\int}\,D\lambda\,\,\,\rho_{X}\left(\mathcal{F}_{\Delta}\right)\left(\lambda\right)\,\,\,\frac{1}{\left\langle a\lambda\right\rangle \left\langle \lambda b\right\rangle },
\end{equation}
which, upon evaluation, yields:
\begin{equation}
\left|X\right|^{\Delta}\,\underset{\mathcal{L}\left(X\right)}{\int}\,D\lambda\,\,\,\rho_{X}\left(\mathcal{F}_{\Delta}\right)\left(\lambda\right)\,\frac{1}{\left\langle a\lambda\right\rangle \left\langle \lambda b\right\rangle }\,=\,\mathcal{C}\left(\Delta\right)\,\frac{\left|X\right|^{\Delta}}{\langle z|X|\bar{z}]^{\Delta}}\,=\,\frac{K_{\Delta}\left(X;z,\bar{z}\right)}{\left\langle a\lambda\right\rangle \left\langle \lambda b\right\rangle },
\end{equation}
where $K_{\Delta}\left(X;z,\bar{z}\right)$ is identified with the
bulk-to-boundary propagator\footnote{See \citet{gelfandgeneralized}, \citet{teschner1997conformal,teschner1999mini,teschner1999structure,teschner2000operator},
\citet{costa2014spinning} and \citet{penedones2017tasi}.} on the hyperboloid $H_{3}^{+}$, with $C\left(\Delta\right)\coloneqq i^{\Delta}\Gamma\left(\Delta\right)$.

We now proceed by employing an inductive argument over $n\in\mathbf{N}$
to generalise this result to an $n$-fold product of Penrose transforms.
Recall that $\left(z,\bar{z}\right)$ label boundary points of $H_{3}^{+}$.
Then, the $n$-fold Penrose transform gives the celestial BMS identity:
\begin{equation}
\underset{\mathcal{L}\left(X\right)^{\times n}}{\int}\,\bigwedge_{i=1}^{n}\,D\lambda_{i}\,\,\,\left|X\right|^{\Delta_{i}}\rho_{X}\left(\mathcal{F}_{\Delta}\right)\left(\lambda\right)\frac{1}{\lambda_{i}\cdot\lambda_{i+1}}=\prod_{i=1}^{n}\frac{K_{\Delta}\left(X;z_{i},\bar{z}_{i}\right)}{z_{i}\cdot z_{i+1}}.\label{eq:-18}
\end{equation}

\subsubsection{Celestial Roiban-Spradlin-Volovich-Witten Integral Identity\label{RSVW}}

We now proceed to establish that the restriction homomorphism $\rho_{X}$
may be enforced by means of a \emph{weighted} Dirac delta function
on minitwistor space. This will allow us to derive an identity analogous
to Eq. (\ref{eq:-18}), expressed as an integral over the minitwistor
space $\mathbf{MT}$.

The theory of distributions was extended to include sections of differential
forms by \citet{de2012differentiable}, building on earlier work by
\citet{schwartz1954espaces,schwartz1957theorie,schwartz1958theorie}.
In this framework, let $\overline{\delta}_{\Delta}$ denote a distributional
form valued in the holomorphic vector bundle $\mathcal{O}\left(\Delta-2,-2\right)$,
defined by the expression:
\begin{equation}
\overline{\delta}_{\Delta}\left(\mu_{\dot{A}},\pi_{\dot{A}}\right)\,\coloneqq\,\frac{1}{\left(2\pi i\right)^{2}}\,\underset{\mathbf{C}^{*}}{\int}\,\frac{dt}{t}\,t^{\Delta}\,\bigwedge_{\dot{A}\in\{\dot{1},\dot{2}\}}\overline{\partial}\,\left(\frac{1}{\mu_{\dot{A}}-t\pi_{\dot{A}}}\right),\label{eq:-40}
\end{equation}
which imposes the projective coincidence condition $\mu_{\dot{A}}\sim\pi_{\dot{A}}$
in the complex projective line. 

To perform the affine integration over $t$, we invoke the analytic
continuation of the Dirac delta function, given by:
\begin{equation}
\overline{\delta}\left(z\right)\,\coloneqq\,\frac{1}{2\pi i}\,\overline{\partial}\,z^{-1}.
\end{equation}
Applying this definition to the integrand of Eq. (\ref{eq:-40}),
we derive the following explicit form for $\overline{\delta}_{\Delta}$:
\begin{equation}
\overline{\delta}_{\Delta}\left(\mu,\pi\right)\,=\,\overline{\delta}\left([\mu\pi]\right)\,\left(\frac{[\mu\bar{\iota}]}{[\pi\bar{\iota}]}\right)^{\Delta-1},
\end{equation}
where $\bar{\iota}^{\dot{A}}$ denotes a fixed reference spinor that
is arbitrarily chosen but non-vanishing.

We now introduce a representative wavefunction on minitwistor space,
\begin{equation}
\widetilde{\mathcal{F}}_{\Delta}\in\Omega^{0,1}\left(\mathbf{MT},\mathcal{O}\left(\Delta,-\Delta\right)\right),
\end{equation}
defined by the expression:
\begin{equation}
\widetilde{\mathcal{F}}_{\Delta}\left(\lambda^{A},\mu_{\dot{A}}\right)\,\coloneqq\,\widehat{\mathsf{f}}_{\Delta,0}\left(\lambda^{A},\mu_{\dot{A}}\right),
\end{equation}
and taking the explicit form:
\begin{equation}
\widetilde{\mathcal{F}}_{\Delta}\left(\lambda,\mu\right)\,=\,\overline{\delta}\left(\left\langle \lambda z\right\rangle \right)\,\left(\frac{\left\langle \lambda\iota\right\rangle }{\left\langle z\iota\right\rangle }\right)^{1+\Delta}\frac{\mathcal{C}\left(\Delta\right)}{\left[\mu\bar{z}\right]^{\Delta}}.
\end{equation}

For each fixed point $X_{A\dot{A}}\in\mathbf{CP}^{3}$, we define
the following distribution on minitwistor space:
\begin{equation}
g_{\Delta}\left(X_{A\dot{A}};\lambda^{A},\mu_{\dot{A}}\right)\,\coloneqq\,\overline{\delta}_{\Delta}\left(\mu_{\dot{A}},\lambda^{A}\frac{X_{A\dot{A}}}{\left|X\right|}\right)\widetilde{\mathcal{F}}_{\Delta}\left(\lambda^{A},\mu_{\dot{A}}\right)\frac{1}{\left\langle a\lambda\right\rangle \left\langle \lambda b\right\rangle },
\end{equation}
which takes values in the holomorphic vector bundle $\mathcal{O}\left(-2,-2\right)$.
This definition enables us to write the following integral over minitwistor
space:
\begin{equation}
\underset{\mathbf{MT}}{\int}D\lambda\wedge D\mu\,\,\,g_{\Delta}\left(X;\lambda,\mu\right),
\end{equation}
where $D\lambda\wedge D\mu$ denotes the canonical volume form on
the non-singular quadric $\mathbf{MT}$, ensuring that the integral
is projectively well-defined.

Utilising the explicit expressions derived above, we obtain the following
form for $g_{\Delta}\left(X;\lambda,\mu\right)$:
\begin{equation}
g_{\Delta}\left(X;\lambda,\mu\right)=\overline{\delta}\left(\left\langle \lambda z\right\rangle \right)\overline{\delta}\left(\frac{\langle\lambda|X|\mu]}{\left|X\right|}\right)\left(\frac{\left\langle \lambda\iota\right\rangle }{\left\langle z\iota\right\rangle }\right)^{1+\Delta}\left(\frac{1}{[\mu\bar{\iota}]}\frac{\langle\lambda|X|\bar{\iota}]}{\left|X\right|}\right)^{1-\Delta}\frac{\mathcal{C}\left(\Delta\right)}{\left[\mu\bar{z}\right]^{\Delta}}\,\,\,\frac{1}{\left\langle a\lambda\right\rangle \left\langle \lambda b\right\rangle }.
\end{equation}
Introducing projective coordinates on minitwistor space via $\mathsf{Z}^{I}\coloneqq\left(\lambda^{A},\mu_{\dot{A}}\right)$,
and denoting the canonical volume form by $D^{2}\mathsf{Z}=D\lambda\wedge D\mu$,
direct evaluation of the integral yields:
\begin{equation}
\underset{\mathbf{MT}}{\int}D^{2}\mathsf{Z}\,\,\,g_{\Delta}\left(X;\lambda,\mu\right)=\frac{K_{\Delta}\left(X;z,\bar{z}\right)}{\left\langle a\lambda\right\rangle \left\langle \lambda b\right\rangle },
\end{equation}
which is identified with the bulk-to-boundary propagator on $H_{3}^{+}$.

Finally, by applying an inductive argument on $n\in\mathbf{N}$, we
derive the \emph{celestial Roiban-Spradlin-Volovich-Witten }(RSVW)
\emph{identity}:
\begin{equation}
\prod_{i=1}^{n}\underset{\mathbf{MT}}{\int}D^{2}\mathsf{Z}_{i}\,\,\,\overline{\delta}_{\Delta_{i}}\left(\mu_{i\dot{A}},\lambda^{A}\frac{X_{A\dot{A}}}{\left|X\right|}\right)\,\widetilde{\mathcal{F}}_{\Delta_{i}}\left(\lambda_{i},\mu_{i};z_{i},\bar{z}_{i}\right)\frac{1}{\lambda_{i}\cdot\lambda_{i+1}}=\prod_{i=1}^{n}\frac{K_{\Delta_{i}}\left(X;z_{i},\bar{z}_{i}\right)}{z_{i}\cdot z_{i+1}}.\label{eq:-49}
\end{equation}

In the subsequent section, we shall employ this identity to reformulate
the celestial leaf amplitudes for gluons in $\mathcal{N}=4$ SYM theory,
and subsequently extend the analysis to gravitons in $\mathcal{N}=8$
Supergravity.

\section{$\mathcal{N}=4$ Supersymmetric Yang-Mills\label{sec:-Supersymmetric-Yang-Mills}}

\subsection{Review}

The starting point of our analysis is the Parke-Taylor formula\footnote{First introduced by \citet{parke1986amplitude} and subsequently provided
with a rigorous derivation by \citet{berends1988recursive} and \citet{kim1997recursion}.
For modern introductions, see \citet{elvang2013scattering,badger2024scattering}.}. Consider a scattering process in four-dimensional Yang-Mills theory
involving $n$ gluons in the MHV configuration $1^{-},2^{-},3^{+},...,n^{+}$.
The corresponding scattering amplitude, denoted $\mathcal{A}_{n}^{a_{1}...a_{n}}$,
is expressed as follows:
\begin{equation}
\mathcal{A}_{n}^{a_{1}...a_{n}}\left(z_{i},\bar{z}_{i},s_{i}\right)\,=\,ig^{n-2}\,\delta^{\left(4\right)}\left(\sum_{i=1}^{n}s_{i}q^{\mu}\left(z_{i},\bar{z}_{i}\right)\right)\,\left\langle \nu_{1}\nu_{2}\right\rangle ^{4}\,\mathsf{Tr}\,\,\,\prod_{i=1}^{n}\frac{\mathsf{T}^{a_{i}}}{\nu_{i}\cdot\nu_{i+1}}.\label{eq:-1}
\end{equation}
Here, $g$ denotes the Yang-Mills coupling constant, $a_{1},...,a_{n}$
are the colour indices associated with the external gluons, and $\mathsf{T}^{a}$
are the generators of the gauge group $\mathbf{G}$. These generators
satisfy the normalisation condition $\mathsf{Tr}(\mathsf{T}^{a}\mathsf{T}^{b})=\frac{1}{2}\boldsymbol{k}^{ab}$,
where $\boldsymbol{k}^{ab}$ denotes the Cartan-Killing form of $\mathbf{G}$,
as well as the Lie algebra commutation relations $[\mathsf{T}^{a},\mathsf{T}^{b}]=if^{abc}\mathsf{T}^{c}$,
where $f^{abc}$ are the structure constants of the Lie algebra $\mathfrak{g}\simeq(T_{e}(\mathbf{G}),[\cdot,\cdot])$.
We also adopt the convention $\nu_{n+1}^{A}\coloneqq\nu_{1}^{A}$,
which ensures cyclic symmetry in the denominator of the Parke-Taylor
formula.

To proceed, we reformulate the amplitude $\mathcal{A}_{n}^{a_{1}...a_{n}}\left(z_{i},\bar{z}_{i},s_{i}\right)$
in terms of the frequencies $s_{i}$ and the normalised spinor basis
$z_{i}^{A}\coloneqq\left(1,z_{i}\right)^{T}$ and $\bar{z}_{i\dot{A}}\coloneqq\left(-\bar{z}_{i},1\right)$,
utilising Eq. (\ref{eq:-1}) along with the integral representation
of the four-dimensional delta-function, 
\begin{equation}
\delta^{\left(4\right)}\left(p\right)=\frac{1}{\left(2\pi\right)^{4}}\,\underset{\mathbf{R}^{4}}{\int}d^{4}x\,\,\,e^{ip\cdot x},\,\,\,\forall\,p^{\mu}\in\mathbf{R}^{4}.
\end{equation}
This yields the expression:
\begin{equation}
\mathcal{A}_{n}^{a_{1}...a_{n}}\left(z_{i},\bar{z}_{i},s_{i}\right)\,=\,\frac{ig^{n-2}}{\left(2\pi\right)^{4}}\,\underset{\mathbf{R}^{4}}{\int}d^{4}x\,\,\,\mathsf{Tr}\,\,\,\prod_{i=1}^{n}\,s_{i}^{e_{i}}\,\exp\left(i\,s_{i}\,\langle z_{i}|x|\bar{z}_{i}]\right)\,\frac{\mathsf{T}^{a_{i}}}{z_{i}\cdot z_{i+1}}.\label{eq:-48}
\end{equation}

\subsubsection{Grassmann-valued Spinors; Berezin-de Witt Integral}

The transition to $\mathcal{N}=4$ SYM theory is achieved via the
on-shell superfield formalism\footnote{We follow the techniques developed by \citet{grisaru1977some}, \citet{brink1977supersymmetric}
and \citet{ferber1977supertwistors}, as reviewed by \citet{wess2020supersymmetry}
and \citet{elvang2013scattering}.}. To this end, we introduce Grassmann-valued two-component spinors
$\theta_{A}^{\alpha}$, where $1\leq\alpha\leq4$ indexes the supersymmetry
generators. These variables satisfy the normalisation condition $\int d^{0|2}\theta\,\,\,\theta_{A}^{\alpha}\theta_{B}^{\alpha}=\varepsilon_{AB},$
where $d^{0|2}\theta$ denotes the Berezin-de Witt integration measure\footnote{Introduced by \citet{berezin2013introduction} and \citet{dewitt1992supermanifolds}.
For a modern and rigorous mathematical exposition, see \citet{manin1997introduction}.} over the fermionic variables. Then, defining the Grassmann-valued
coefficients $\xi_{i}^{\alpha}\coloneqq z_{i}^{A}\theta_{A}^{\alpha}$,
we note the identity:
\begin{equation}
(z_{i}\cdot z_{j})^{4}=\int d^{0|8}\theta\,\,\,\prod_{\alpha=1}^{4}\,\xi_{i}^{\alpha}\xi_{j}^{\alpha},\,\,\,\text{for all}\,1\leq i,j\leq n.
\end{equation}

In addition, we must specify both the integration measure and the
domain over which the super-amplitudes are defined. Thus, let $\mathbf{R}^{4|8}$
denote the supersymmetric extension of real four-dimensional Euclidean
space, augmented with the Grassmann-valued coordinates $\theta_{A}^{\alpha}$
$\left(a=1,...,8\right)$. The Berezin-de Witt integration measure
on $\mathbf{R}^{4|8}$ is defined as $d^{4|8}x\coloneqq d^{4}x\wedge d^{0|8}\theta$.

Therefore, the amplitude $\mathcal{A}_{n}^{a_{1}...a_{n}}$ obtained
in Eq. (\ref{eq:-48}) can be reformulated as:
\begin{equation}
\mathcal{A}_{n}^{a_{1}...a_{n}}\left(z_{i},\bar{z}_{i},s_{i}\right)\,=\,\frac{ig^{n-2}}{\left(2\pi\right)^{4}}\,\underset{\mathbf{R}^{4|8}}{\int}d^{4|8}x\,\prod_{\alpha=1}^{4}\,\xi_{1}^{\alpha}\xi_{2}^{\alpha}\,\,\,\mathsf{Tr}\,\,\,\prod_{i=1}^{n}\,s_{i}^{e_{i}}\,\exp\left(i\,s_{i}\,\langle z_{i}|x|\bar{z}_{i}]\right)\,\frac{\mathsf{T}^{a_{i}}}{z_{i}\cdot z_{i+1}}.
\end{equation}

\subsubsection{$\mathcal{N}=4$ SYM Super-amplitude}

To extend this to \emph{supersymmetric }amplitudes, we define the
superfield encoding the particle spectrum of $\mathcal{N}=4$ SYM
theory:
\begin{equation}
\varphi\left(\xi^{\alpha}\right)\,\coloneqq\,\mathsf{a}^{-}+\xi^{\alpha}\,\lambda_{\alpha}+\frac{1}{2!}\,\xi^{\alpha}\xi^{\beta}\,\,\,\phi_{\alpha\beta}+\frac{1}{3!}\,\xi^{\alpha}\xi^{\beta}\xi^{\gamma}\varepsilon_{\alpha\beta\gamma\delta}\,\,\,\eta^{\delta}+\mathsf{a}^{+}\,\prod_{\alpha=1}^{4}\xi^{\alpha}.
\end{equation}
Here, $\mathsf{a}_{i}^{\pm}$ represent the classical expectation
values associated with the annihilation operators for gluons of positive
and negative helicities.

Thus, the gluonic super-amplitude in $\mathcal{N}=4$ SYM theory can
be expressed as:
\begin{equation}
\mathscr{A}_{n}^{a_{1}...a_{n}}\left(z_{i},\bar{z}_{i},s_{i}\right)=\frac{ig^{n-2}}{\left(2\pi\right)^{4}}\,\underset{\mathbf{R}^{4|8}}{\int}d^{4|8}x\,\,\,\mathsf{Tr}\,\,\,\prod_{i=1}^{n}\,s_{i}^{e_{i}}\,\varphi\left(z_{i}\cdot\theta^{\alpha}\right)\,\exp\left(i\,s_{i}\,\langle z_{i}|x|\bar{z}_{i}]\right)\,\frac{\mathsf{T}^{a_{i}}}{z_{i}\cdot z_{i+1}}.\label{eq:-2}
\end{equation}

\subsection{Minitwistor Amplitudes for Gluons}

We are now positioned to derive the main result of this section. We
shall demonstrate that the celestial leaf amplitude for gluons in
$\mathcal{N}=4$ SYM theory admits a representation as an integral
over the moduli space of minitwistor lines in the non-singular quadric
$\mathbf{MT}$.

\textcompwordmark{}

\subsubsection{Mellin Transform and Leaf Amplitudes}

We begin by recalling the definition of the celestial amplitude\footnote{Cf. \citet{pasterski2017conformal,pasterski2017gluon,pasterski2017flat,arkani2021celestial,banerjee2020modified,banerjee2021mhv}.
For recent pedagogical reviews, see \citet{pasterski2021lectures,raclariu2021lectures,strominger2018lectures,aneesh2022celestial,pasterski2023chapter}.} as the $\varepsilon$-regulated Mellin transform:
\begin{equation}
\widehat{\mathscr{A}}_{n}^{a_{1}...a_{n}}\left(z_{i},\bar{z}_{i},\Delta_{i}\right)\,\,\,\coloneqq\,\,\,\prod_{i=1}^{n}\,\,\,\underset{\mathbf{R}_{+}^{\times}}{\int}\,\,\,\frac{ds_{i}}{s_{i}}\,s_{i}^{\Delta i}\,e^{-\varepsilon s_{i}}\,\,\,\mathscr{A}_{n}^{a_{1}...a_{n}}\left(z_{i},\bar{z}_{i},s_{i}\right),\label{eq:-12}
\end{equation}
where $\mathbf{R}_{\times}^{+}\coloneqq(\mathbf{R}^{+},\cdot)$ denotes
the multiplicative group of positive real numbers, and $\frac{ds_{i}}{s_{i}}$
the Haar measure on $\mathbf{R}_{\times}^{+}$.

Substituting the expression for the gluon super-amplitude, obtained
in Eq. (\ref{eq:-2}), into Definition (\ref{eq:-12}) yields:
\begin{equation}
\widehat{\mathscr{A}}_{n}^{a_{1}...a_{n}}\left(z_{i},\bar{z}_{i},\Delta_{i}\right)\,=\,\frac{ig^{n-2}}{\left(2\pi\right)^{4}}\,\,\,\underset{\mathbf{R}^{4|8}}{\int}d^{4|8}x\,\,\,\mathsf{Tr}\,\,\,\prod_{i=1}^{n}\,\boldsymbol{\varphi}_{2h_{i}}\left(z_{i}\cdot\theta^{\alpha}\right)\,\phi_{2h_{i}}\left(x;z_{i},\bar{z}_{i}\right)\,\frac{\mathsf{T}^{a_{i}}}{z_{i}\cdot z_{i+1}}.\label{eq:-14}
\end{equation}
Here, $\phi_{\Delta}\left(x;z,\bar{z}\right)$ represents the celestial
wavefunction\footnote{See \citet{pasterski2017conformal} and our Appendix A.}
for massless scalars, which is defined by:
\begin{equation}
\phi_{\Delta}\left(x;z,\bar{z}\right)\coloneqq\frac{\mathcal{C}\left(\Delta\right)}{\left(i\varepsilon+\langle z|x|\bar{z}]\right)^{\Delta}},\,\,\,\mathcal{C}\left(\Delta\right)\coloneqq i^{\Delta}\Gamma\left(\Delta\right).\label{eq:-46}
\end{equation}
Additionally, $\boldsymbol{\varphi}_{\Delta}\left(\xi\right)$ is
the superfield expansion for the Mellin-transformed mode coefficients:
\begin{equation}
\boldsymbol{\varphi}_{\Delta}\left(\xi^{\alpha}\right)\,=\,\widehat{\mathsf{a}}_{\Delta}^{-}\,+\,\xi^{\alpha}\,\widehat{\lambda}_{\Delta,\alpha}+\frac{1}{2!}\,\xi^{\alpha}\xi^{\beta}\,\,\,\widehat{\phi}_{\Delta,\alpha\beta}\,+\,\frac{1}{3!}\,\xi^{\alpha}\xi^{\beta}\xi^{\gamma}\,\varepsilon_{\alpha\beta\gamma\delta}\,\,\,\widehat{\eta}_{\Delta}^{\delta}\,+\,\widehat{\mathsf{a}}_{\Delta}^{+}\,\,\,\prod_{\alpha=1}^{4}\xi^{\alpha}.\label{eq:-52}
\end{equation}
We note that the mode functions $\widehat{\mathsf{a}}_{\Delta}^{\pm}=\widehat{\mathsf{a}}_{\Delta}^{\pm}\left(z,\bar{z}\right)$,
$\widehat{\lambda}_{\Delta,\alpha}=\widehat{\lambda}_{\Delta,\alpha}\left(z,\bar{z}\right)$
etc. depends on the normalised spinor basis $\{z_{i}^{A},\bar{z}_{i\dot{A}}\}$,
parametrising the insertion points of the gluons on the celestial
sphere, and the conformal weights $\Delta_{i}$.

\paragraph*{Leaf Representation.}

We now proceed to implement the formalism of celestial leaf amplitudes\footnote{See Appendix \ref{sec:Leaf-Amplitudes-Review} or \citet{melton2023celestial}.},
requiring a specification of both the integration measure and the
domains over which the gluonic amplitudes will be constructed. To
that end, we recall that, within the embedding-space formalism, spacetime
points are parametrised by homogeneous coordinates $X_{A\dot{A}}$,
where the real three-dimensional Euclidean hyperboloid $H_{3}^{+}$
is realised as the projective space $\mathbf{RP}^{3}$. The geometry
of the latter is given by the projectively invariant line-element
$ds^{2}=\boldsymbol{g}_{A\dot{A}B\dot{B}}dX^{A\dot{A}}\otimes dX^{B\dot{B}}$,
as defined in Eq. (\ref{eq:-21}). The orientation of $\mathbf{RP}^{3}$
is induced by the natural volume form\footnote{As defined in \citet{gelfandgeneralized,gel_fand2003selected}.}:
\begin{equation}
D^{3}X\coloneqq\varepsilon_{A\dot{A}}\,\varepsilon_{B\dot{B}}\,\varepsilon_{C\dot{C}}\,\varepsilon_{D\dot{D}}\,\,\,X^{A\dot{A}}\,dX^{B\dot{B}}\wedge dX^{C\dot{C}}\wedge dX^{D\dot{D}}.
\end{equation}

Extending this construction to the supersymmetric case requires promoting
the bosonic coordinates $X_{A\dot{A}}$ to their supersymmetric counterparts
$\mathbb{X}^{\hat{I}}\coloneqq(X_{A\dot{A}},\theta_{A}^{\alpha})$,
where $\theta_{A}^{\alpha}$ $\left(1\leq\alpha\leq4\right)$ are
Grassmann-valued spinorial coordinates parametrising the ``fermionic
dimensions.'' The corresponding Berezin-de Witt integration measure
on the projective superspace $\mathbf{RP}^{3|8}$ is then given by:
\begin{equation}
D^{3|8}\mathbb{X}\,\coloneqq\,\frac{D^{3}X}{\left|X\right|^{4}}\,\wedge\,d^{0|8}\theta.
\end{equation}

With these preliminaries established, the celestial superamplitude
in the leaf representation takes the form:
\begin{equation}
\widehat{\mathscr{A}}_{n}^{a_{1}...a_{n}}\left(z_{i},\bar{z}_{i},\Delta_{i}\right)=\frac{ig^{n-2}}{\left(2\pi\right)^{3}}\,\delta\left(\beta\right)\,\times\,A_{n}^{a_{1}...a_{n}}\left(z_{i},\bar{z}_{i},\Delta_{i}\right)+\big(\bar{z}_{i}\rightarrow i\bar{z}_{i}\big),
\end{equation}
where $\beta\coloneqq4-2\sum_{i=1}^{n}h_{i}$ enforces the scaling
constraint, and the symbol $(\bar{z}_{i}\rightarrow i\bar{z}_{i})$
signifies the repetition of the first term under the implied replacement.
The function $A_{n}^{a_{1}...a_{n}}\left(z_{i},\bar{z}_{i},\Delta_{i}\right)$
represents the gluonic leaf super-amplitude, which is defined as the
following supersymmetric integral over $\mathbf{RP}^{3|8}$:
\begin{equation}
A_{n}^{a_{1}...a_{n}}\left(z_{i},\bar{z}_{i},\Delta_{i}\right)\,=\underset{\mathbf{RP}^{3|8}}{\int}D^{3|8}\mathbb{X}\,\,\,\mathsf{Tr}\,\,\,\prod_{i=1}^{n}\,\boldsymbol{\varphi}\left(z_{i}\cdot\theta^{\alpha}\right)\,K_{2h_{i}}\left(X;z_{i},\bar{z}_{i}\right)\,\frac{\mathsf{T}^{a_{i}}}{z_{i}\cdot z_{i+1}}.\label{eq:-10}
\end{equation}
The kernel $K_{\Delta}\left(X;z,\bar{z}\right)$ is the bulk-to-boundary
Green's function for the covariant Laplacian $\square_{\mathbf{H}_{3}}$
with conformal weight $\Delta$, given in homogeneous coordinates
$X^{A\dot{A}}\in\mathbf{H}_{3}$ by: 
\begin{equation}
K_{\Delta}\left(X;z,\bar{z}\right)=\mathcal{C}\left(\Delta\right)\frac{\left|X\right|^{\Delta}}{\langle z|X|\bar{z}]^{\Delta}}.
\end{equation}
For a concise review of leaf amplitudes, we refer the reader to Appendix
\ref{sec:Leaf-Amplitudes-Review}. For further mathematical details,
see \citet{gelfandgeneralized}, \citet{teschner1999mini} and \citet{costa2014spinning,penedones2017tasi}.

\subsubsection{Celestial RSVW Formalism}

Now we reformulate Eq. (\ref{eq:-10}) to derive the \emph{celestial
RSVW amplitude}. First, the procedure requires the resolution of the
identity for the $\mathcal{N}=4$ superfield $\boldsymbol{\varphi}$:
\begin{equation}
\boldsymbol{\varphi}\left(z_{r}\cdot\theta^{\alpha}\right)=\underset{\mathbf{CP}^{0|4}}{\int}d^{0|4}\xi_{r}\,\,\,\delta^{\left(4\right)}\left(\xi_{r}^{\alpha}-z_{r}^{A}\theta_{A}^{\alpha}\right)\,\boldsymbol{\varphi}\left(\xi_{i}^{\alpha}\right).
\end{equation}
This identity allows us to recast the leaf amplitude $A_{n}^{a_{1}...a_{n}}\left(z_{i},\bar{z}_{i},\Delta_{i}\right)$
as follows:
\begin{equation}
A_{n}^{a_{1}...a_{n}}\,=\prod_{r=1}^{n}\underset{\mathbf{CP}^{0|4}}{\int}d^{0|4}\xi_{r}\underset{\mathbf{RP}^{3|8}}{\int}D^{3|8}\mathbb{X}\,\,\,\mathsf{Tr}\,\,\,\prod_{i=1}^{n}\delta^{\left(4\right)}\left(\xi_{i}^{\alpha}-z_{i}^{A}\theta_{A}^{\alpha}\right)\,\boldsymbol{\varphi}\left(\xi_{i}^{\alpha}\right)\,K_{2h_{i}}\left(X;z_{i},\bar{z}_{i}\right)\,\frac{\mathsf{T}^{a_{i}}}{z_{i}\cdot z_{i+1}}\label{eq:-50}
\end{equation}

Our subsequent task is to define the integration measure over which
the gluonic leaf amplitude $A_{n}^{a_{1}...a_{n}}\left(z_{i},\bar{z}_{i},\Delta_{i}\right)$
will be expressed as an integral transform. As will be explained in
Section \ref{sec:Minitwistor-String-Theories}, the minitwistor superspace\footnote{Cf. \citet{samann2009mini}. }
$\mathbf{MT}^{2|4}$ may be identified with the trivial vector superbundle\footnote{See \citet{rogers2007supermanifolds}.}
$\mathbf{MT}^{2|4}\simeq\mathbf{MT}\times\mathbf{CP}^{0|4}$, where
the base manifold $\mathbf{MT}$ corresponds to the bosonic minitwistor
space, while the typical fibres are isomorphic to the $\mathbf{Z}_{2}$-graded
vector space spanned by the Grassmann-odd projective coordinates $\xi^{\alpha}$,
with $\alpha=1,...,4$.

The vector superbundle $\mathbf{MT}^{2|4}$ admits a local trivialisation
in terms of homogeneous coordinates:
\begin{equation}
\mathsf{Z}^{I}\,\coloneqq\,\left(\lambda^{A},\mu_{\dot{A}},\xi^{\alpha}\right)\in\mathbf{CP}^{1|0}\times\mathbf{CP}^{1|0}\times\mathbf{CP}^{0|4}.
\end{equation}
The natural orientation of the total space $\mathbf{MT}^{2|4}$ is
induced by the Berezin-de Witt volume superform, given by:
\begin{equation}
D^{2|4}\mathsf{Z}\,\coloneqq\,D\lambda\,\wedge\,D\mu\,\wedge\,d^{0|4}\xi.
\end{equation}

Incorporating the celestial RSVW identity established in Eq. (\ref{eq:-49}),
the gluonic leaf amplitude $A_{n}^{a_{1}...a_{n}}\left(z_{i},\bar{z}_{i},\Delta_{i}\right)$
admis the following representation in terms of an integral transform
over minitwistor superspace:
\begin{align}
 & A_{n}^{a_{1}...a_{n}}\left(z_{i},\bar{z}_{i},\Delta_{i}\right)=\prod_{r=1}^{n}\underset{\mathbf{MT}^{2|4}}{\int}D^{2|4}\mathsf{Z}_{i}\,\,\,\boldsymbol{\varphi}_{2h_{i}}\left(\xi_{i}^{\alpha}\right)\,\widetilde{\mathcal{F}}_{2h_{i}}\left(\lambda_{i},\mu_{i};z_{i},\bar{z}_{i}\right)\,\underset{\mathbf{RP}^{3|8}}{\int}D^{3|8}\mathbb{X}\\
 & \mathsf{Tr}\,\,\,\prod_{i=1}^{n}\,\,\,\overline{\delta}_{2h_{i}}\left(\mu_{i\dot{A}},\lambda_{i}^{A}\frac{X_{A\dot{A}}}{\left|X\right|}\right)\,\delta^{\left(4\right)}\left(\xi_{i}^{\alpha}-\lambda_{i}^{A}\theta_{A}^{\alpha}\right)\,\,\,\frac{\mathsf{T}^{a_{i}}}{\lambda_{i}\cdot\lambda_{i+1}}
\end{align}

Following the key insight of \citet{witten2004perturbative,edward2004parity},
we identify from the preceding expression the \emph{minitwistor gluon
wavefunction }with \emph{celestial conformal weight }$\Delta$ as:
\begin{equation}
\Phi_{\Delta}\left(\mathsf{Z}^{I};z^{A},\bar{z}_{\dot{A}}\right)\,\coloneqq\,\boldsymbol{\varphi}_{\Delta}\left(\xi^{\alpha}\right)\,\widetilde{\mathcal{F}}_{\Delta}\left(\lambda^{A},\mu_{\dot{A}};z^{A},\bar{z}_{\dot{A}}\right).
\end{equation}
We thus conclude that the leaf amplitude can be interpreted as a Fourier-transform
over $\mathbf{MT}^{2|4}$,
\begin{equation}
A_{n}^{a_{1}...a_{n}}\left(z_{i},\bar{z}_{i},\Delta_{i}\right)\,=\,\underset{\left(\mathbf{MT}^{2|4}\right)^{\times n}}{\int}\,\bigwedge_{i=1}^{n}\,D^{2|4}\mathsf{Z}_{i}\,\,\,\Phi_{2h_{i}}\left(\mathsf{Z}_{i};z_{i},\bar{z}_{i}\right)\,\widetilde{A}_{n}^{a_{1}...a_{n}}\left(\mathsf{Z}_{i}\right),
\end{equation}
where the Fourier-transformed amplitude is:
\begin{equation}
\widetilde{A}_{n}^{a_{1}...a_{n}}\left(\mathsf{Z}_{i}^{I}\right)\,=\,\underset{\mathbf{RP}^{3|8}}{\int}\,D^{3|8}\mathbb{X}\,\,\,\mathsf{Tr}\,\,\,\prod_{i=1}^{n}\,\overline{\delta}_{2h_{i}}\left(\mu_{i\dot{A}},\lambda_{i}^{A}\frac{X_{A\dot{A}}}{\left|X\right|}\right)\,\delta^{\left(4\right)}\left(\xi_{i}^{\alpha}-\lambda_{i}^{A}\theta_{A}^{\alpha}\right)\,\,\,\frac{\mathsf{T}^{a_{i}}}{\lambda_{i}\cdot\lambda_{i+1}}.\label{eq:-51}
\end{equation}

Finally we can formulate an important geometrical and physical interpretation
of the result obtained. The (bosonic) spinor delta-functions appearing
in the integrand serve to localise the integration over $X$ to the
locus of incidence curves $\mathcal{L}\left(X\right)$, which, as
previously discussed, correspond to minitwistor lines. These incidence
curves are precisely the conic curves in $\mathbf{MT}$ that are associated
with a fixed point $X^{A\dot{A}}\in\mathbf{H}_{3}$ on the hyperboloid,
through the incidence relation $\mu_{\dot{A}}=\lambda^{A}X_{A\dot{A}}$.
\emph{The gluonic leaf amplitudes are identically null in all instances
wherein the insertion points of the gluons fail to lie on a curve
satisfying the defining incidence relation of a minitwistor line.}

Therefore, the leaf amplitudes acquires an elegant geometrical interpretation.
\emph{It can be viewed as an integral over the moduli space of minitwistor
lines $\mathcal{L}\left(X\right)$ within the supersymmetric, non-singular
quadric $\mathbf{MT}^{2|4}$. }However, it must be stressed that this
integral cannot be interpreted as an ordinary volume form over the
moduli space owing to the scaling dimensions associated with the delta
functions $\overline{\delta}_{2h_{i}}$. It would be interesting to
find an interpretation for Eq. (\ref{eq:-51}) in the formalism developed
by \citet{movshev2006berezinian} and \citet{adamo2013moduli}.

\subsection{Generating Functional\label{subsec:Generating-Functional}}

The geometric reformulation of gluonic leaf amplitudes enables the
construction of a generating functional by invoking the Quillen determinant
line bundle\footnote{Introduced by \citet{quillen1985determinants} and further developed
by \citet{biswas1999determinant,brylinski1999geometric}. For a review
emphasising physical applications, see \citet{freed1987determinant}.} and the celestial BMS identity.

As we shall explain in Section \ref{sec:Minitwistor-String-Theories},
to each point in projective superspace $\mathbb{X}^{\hat{I}}=\left(X_{A\dot{A}},\theta_{A}^{\alpha}\right)\in\mathbf{RP}^{3|8}$,
there correspond a \emph{minitwistor superline} embedded in $\mathbf{MT}^{2|4}$,
defined by the locus of incidence:
\begin{equation}
\mathcal{L}\left(X,\theta\right)\,\coloneqq\,\left\{ \,\big(\lambda^{A},\mu_{\dot{A}},\xi^{\alpha}\big)\in\mathbf{MT}^{2|4}\,\Big|\,\mu_{\dot{A}}=\lambda^{A}\frac{X_{A\dot{A}}}{\left|X\right|},\,\xi^{\alpha}=\lambda^{A}\theta_{A}^{\alpha}\,\right\} .
\end{equation}
In what follows, we shall adopt the conventional notation for the
restriction homomorphism associated with cohomological classes on
$\mathbf{MT}^{2|4}$. For example, let:
\begin{equation}
[g]\in H^{0,1}\big(\mathbf{MT}^{2|4},\mathcal{O}\left(-2,-\Delta\right)\big),
\end{equation}
be a representative of the Dolbeault cohomology class. The restriction
of any such representative to the supercurve $\mathcal{L}\left(X,\theta\right)$
will be denoted by:
\begin{equation}
g\,\big|_{\mathcal{L}\left(X,\theta\right)}\,\left(\lambda^{A}\right)\,\coloneqq\,\rho_{\mathcal{L}\left(X,\theta\right)}(g)\left(\lambda^{A}\right)\,=\,g\left(\lambda^{A},\lambda^{A}\frac{X_{A\dot{A}}}{\left|X\right|},\lambda^{A}\theta_{A}^{\alpha}\right).
\end{equation}

We introduce the \emph{minitwistor background potential }$\boldsymbol{\omega}$
as a $\left(0,1\right)$-form on the holomorphic vector superbundle
$\mathcal{O}\left(-2,-\Delta\right)\longrightarrow\mathbf{MT}^{2|4}$,
taking values in the Lie algebra $\mathfrak{g}$ associated with the
gauge group $\mathbf{G}$. Formally, this background potential is
an element of the space:
\begin{equation}
\boldsymbol{\omega}\in\mathfrak{g}\otimes\Omega^{0,1}\big(\mathbf{MT}^{2|4},\mathcal{O}\left(-2,-\Delta\right)\big).
\end{equation}
Our objective is now to demonstrate that:
\begin{equation}
\digamma\left[\boldsymbol{\omega}\right]\,\coloneqq\,\underset{\mathbf{RP}^{3|8}}{\int}D^{3|8}\mathbb{X}\,\,\,\log\det\left(\overline{\partial}\,+\boldsymbol{\omega}\right)\big|_{\mathcal{L}\left(X,\theta\right)},
\end{equation}
serves as a \emph{generating functional }for the gluonic leaf amplitudes
$A_{n}^{a_{1}...a_{n}}\left(z_{i},\bar{z}_{i},\Delta_{i}\right)$,
when evaluated at the specific background configuration:
\[
\boldsymbol{\omega}\left(\lambda^{A},\mu_{\dot{A}},\xi^{\alpha};z^{A},\bar{z}_{\dot{A}}\right)\,=\,\mathsf{T}^{a}\underset{\mathcal{P}\times\mathbf{C}^{2}}{\int}d\Delta\wedge dz\wedge d\bar{z}\,\,\,\boldsymbol{\varphi}_{\Delta}\left(\xi^{\alpha}\right)\mathcal{F}_{\Delta}\left(\lambda^{A},\mu_{\dot{A}};z^{A},\bar{z}_{\dot{A}}\right).
\]
In this expression, $\overline{\partial}\,\big|_{\mathcal{L}\left(X,\theta\right)}$
denotes the restriction of the Dolbeault operator to the minitwistor
superline $\mathcal{L}\left(X,\theta\right)$, $\mathsf{T}^{a}$ represents
a generator of the Lie algebra $\mathfrak{g}$, the domain of integration
$\mathcal{P}\coloneqq1+i\mathbf{R}$ corresponds to the continuous
principal series, where the celestial conformal weight $\Delta$ resides,
$\boldsymbol{\varphi}_{\Delta}\left(\xi^{\alpha}\right)$ is the Mellin-transformed
$\mathcal{N}=4$ superfield introduced in Eq. (\ref{eq:-52}), and
$\mathcal{F}_{\Delta}\in\Omega^{0,1}\left(\mathbf{MT},\mathcal{O}\left(-2,-\Delta\right)\right)$
the minitwistor wavefunction defined in Eq. (\ref{eq:-22}).

The first step in our derivation consists in observing that the Penrose
integral-geometric transform of the minitwistor background potential
$\boldsymbol{\omega}$ is expressed as:
\begin{equation}
\underset{\mathcal{L}\left(X,\theta\right)}{\int}\,\boldsymbol{\omega}\,\big|_{\mathcal{L}\left(X,\theta\right)}\left(\lambda\right)\,=\,\mathsf{T}^{a}\underset{\mathcal{P}\times\mathbf{C}^{2}}{\int}d\Delta\wedge dz\wedge d\bar{z}\,\boldsymbol{\varphi}_{\Delta}\left(\xi^{\alpha}\right)\,K_{\Delta}\left(X;z,\bar{z}\right),
\end{equation}
where $K_{\Delta}\left(X;z,\bar{z}\right)$ denotes the bulk-to-boundary
propagator on the hyperboloid $H_{3}^{+}$ expressed in terms of homogeneous
coordinates $X_{A\dot{A}}$ that chart real projective superspace
$\mathbf{RP}^{3}$. Accordingly, we can apply the celestial BMS identity
established in Eq. (\ref{eq:-18}) to $\boldsymbol{\omega}$.

To proceed, we employ the expansion of the Quillen determinant, as
explained by \citet{boels2007twistor}.\footnote{See also \citet{mason2010complete,mason2010gravity,mason2010scattering,bullimore2010mhv}}
Upon applying the celestial BMS identity, we find that the generating
functional admits the expansion:
\begin{align}
 & \digamma[\boldsymbol{\omega}]=\sum_{m\geq2}\frac{\left(-1\right)^{m+1}}{m}\underset{\mathbf{RP}^{3|8}}{\int}D^{3|8}\mathbb{X}\,\,\,\mathsf{Tr}\underset{(\mathcal{L}\left(X,\theta\right))^{\times n}}{\int}\,\bigwedge_{i=1}^{m}D\lambda_{i}\,\,\,\boldsymbol{\omega}\,\big|_{\mathcal{L}\left(X,\theta\right)}\,\,\,\frac{1}{\lambda_{i}\cdot\lambda_{i+1}}\\
 & =\sum_{m\geq2}\frac{\left(-1\right)^{m+1}}{m}\underset{\left(\mathcal{P}\times\mathbf{C}^{2}\right)^{\times m}}{\int}\,\bigwedge_{r=1}^{m}\,\boldsymbol{\beta}_{r}\,\,\,\underset{\mathbf{RP}^{3|8}}{\int}\,D^{3|8}\mathbb{X}\,\,\,\mathsf{Tr}\,\,\,\prod_{i=1}^{n}\boldsymbol{\varphi}_{\Delta_{i}}\left(\xi_{i}^{\alpha}\right)\,K_{\Delta_{i}}\left(X;z_{i},\bar{z}_{i}\right)\,\,\,\frac{\mathsf{T}^{a_{i}}}{z_{i}\cdot z_{i+1}},
\end{align}
where:
\begin{equation}
\boldsymbol{\beta}_{r}\,\coloneqq\,d\Delta_{r}\wedge dz_{r}\wedge d\bar{z}_{r}.
\end{equation}
To complete the derivation, we perform functional differentiation
of $\digamma\left[\boldsymbol{\omega}\right]$ with respect to the
classical expectation values of the annihilation operators $\mathsf{a}_{\Delta_{i}}^{\ell_{i}}\left(z_{i},\bar{z}_{i}\right)$,
which correspond to gluon states with helicity $\ell_{i}$ and celestial
conformal weight $\Delta_{i}$,
\begin{equation}
\frac{\delta}{\delta\mathsf{a}_{2h_{i}}^{\ell_{1}}\left(z_{i}\right)}...\frac{\delta}{\delta\mathsf{a}_{\Delta_{n}}^{\ell_{n}}\left(z_{n},\bar{z}_{n}\right)}\digamma\left[\boldsymbol{\omega}\right]\Bigg|_{\boldsymbol{\varphi}=0}=\underset{\mathbf{RP}^{3|8}}{\int}D^{3|8}\mathbb{X}\,\,\,\mathsf{Tr}\,\,\,\prod_{i=1}^{n}\,\boldsymbol{\varphi}\left(z_{i}\cdot\theta^{\alpha}\right)\,K_{2h_{i}}\left(X;z_{i},\bar{z}_{i}\right)\,\frac{\mathsf{T}^{a_{i}}}{z_{i}\cdot z_{i+1}}.
\end{equation}
The resulting expression is immediately recognised as the gluonic
leaf amplitude $A_{n}^{a_{1}...a_{n}}\left(z_{i},\bar{z}_{i},\Delta_{i}\right)$.
We conclude that the formalism constructed in this subsection in terms
of the minitwistor background potential $\boldsymbol{\omega}$ provides
a generating functional for tree-level MHV scattering amplitudes.

\paragraph*{Geometrical Interpretation.}

Let $\pi_{\left(X,\theta\right)}:\mathcal{L}\left(X,\theta\right)\longrightarrow\mathbf{CP}^{1}$
be the canonical projection. The embedding of the celestial sphere,
modelled by the holomorphic Riemann sphere, into minitwistor superspace
$\mathbf{MT}^{2|4}$ as a minitwistor superline $\mathcal{L}\left(X,\theta\right)$
can be interpreted as a section $\sigma_{\left(X,\theta\right)}$
of the fibration. In fact, by trivialising the fibration $\pi_{\left(X,\theta\right)}$
using homogeneous coordinates $\lambda^{A}$, we define the map:
\begin{equation}
\sigma:\mathbf{CP}^{1}\longrightarrow\mathcal{L}\left(X,\theta\right),
\end{equation}
which acts as:
\begin{equation}
\sigma:\lambda^{A}\mapsto\left(\lambda^{A},\lambda^{A}\frac{X_{A\dot{A}}}{\left|X\right|},\lambda^{A}\theta_{A}^{\alpha}\right).
\end{equation}
It is obvious that this mapping satisfies the section condition $\pi_{\left(X,\theta\right)}\circ\sigma_{\left(X,\theta\right)}=id_{\mathbf{CP}^{1}}$
by construction. However, $\sigma_{\left(X,\theta\right)}$ is also
an embedding of the celestial sphere into the Hitchin-special supercurve
$\mathcal{L}\left(X,\theta\right)\subset\mathbf{MT}^{2|4}$.

This simple observation has an important implication for the interpretation
of the generating functional $\digamma\left[\boldsymbol{\omega}\right]$,
which involves an integral of the form:
\begin{equation}
\underset{\mathbf{RP}^{3|8}}{\int}D^{3|8}\mathbb{X}\,\,\,\underset{\mathcal{L}\left(X,\theta\right)}{\int}(...)\,\,\,.\label{eq:-53}
\end{equation}
A careful analysis shows that this expression \emph{should not} be
interpreted as integrating over celestial \emph{spheres} at each point
$\mathbb{X}^{\hat{I}}\in\mathbf{RP}^{3|8}$. Instead, the correct
interpretation is that the integral is taken over the space of all
\emph{embeddings} of the celestial sphere into the minitwistor space
$\mathbf{MT}^{2|4}$. In other words, (\ref{eq:-53}) should be regarded
as an integral over the moduli space of embeddings of the celestial
sphere. This interpretation has important consequences for the construction
of the on-shell effective action of our sigma model.

\section{$\mathcal{N}=8$ Supergravity\label{sec:-Supergravity}}

\subsection{Review}

Our analysis begins with the Berends-Giele-Kuijf (BGK) formula\footnote{Originally derived in \citet{berends1988relations}.},
which describes the tree-level gravitational scattering amplitudes
for configurations characterised by maximal helicity violation. Consider
a scattering process involving $n$ gravitons in the MHV configuration
$1^{--},2^{--},3^{++},...,n^{++}$, for which the amplitude is expressed
as:
\begin{equation}
\mathcal{M}_{n}\left(z_{i},\overline{z}_{i},s_{i}\right)\,=\,\left(\frac{\kappa}{2}\right)^{n-2}\,\delta^{\left(4\right)}\left(\sum_{i=1}^{n}\nu_{i}^{A}\bar{\nu}_{i}^{\dot{A}}\right)\,\,\,BGK_{n},
\end{equation}
where the $BGK_{n}$ factor is defined by:
\begin{equation}
BGK_{n}\coloneqq\frac{\left\langle \nu_{1}\nu_{2}\right\rangle ^{8}}{\prod_{r=1}^{n}\nu_{r}\cdot\nu_{r+1}}\frac{1}{\left\langle \nu_{n}\nu_{1}\right\rangle \left\langle \nu_{1}\nu_{n-1}\right\rangle \left\langle \nu_{n-1}\nu_{n}\right\rangle }\prod_{i=2}^{n-2}\frac{[\bar{\nu}_{i}|p_{i+1}+...+p_{n-1}|\nu_{n}\rangle}{\nu_{i}\cdot\nu_{n}}.
\end{equation}
A specially enlightening modern derivation of this result, using Plebanski's
second heavenly equation, was given by \citet{miller2024proof}.

Our objective is to apply the formalism of celestial leaf amplitudes
to $\mathcal{M}_{n}$. Using the minitwistor wavefunctions developed
in Subsection (\ref{subsec:Minitwistor-Wavefunctions}), we seek to
recast the modified Mellin transform of $\mathcal{M}_{n}$ into a
form that admits an interpretation as an integral over a moduli space
of minitwistor lines. 

\subsubsection{$\mathbf{CP}^{1}$ Fermionic Doublet}

The factorisation procedure to be employed in this section is based
on the method developed by \citet{nair2005note}. In this formalism,
tree-level MHV gravitational scattering amplitudes are expressed in
terms of a correlator of ``vertex operators.'' These operators are
constructed with the aid of an auxiliary fermionic doublet $(\hat{\boldsymbol{\chi}},\hat{\boldsymbol{\chi}}^{\dagger})$
defined on $\mathbf{CP}^{1}$, and their explicit forms are determined
by the following mode expansions:
\begin{equation}
\hat{\boldsymbol{\chi}}\left(z\right)\,\coloneqq\,\sum_{p\geq0}\,\frac{\mathsf{b}_{p}}{z^{1+p}},\,\,\,\hat{\boldsymbol{\chi}}^{\dagger}\left(z\right)\,\coloneqq\,\sum_{p\geq0}\,z^{p}\,\mathsf{b}_{p}^{\dagger}.
\end{equation}
Here, the fermionic annihilation and creation operators $\mathsf{b}_{p}$
and $\mathsf{b}_{p}^{\dagger}$ satisfy the anti-commutation relations
$\{\mathsf{b}_{p},\mathsf{b}_{q}^{\dagger}\}=\delta_{pq}$ for all
$p,q\geq0$, and their action on the vacuum state $|0\rangle$ is
given by $\mathsf{b}_{p}|0\rangle=0$ for all $p\geq0$. 

For notational convenience, we define $\hat{\boldsymbol{\chi}}_{i}\coloneqq\hat{\boldsymbol{\chi}}\left(z_{i}\right)$
and $\hat{\boldsymbol{\chi}}_{i}^{\dagger}\coloneqq\hat{\boldsymbol{\chi}}^{\dagger}\left(z_{i}\right)$,
where the sequence $\{z_{i}\}$ corresponds to the holomorphic coordinates
that parametrise the momenta of the gravitons participating in the
scattering process under consideration. Boldface symbols are employed
for the fermionic fields to emphasise their operator character, and,
as will be demonstrated, their correlators will serve to reconstruct
the graviton amplitudes.

The anti-commutation relations imply that the two-point correlation
function of the doublet $(\hat{\boldsymbol{\chi}},\hat{\boldsymbol{\chi}}^{\dagger})$
assumes the form:
\begin{equation}
\langle0|\,\hat{\boldsymbol{\chi}}_{i}\,\hat{\boldsymbol{\chi}}_{j}^{\dagger}\,|0\rangle\,=\,\frac{1}{z_{ij}},\,\,\,z_{ij}\,\coloneqq\,z_{i}-z_{j}.\label{eq:-5}
\end{equation}

In order to introduce the graviton vertex operators, it is useful
to reformulate the fermionic system $(\hat{\boldsymbol{\chi}},\hat{\boldsymbol{\chi}}^{\dagger})$
in terms of an alternative representation, denoted $(\boldsymbol{\chi},\boldsymbol{\chi}^{\dagger})$,
which is defined on the holomorphic vector bundle $\mathcal{O}\left(1\right)\oplus\mathcal{O}\left(1\right)$
over $\mathbf{CP}^{1}$ of undotted two-component spinors. The precise
correspondence between these two representations is specified as follows.
Let $\nu^{A}$ denote a two-component spinor, and consider a local
trivialisation $\mathcal{\mathscr{U}}$ of the bundle $\mathcal{O}\left(1\right)\oplus\mathcal{O}\left(1\right)$
such that $(\nu^{A})_{\mathscr{U}}=\left(\alpha,\beta\right)^{T}$,
where $\alpha\in\mathbf{C}$ and $\beta\in\mathbf{C}^{*}$. Within
the trivialisation $\mathscr{U}$, it is natural to projectively represent
$\nu^{A}$ on the complex projective line by introducing the local
affine coordinate $z$, such that $[z]\coloneqq[\alpha\beta^{-1}]\in\mathbf{CP}^{1}$.
With these preliminary observations in place, we define:
\begin{equation}
\boldsymbol{\chi}\left(\nu\right)\,\coloneqq\,\beta^{-1}\,\hat{\boldsymbol{\chi}}\left(z\right),\,\,\,\hat{\boldsymbol{\chi}}^{\dagger}\left(\nu\right)\,\coloneqq\,\beta^{-1}\,\hat{\boldsymbol{\chi}}^{\dagger}\left(z\right).
\end{equation}

For notational convenience, we introduce the shorthand $\boldsymbol{\chi}_{i}\coloneqq\boldsymbol{\chi}\left(z_{i}\right)$
and $\boldsymbol{\chi}_{i}^{\dagger}\coloneqq\boldsymbol{\chi}^{\dagger}\left(z_{i}\right)$,
where $\{z_{i}^{A},\overline{z}_{i\dot{A}}\}$ constitutes a normalised
spinor basis, defined explicitly by:
\begin{equation}
z_{i}^{A}\,\coloneqq\,\begin{pmatrix}1\\
z_{i}
\end{pmatrix},\,\,\,\overline{z}_{i\dot{A}}\,\coloneqq\,\begin{pmatrix}\,\overline{z}_{i}\,\,\, & \,\,\,-1\,\end{pmatrix}
\end{equation}
Similarly, we employ the notation $\boldsymbol{\chi}_{\nu_{i}}\coloneqq\boldsymbol{\chi}\left(\nu_{i}\right)$
and $\boldsymbol{\chi}_{\nu_{i}}^{\dagger}\coloneqq\boldsymbol{\chi}^{\dagger}\left(\nu_{i}\right)$,
where the spinors $\nu_{i}^{A}$ $\left(1\leq i\leq n\right)$ parametrise
the kinematical data associated with the momenta of the graviton states.

Given these definitions, the two-point correlation function of the
fermionic doublet $(\boldsymbol{\chi},\boldsymbol{\chi}^{\dagger})$
may be expressed simply as:
\begin{equation}
\langle0|\,\boldsymbol{\chi}_{\nu_{i}}\,\boldsymbol{\chi}_{\nu_{j}}^{\dagger}\,|0\rangle\,=\,\frac{1}{\nu_{i}\cdot\nu_{j}}.
\end{equation}

\subsubsection{$BGK_{n}$ Formula from $\mathbf{CP}^{1}$ Correlators}

The graviton ``vertex operators'' are defined as:
\begin{equation}
\mathcal{G}_{i}\,\coloneqq\,\exp\big(\,i\,\langle\nu_{i}|\,x\,|\overline{\nu}_{i}]\,\big)\,\boldsymbol{\chi}_{\nu_{i}}^{\dagger}\,\boldsymbol{\chi}_{\nu_{i}},\,\,\:\mathcal{H}_{i}\,\coloneqq\,\frac{[\overline{\nu}_{i}|\left(-i\boldsymbol{\partial}\right)|\omega\rangle}{\nu_{i}\cdot\omega}\,\exp\big(\,i\,\langle\nu_{i}|\,x\,|\overline{\nu}_{i}]\,\big).\label{eq:-19}
\end{equation}
In the above expressions, $(\boldsymbol{\partial})_{A\dot{A}}\coloneqq(\sigma^{\mu})_{A\dot{A}}\frac{\partial}{\partial x^{\mu}}$
and $\omega_{A}\,\coloneqq\,\left(\frac{i}{2\pi}\right)^{1/2}\,\left(\,\omega,\,-1\,\right)$
is an auxiliary two-component spinor parametrised by $\left[\omega\right]\in\mathbf{CP}^{1}$.
This spinor serves as a reference variable over which integration
is performed in the final expressions involving the correlation functions
of $\mathcal{G}_{i}$ and $\mathcal{H}_{i}$. Furthermore, the spinor
$\omega_{A}$ defines a state-vector within the $\mathbf{CP}^{1}$
fermionic system $(\boldsymbol{\chi},\boldsymbol{\chi}^{\dagger})$,
which is given by $|\omega\rangle\coloneqq\chi^{\dagger}\left(\omega\right)|0\rangle.$

Through an inductive argument on $n\in\mathbf{N}$, it can be established
that the operators $\mathcal{G}_{i}$ and $\mathcal{H}_{i}$ satisfy
the identity:
\begin{align}
 & \mathcal{G}_{1}\,\left(\prod_{i=2}^{n-2}\mathcal{H}_{i}\right)\,\mathcal{G}_{n-1}\,\mathcal{G}_{n}\\
 & =\,\,\,e^{i\left(p_{1}+...+p_{n}\right)\cdot x}\,\,\,\prod_{i=2}^{n-2}\,\frac{[\overline{\nu}_{i}|p_{i+1}+...+p_{n}|\omega\rangle}{\nu_{i}\cdot\omega}\,\,\,\big(\boldsymbol{\chi}_{\nu_{1}}^{\dagger}\,\boldsymbol{\chi}_{\nu_{1}}\big)\,\big(\boldsymbol{\chi}_{\nu_{n-1}}^{\dagger}\,\boldsymbol{\chi}_{\nu_{n-1}}\big)\,\big(\boldsymbol{\chi}_{\nu_{n}}^{\dagger}\,\boldsymbol{\chi}_{\nu_{n}}\big).
\end{align}

This property implies an integral relation satisfied by the correlation
functions of the operators $\mathcal{G}_{i}$ and $\mathcal{H}_{i}$,
which is expressed as follows:
\begin{align}
 & \frac{1}{\left(2\pi\right)^{4}}\,\,\,\underset{\mathbf{R}^{4}}{\int}\,\,\,d^{4}x\,\,\,\underset{\mathscr{C}_{n}}{\oint}\,\,\,\left\langle \omega d\omega\right\rangle \,\,\,\langle\omega|\,\mathcal{G}_{1}\left(\prod_{i=2}^{n-2}\mathcal{H}_{i}\right)\mathcal{G}_{n-1}\,\mathcal{G}_{n}\,|\omega\rangle\\
 & =\,\,\,\delta^{\left(4\right)}\left(\sum_{i=1}^{n}\nu_{i}^{A}\bar{\nu}_{i}^{\dot{A}}\right)\,\frac{1}{\left\langle \nu_{n}\nu_{1}\right\rangle \left\langle \nu_{1}\nu_{n-1}\right\rangle \left\langle \nu_{n-1}\nu_{n}\right\rangle }\,\prod_{j=2}^{n-2}\,\frac{[\bar{\nu}_{j}|p_{j+1}+...+p_{n-1}|\nu_{n}\rangle}{\nu_{j}\cdot\nu_{n}},
\end{align}
where $\mathscr{C}_{n}$ denotes a small contour centred at the insertion
point on $\mathbf{CP}^{1}$ representing the $n$-th graviton. 

Consequently, the scattering amplitude $\mathcal{M}_{n}$ can be expressed
in a form structurally reminiscent of the correlation functions of
vertex operators in conventional twistor string theory\footnote{See, e.g., \citet{abou2008einstein,adamo2013twistor,mason2014ambitwistor}.},
\begin{equation}
\mathcal{M}_{n}\,=\,\left(\frac{\kappa}{2}\right)^{2}\,\,\,\left\langle \nu_{1}\nu_{2}\right\rangle ^{8}\,\,\,\underset{\mathbf{R}^{4}}{\int}\,\,\,\frac{d^{4}x}{\left(2\pi\right)^{4}}\,\,\,\underset{\mathscr{C}_{n}}{\oint}\,\left\langle \omega d\omega\right\rangle \,\,\,\langle\omega|\,\mathcal{G}_{1}\,\left(\prod_{i=2}^{n-2}\mathcal{H}_{i}\right)\,\mathcal{G}_{n-1}\,\mathcal{G}_{n}\,|\omega\rangle\,\,\,\prod_{j=1}^{n}\frac{1}{\nu_{j}\cdot\nu_{j+1}}.\label{eq:-7}
\end{equation}
The precise nature of this correspondence will be clarified in the
subsequent sections. 

\subsubsection{Frequency Dependency}

To proceed with our aim of performing the Mellin transform of $\mathcal{M}_{n}$,
thereby yielding the corresponding celestial amplitude, it is necessary
to re-express Eq. (\ref{eq:-7}) explicitly in terms of the graviton
frequencies $s_{i}$ and the normalised spinor basis $\{z_{i}^{A},\overline{z}_{i\dot{A}}\}$.
This reformulation takes the form:
\begin{equation}
\mathcal{M}_{n}\,=\,\left(\frac{\kappa}{2}\right)^{n-2}\,\,\,\left(z_{1}\cdot z_{2}\right)^{8}\,\,\,\underset{\mathbf{R}^{4}}{\int}\,\frac{d^{4}x}{\left(2\pi\right)^{4}}\,\,\,\underset{\mathscr{C}_{n}}{\oint}\,\left\langle \omega d\omega\right\rangle \,\,\,\langle\omega|\,\mathcal{G}_{1}\left(\prod_{i=2}^{n-2}\mathcal{H}_{i}\right)\mathcal{G}_{n-1}\,\mathcal{G}_{n}\,|\omega\rangle\,\,\,\prod_{j=1}^{n}\frac{s_{j}^{e_{j}}}{z_{j}\cdot z_{j+1}}.\label{eq:-6}
\end{equation}
In this expression, the sequence $e_{i}$ $\left(1\leq i\leq n\right)$
denotes the set of exponents characterising the powers in which the
frequencies $s_{i}$ appear in $\mathcal{M}_{n}$. For the MHV configuration
$1^{--},2^{--},3^{++},...,n^{++}$, these exponents are given by $e_{1}=e_{2}=3$
and $e_{3}=...=e_{n}=-1$. 

As we proceed, the analysis transitions from the MHV amplitude $\mathcal{M}_{n}$
to the graviton super-amplitude $\mathscr{M}_{n}$ within the onshell
superspace formalism of $\mathcal{N}=8$ Supergravity. To maintain
generality throughout the derivation, we shall retain the explicit
dependence on the sequence $e_{i}$ in all subsequent equations. 

\subsubsection{$\mathcal{N}=8$ Supergravity}

The transition from Einstein's gravity to $\mathcal{N}=8$ Supergravity
is facilitated by adopting the on-shell superfield formalism, as reviewed
in \citet{wess2020supersymmetry}. 

\paragraph{On-shell Superfield Expansion.}

This formalism introduces Grassmann-valued two-component spinors,
denoted $\eta_{A}^{\alpha}$ $\left(1\leq\alpha\leq8\right)$ and
normalised according to the relation:
\begin{equation}
\underset{\mathbf{R}^{0|2}}{\int}\,d^{0|2}\eta\,\,\,\eta_{A}^{\alpha}\,\eta_{B}^{\alpha}=\varepsilon_{AB}.
\end{equation}
To construct the requisite supermultiplets, we define the set of Grassmann-valued
coefficients:
\begin{equation}
\zeta_{i}^{\alpha}\coloneqq z_{i}\cdot\eta^{\alpha}\,\,\,\left(1\leq\alpha\leq8\right).\label{eq:-39}
\end{equation}
 By employing the properties of the Berezin-de Witt integral, the
following identity is established:
\begin{equation}
\big(z_{i}\cdot z_{j}\big)^{8}\,=\,\underset{\mathbf{R}^{0|16}}{\int}\,d^{0|16}\eta\,\prod_{\alpha=1}^{8}\,\zeta_{i}^{\alpha}\zeta_{j}^{\alpha}.
\end{equation}
which permits the following reformulation of Eq. (\ref{eq:-6}):
\begin{equation}
\mathcal{M}_{n}=\left(\frac{\kappa}{2}\right)^{n-2}\int\,\frac{d^{4}x}{\left(2\pi\right)^{4}}\,\wedge\,d^{0|16}\eta\,\prod_{\alpha=1}^{8}\zeta_{1}^{\alpha}\zeta_{2}^{\alpha}\underset{\mathscr{C}_{n}}{\oint}\left\langle \omega d\omega\right\rangle \langle\omega|\,\mathcal{G}_{1}\left(\prod_{i=2}^{n-2}\mathcal{H}_{i}\right)\mathcal{G}_{n-1}\,\mathcal{G}_{n}\,|\omega\rangle\prod_{j=1}^{n}\frac{s_{j}^{e_{j}}}{z_{j}\cdot z_{j+1}}.
\end{equation}

The particle spectrum of $\mathcal{N}=8$ Supergravity is described
by the superfield $\psi$, which parametrises all one-particle states
of the theory. This superfield may be expanded as follows:
\begin{equation}
\psi_{i}(\zeta_{i}^{\alpha})\,=\,\mathsf{h}^{-}\,+\,\zeta_{i}^{\alpha}\,\widetilde{\lambda}_{\alpha}\,+\,\frac{1}{2!}\,\zeta_{i}^{\alpha}\,\zeta_{i}^{\beta}\,\,\,\widetilde{\phi}_{\alpha\beta}\,+\,...\,+\,\mathsf{h}_{i}^{+}\,\,\,\prod_{\alpha=1}^{8}\,\zeta_{i}^{\alpha}.
\end{equation}
Here, $\mathsf{h}_{i}^{-}\coloneqq\mathsf{h}^{-}\left(z,\bar{z}_{i},s_{i}\right)$
represents the classical expectation value associated with the annihilation
operator of a graviton of negative helicity, with momentum parametrised
by $p^{\mu}=s_{i}q\left(z_{i},\bar{z}_{i}\right)$, and $\mathsf{h}_{i}^{+}\coloneqq\mathsf{h}^{+}\left(z_{i},\overline{z}_{i},s_{i}\right)$
denotes the analogous quantity for a graviton of positive helicity.
The intermediate terms in this expansion describe the remaining one-particle
states in the supermultiplet.

\paragraph*{Superspace.}

At this stage, it is pertinent to specify the structure of the superspace
on which the graviton super-amplitude is formulated. Thus, let $\mathbf{R}^{4|16}$
denote the supersymmetric extension of the four-dimensional Euclidean
space $\mathbf{R}^{4}$, obtained by appending to the bosonic coordinates
$x^{\mu}$ the set of Grassmann-odd spinorial coordinates $\theta_{A}^{\alpha}$
$\left(\alpha=1,...,8\right)$. These fermionic ``dimensions,'' together
with the bosonic ones, are collectively encoded in the supercoordinates\footnote{We follow standard conventions in abstract index notation. In addition,
indices associated with supercoordinates, such as $\hat{\text{I}},\hat{\text{J}},...$,
are adorned with a hat to distinguish them from the minitwistor indices
$I,J,...$ appearing in expressions involving the minitwistor representatives
$\mathsf{Y}_{I},\mathsf{Z}^{J},$ etc.} $\mathsf{x}^{\hat{\text{I}}}\coloneqq\left(x^{\mu},\eta_{A}^{\alpha}\right)$.
The supermanifold $\mathbf{R}^{4|16}$ is endowed with the canonical
orientation defined by the Berezin-de Witt volume superform:
\begin{equation}
d^{4|16}\,\mathsf{x}\,\coloneqq\,d^{4}x\,\wedge\,d^{0|16}\theta.
\end{equation}

Accordingly, the graviton super-amplitude $\mathscr{M}_{n}$ is expressed
in the on-shell superfield formalism as:
\begin{equation}
\mathscr{M}_{n}\,=\,\frac{1}{\left(2\pi\right)^{4}}\left(\frac{\kappa}{2}\right)^{n-2}\underset{\mathbf{R}^{4|16}}{\int}d^{4|16}\mathsf{x}\,\underset{\mathscr{C}_{n}}{\oint}\,\left\langle \omega d\omega\right\rangle \,\langle\omega|\,\mathcal{G}_{1}\left(\prod_{i=2}^{n-2}\mathcal{H}_{i}\right)\mathcal{G}_{n-1}\,\mathcal{G}_{n}\,|\omega\rangle\,\prod_{j=1}^{n}\psi\left(z_{i}\cdot\theta^{\alpha}\right)\,\frac{s_{j}^{e_{j}}}{z_{j}\cdot z_{j+1}}.\label{eq:-8}
\end{equation}

\subsection{Celestial Leaf Amplitudes for Gravitons}

With the requisite mathematical preliminaries established, we now
begin our analysis of graviton celestial amplitudes. The main objective
it to demonstrate that the gravitational leaf amplitudes can be interpreted
as an integral over the moduli space of minitwistor lines on the $\mathcal{N}=8$
supersymmetric extension of minitwistor space, $\mathbf{MT}^{2|8}$.

\subsubsection{Mellin Transform}

Let us first recall that the celestial superamplitude $\widehat{\mathscr{M}}_{n}\left(z_{i},\bar{z}_{i},\Delta_{i}\right)$
for gravitons in $\mathcal{N}=8$ Supergravity is defined as the $\varepsilon$-regulated
Mellin transform of the corresponding scattering amplitude $\mathscr{M}_{n}\left(z_{i},\bar{z}_{i},s_{i}\right)$,
as follows:
\begin{equation}
\widehat{\mathscr{M}}_{n}\left(z_{i},\overline{z}_{i},\Delta_{i}\right)\,\,\,\coloneqq\,\,\,\prod_{i=1}^{n}\,\underset{\mathbf{R}_{+}^{\times}}{\int}\,\frac{ds_{i}}{s_{i}}\,s_{i}^{\Delta_{i}}\,e^{-\varepsilon s_{i}}\,\,\,\mathscr{M}_{n}\left(z_{i},\overline{z}_{i},s_{i}\right),
\end{equation}
where $\mathbf{R}_{+}^{\times}$ denotes the multiplicative group
of positive real numbers, and the integral is taken with respect to
the Haar measure $\frac{ds_{i}}{s_{i}}$.

We now introduce the \emph{Mellin-transformed graviton vertex operators}:
\begin{equation}
\widehat{\mathcal{G}}_{i}\,\coloneqq\,\underset{\mathbf{R}_{+}^{\times}}{\int}\,\frac{ds_{i}}{s_{i}^{1-e_{i}-\Delta_{i}}}\,\,\,e^{-\varepsilon s_{i}}\,\mathcal{G}_{i},\,\,\,\widehat{\mathcal{H}}_{i}\coloneqq\underset{\mathbf{R}_{+}^{\times}}{\int}\,\frac{ds_{i}}{s_{i}^{1-e_{i}-\Delta_{i}}}\,\,\,e^{-\varepsilon s_{i}}\,\mathcal{H}_{i}.\label{eq:-20}
\end{equation}
In these definitions, $e_{i}$ $\left(1\leq i\leq n\right)$ denotes
exponents characterising the frequency dependence of the super-amplitude
$\mathscr{M}_{n}$.

In addition to these vertex operators, we define the \emph{Mellin-transformed
$\mathcal{N}=8$ superfield }$\boldsymbol{\psi}_{\Delta_{i}}\left(\zeta_{i}^{\alpha}\right)$
as:
\begin{equation}
\boldsymbol{\psi}_{\Delta_{i}}\left(\zeta_{i}^{\alpha}\right)\,\coloneqq\,\widehat{\mathsf{h}}_{\Delta_{i}}^{-}\left(z_{i},\overline{z}_{i}\right)\,+\,\zeta_{i}^{\alpha}\,\widehat{\lambda}_{\alpha,\Delta_{i}}\left(z_{i},\overline{z}_{i}\right)\,+\,\zeta_{i}^{\alpha}\zeta_{i}^{\beta}\,\widehat{\phi}_{\alpha\beta,\Delta_{i}}\,+\,...\,+\widehat{\mathsf{h}}_{\Delta_{i}}^{+}\left(z_{i},\overline{z}_{i}\right)\,\,\,\prod_{\alpha=1}^{8}\zeta_{i}^{\alpha}.\label{eq:-45}
\end{equation}
The modes of the superfield expansion now explicitly depend on the
celestial conformal weight $\Delta_{i}$.

By substituting the expressions from Eqs. (\ref{eq:-19}) and (\ref{eq:-20}),
and subsequently performing the integral transforms, we obtain the
following explicit forms of the Mellin-transformed vertex operators:
\begin{equation}
\widehat{\mathcal{G}}_{i}\,=\,\phi_{2h_{i}}\left(x;z_{i},\overline{z}_{i}\right)\,\hat{\boldsymbol{\chi}}_{i}^{\dagger}\,\hat{\boldsymbol{\chi}}_{i},\,\,\,\widehat{\mathcal{H}}_{i}\,=\,\frac{[\overline{z}_{i}\big|\left(-i\boldsymbol{\partial}\right)\big|\omega\rangle}{z_{i}\cdot\omega}\,\phi_{2h_{i}}\left(x;z_{i},\overline{z}_{i}\right),\label{eq:-54}
\end{equation}
where $\phi_{\Delta}\left(x;z,\overline{z}\right)$ is the celestial
wavefunction defined in Eq. (\ref{eq:-46}). 

The scaling dimensions of these operators are given by:
\begin{equation}
h_{i}=\frac{\Delta_{i}+e_{i}-1}{2},\,\,\,h_{j}\coloneqq\frac{\Delta_{j}+e_{j}}{2}.
\end{equation}

Consequently, the celestial superamplitude for gravitons can be expressed
in the following integral form:
\begin{equation}
\widehat{\mathscr{M}}_{n}=\frac{1}{\left(2\pi\right)^{4}}\left(\frac{\kappa}{2}\right)^{n-2}\underset{\mathbf{R}^{4|16}}{\int}d^{4|16}\mathsf{w}\,\,\,\underset{\mathscr{C}_{n}}{\oint}\,\left\langle \omega d\omega\right\rangle \,\langle\omega|\,\widehat{\mathcal{G}}_{1}\left(\prod_{i=2}^{n-2}\widehat{\mathcal{H}}_{i}\right)\widehat{\mathcal{G}}_{n-1}\,\widehat{\mathcal{G}}_{n}\,|\omega\rangle\,\prod_{j=1}^{n}\frac{\boldsymbol{\psi}_{2h_{j}}\left(z_{i}\cdot\theta^{\alpha}\right)}{z_{j}\cdot z_{j+1}}.\label{eq:-8-1}
\end{equation}

\subsubsection{Gravitational Celestial Leaf Amplitudes}

To reformulate the graviton celestial amplitude as an integral over
the moduli space of minitwistor lines, we employ the formalism of
leaf amplitudes. However, a technical complication immediately presents
itself: the celestial wavefunctions $\phi_{2h_{i}}\left(x;z_{i},\overline{z}_{i}\right)$,
which appear as multiplicative factors in the vertex operators $\widehat{\mathcal{G}}_{i}$
and $\widehat{\mathcal{H}}_{i}$, are currently enclosed within the
fixed-ordering correlator $\langle\omega|\cdot\cdot\cdot|\omega\rangle$.
It is necessary to extract these wavefunctions from the correlator,
as this step permits the separation of the spacetime integral from
the contour integral $\oint_{\mathscr{C}_{n}}\left\langle \omega d\omega\right\rangle $,
thus making it possible to apply the leaf amplitude formalism.

We proceed by implementing the following strategy. Let $\mathsf{P}_{i}$
denote the \emph{weight-shifting operator} acting on the Hilbert space
of celestial wavefunctions $\mathscr{H}_{c}$. The action of $\mathsf{P}_{i}$
on a given wavefunction $\phi_{\Delta_{j}}\left(x;z_{j},\bar{z}_{j}\right)$
is:
\begin{equation}
\mathsf{P}_{i}\,\cdot\,\phi_{\Delta_{j}}\left(x;z_{i},\bar{z}_{i}\right)\,\coloneqq\,\phi_{\Delta_{j}+\delta_{ij}}\left(x;z_{j},\bar{z}_{j}\right).
\end{equation}
Consequently, we define a new set of graviton vertex operators:
\begin{equation}
\mathcal{U}_{i}\coloneqq\exp\big(-\langle z_{i}\big|y\big|\overline{z}_{i}]\,\mathsf{P}_{i}\big)\,\,\,\hat{\boldsymbol{\chi}}_{i}^{\dagger}\,\hat{\boldsymbol{\chi}}_{i},\,\,\,\mathcal{V}_{i}\coloneqq\frac{[\bar{z}_{i}\big|\widetilde{\boldsymbol{\partial}}\big|\omega\rangle}{z_{i}\cdot\omega}\,\,\,\exp\big(-\langle z_{i}\big|y\big|\overline{z}_{i}]\,\mathsf{P}_{i}\big).\label{eq:-25}
\end{equation}
In the above expression, $y^{\mu}\in\mathbf{R}^{4}$ denotes an \emph{auxiliary
}four-vector that parametrises the operator family $\{\mathcal{U}_{i},\mathcal{V}_{i}\}$.
The partial derivative $\widetilde{\boldsymbol{\partial}}$ acts with
respect to $y^{\mu}$, and is defined by:
\begin{equation}
\big(\,\widetilde{\boldsymbol{\partial}}\,\big)_{A\dot{A}}\coloneqq(\boldsymbol{\sigma}^{\mu})_{A\dot{A}}\frac{\partial}{\partial y^{\mu}}\,\,\,(y^{\mu}\in\mathbf{R}^{4}),
\end{equation}
where $(\boldsymbol{\sigma}^{\mu})_{A\dot{A}}$ represents the sigma
matrices in the Kleinian signature, defined in Appendix \ref{sec:Leaf-Amplitudes-Review}.

It is important to emphasise that the four-vector $y^{\mu}$ should
be regarded solely as a continuous label indexing the operator family.
It does not carry the interpretation of a spacetime coordinate. Moreover,
in all subsequent expressions where $y^{\mu}$ appears, the final
evaluation is to be performed at $y^{\mu}=0$.

We now establish, through an inductive argument on $n\in\mathbf{N}$,
that the vertex operators satisfy the following algebraic identity:
\begin{equation}
\widehat{\mathcal{G}}_{1}\,\left(\,\prod_{i=2}^{n-2}\,\widehat{\mathcal{H}}_{i}\,\right)\,\widehat{\mathcal{G}}_{n-1}\,\widehat{\mathcal{G}}_{n}\,\,\,=\,\,\,\mathcal{U}_{1}\,\left(\,\prod_{i=2}^{n-2}\,\mathcal{V}_{i}\,\right)\,\mathcal{U}_{n-1}\mathcal{U}_{n}\,\,\,\prod_{j=1}^{n}\phi_{2h_{j}}\left(x;z_{i},\bar{z}_{i}\right).
\end{equation}

From this equality, it follows that the celestial super-amplitude
$\widehat{\mathscr{M}}_{n}\left(z_{i},\bar{z}_{i},\Delta_{i}\right)$
can be reformulated as a correlator involving the newly introduced
graviton vertex operators:
\begin{align}
\widehat{\mathscr{M}}_{n}\left(z_{i},\bar{z}_{i},\Delta_{i}\right)\, & =\,\frac{1}{\left(2\pi\right)^{4}}\,\left(\frac{\kappa}{2}\right)^{n-2}\,\underset{\mathscr{C}_{n}}{\oint}\,\left\langle \omega d\omega\right\rangle \,\,\,\langle\omega|\,\mathcal{U}_{1}\left(\prod_{r=2}^{n-2}\mathcal{V}_{r}\right)\mathcal{U}_{n-1}\mathcal{U}_{n}\,|\omega\rangle\,\\
 & \times\,\underset{\mathbf{R}^{4|16}}{\int}\,\,\,d^{4|16}\mathsf{x}\,\,\,\prod_{i=1}^{n}\,\,\,\boldsymbol{\psi}_{2h_{i}}\left(z_{i}\cdot\theta^{\alpha}\right)\,\phi_{2h_{i}}\left(x;z_{i},\bar{z}_{i}\right)\,\,\,\frac{1}{z_{i}\cdot z_{i+1}}.
\end{align}

The final step required prior to the application of the leaf amplitude
formalism is to specify the integration domain over which it is defined.
In Appendix \ref{sec:Minitwistor-Geometry}, we review how the Riemannian
geometry of the hyperboloid $H_{3}^{+}$ arises from the projective
geometry of $\mathbf{RP}^{3}$. Thus, let $X_{A\dot{A}}$ denote homogeneous
coordinates charting the real projective space $\mathbf{RP}^{3}$.
The natural orientation of $\mathbf{RP}^{3}$ is given by the volume
form:
\begin{equation}
D^{3}X\,=\,\varepsilon_{A\dot{A}}\varepsilon_{B\dot{B}}\varepsilon_{C\dot{C}}\varepsilon_{D\dot{D}}\,X^{A\dot{A}}dX^{B\dot{B}}\wedge dX^{C\dot{C}}\wedge dX^{D\dot{D}}.
\end{equation}
The $\left(3,16\right)$-dimensional real projective \emph{superspace}
$\mathbf{RP}^{3|16}$ is obtained by augmenting the bosonic coordinates
$X_{A\dot{A}}$ with Grassmann-valued spinorial variables $\theta_{A}^{\alpha}$,
with $\alpha=1,...,8$. Thus, the coordinates of the resulting superchart\footnote{We follow the terminology of \citet{rogers2007supermanifolds}. See
also \citet{leites1980introduction} and \citet{manin1997introduction}.} are given by $\mathbb{X}^{\hat{I}}\coloneqq\left(X_{A\dot{A}},\theta_{A}^{\alpha}\right)\in\mathbf{RP}^{3}\times\mathbf{R}^{0|16}$.
The natural orientation of the supermanifold $\mathbf{RP}^{3|16}$
is defined by the Berezin-de Witt volume superform:
\begin{equation}
D^{3|16}\mathbb{X}\,\coloneqq\,\frac{D^{3}X}{\left|X\right|^{4}}\wedge d^{0|16}\zeta.
\end{equation}

We are now prepared to employ the celestial leaf amplitude formalism,
reviewed in Appendix \ref{sec:Leaf-Amplitudes-Review}, which permits
to recast $\widehat{\mathscr{M}}_{n}\left(z_{i},\bar{z}_{i},\Delta_{i}\right)$
as:
\begin{equation}
\widehat{\mathscr{M}}_{n}\left(z_{i},\bar{z}_{i},\Delta_{i}\right)\,=\,\frac{1}{\left(2\pi\right)^{3}}\,\left(\frac{\kappa}{2}\right)^{n-2}\delta\left(\beta\right)\,\times\,M_{n}\left(z_{i},\bar{z}_{i},\Delta_{i}\right)\,+\,(\bar{z}_{i\dot{A}}\longrightarrow i\bar{z}_{i\dot{A}}),
\end{equation}
where $\beta\coloneqq4-2\sum_{i=1}^{n}h_{i}$, and the graviton leaf
amplitude $M_{n}\left(z_{i},\bar{z}_{i},\Delta_{i}\right)$ is given
by:
\begin{equation}
M_{n}\,=\,\underset{\mathscr{C}_{n}}{\oint}\left\langle \omega d\omega\right\rangle \,\,\,\langle\omega|\,\mathcal{U}_{1}\,\left(\,\prod_{r=2}^{n-2}\,\mathcal{V}_{r}\,\right)\,\mathcal{U}_{n-1}\mathcal{U}_{n}\,|\omega\rangle\,\underset{\mathbf{RP}^{3|16}}{\int}\,D^{3|16}\,\mathbb{X}\,\,\,\prod_{i=1}^{n}\,K_{2h_{i}}\left(X;z_{i},\bar{z}_{i}\right)\,\,\,\frac{\boldsymbol{\psi}_{2h_{i}}\left(z_{i}\cdot\theta^{\alpha}\right)}{z_{i}\cdot z_{i+1}}
\end{equation}

\subsection{Celestial RSVW Formalism of $\mathcal{N}=8$ Supergravity}

We wish to find a geometrical interpretation for the gravitational
leaf amplitude $M_{n}\left(z_{i},\bar{z}_{i},\Delta_{i}\right)$.
To achieve this, it proves useful to employ the following decomposition:
\begin{equation}
M_{n}\left(z_{i},\bar{z}_{i},\Delta_{i}\right)\,\,\,=\,\,\,\fintop\,dh_{1}...dh_{n}\,\,\,\text{m}_{n}\left(z_{i},\bar{z}_{i},h_{i}\right),
\end{equation}
where $\,\fintop\,dh_{1}...dh_{n}$ denotes a weight-shifting ``integral''
operator, defined by:
\begin{equation}
\fintop\,dh_{1}...dh_{n}\,\,\coloneqq\,\underset{\mathscr{C}_{n}}{\oint}\left\langle \omega d\omega\right\rangle \,\,\,\langle\omega|\,\mathcal{U}_{1}\,\left(\,\prod_{r=2}^{n-2}\,\mathcal{V}_{r}\,\right)\,\mathcal{U}_{n-1}\mathcal{U}_{n}\,|\omega\rangle\,,\label{eq:-47}
\end{equation}
and the ``reduced'' leaf amplitude $\text{m}_{n}\left(z_{i},\bar{z}_{i},h_{i}\right)$
is given by
\begin{equation}
\text{m}_{n}\left(z_{i},\bar{z}_{i},h_{i}\right)\,\coloneqq\,\underset{\mathbf{RP}^{3|16}}{\int}\,D^{3|16}\,\mathbb{X}\,\,\,\prod_{i=1}^{n}\,\boldsymbol{\psi}_{2h_{i}}\left(z_{i}\cdot\theta^{\alpha}\right)\,K_{2h_{i}}\left(X;z_{i},\bar{z}_{i}\right)\,\,\,\frac{1}{z_{i}\cdot z_{i+1}}.\label{eq:-15}
\end{equation}

Using Eq. (\ref{eq:-25}), which defines the operators $\mathcal{U}_{i}$
and $\mathcal{V}_{i}$, and applying it to the contour integral in
Eq. (\ref{eq:-47}), we deduce the following expression for the weight-shifting
``integral'' operator:
\begin{equation}
\fintop\,dh_{1}...dh_{n}\,=\,\left(-1\right)^{n+1}\,\left(\prod_{i=2}^{n-2}\sum_{j_{i}=i+1}^{n-1}\right)\,\prod_{k=2}^{n-2}\,\frac{[\bar{z}_{k}\bar{z}_{j_{k}}]\langle z_{j_{k}}z_{n}\rangle}{z_{k}\cdot z_{n}}\,\mathsf{P}_{j_{k}}.
\end{equation}
This equation establishes that $\fintop\,dh_{1}...dh_{n}$ is a polynomial
constructed from a sequence of weight-shifting operators $\mathsf{P}_{i}$,
acting on the Hilbert space $\mathscr{H}_{c}$ of celestial wavefunctions.

\subsubsection*{Celestial RSVW Formula for Gravitons}

Our derivation of the celestial RSVW formula for gravitational leaf
amplitudes begins with the resolution of the identity for the $\mathcal{N}=8$
superfield:
\begin{equation}
\boldsymbol{\psi}_{2h_{i}}\left(z_{i}\cdot\theta^{\alpha}\right)\,\,\,=\,\,\,\underset{\mathbf{CP}^{0|8}}{\int}\,d^{0|8}\zeta_{i}\,\,\,\delta^{\left(8\right)}\left(\zeta_{i}^{\alpha}-z_{i}^{A}\theta_{A}^{\alpha}\right)\,\boldsymbol{\psi}_{2h_{i}}\left(\zeta_{i}\right),
\end{equation}
where $d^{0|8}\zeta$ denotes the Berezin-de Witt integral over the
fermionic coordinates $\zeta^{\alpha}$. This identity allows us to
reformulate the ``reduced'' leaf amplitude $\text{m}_{n}\left(z_{i},\bar{z}_{i},\Delta_{i}\right)$
defined in Eq. (\ref{eq:-15}) as follows:
\begin{align}
\text{m}_{n}\left(z_{i},\bar{z}_{i},\Delta_{i}\right) & \,\,\,=\,\,\,\prod_{r=1}^{n}\,\,\,\underset{\mathbf{CP}^{0|8}}{\int}\,\,\,d^{0|8}\zeta_{r}\,\,\,\underset{\mathbf{RP}^{3|16}}{\int}\,\,\,D^{3|16}\mathbb{X}\\
 & \prod_{i=1}^{n}\,\,\,\delta^{\left(8\right)}\left(\zeta_{i}^{\alpha}-z_{i}^{A}\theta_{A}^{\alpha}\right)\,\boldsymbol{\psi}_{2h_{i}}\left(\zeta_{i}^{\alpha}\right)\,K_{2h_{i}}\left(X;z_{i},\bar{z}_{i}\right)\,\,\,\frac{1}{z_{i}\cdot z_{i+1}}.
\end{align}

To proceed, we introduce local coordinates charting the minitwistor
space $\mathbf{MT}$. Let $p\in\mathbf{MT}\mapsto\mathsf{Z}^{I}\left(p\right)\coloneqq\left(\lambda^{A}\left(p\right),\mu_{\dot{A}}\left(p\right)\right)\in\mathbf{CP}^{1}\times\overline{\mathbf{CP}}^{1}$,
where $\lambda^{A}$ and $\mu_{\dot{A}}$ are homogeneous coordinates
on the complex projective lines $\mathbf{CP}^{1}$ and $\overline{\mathbf{CP}}^{1}$,
respectively. Recall that the minitwistor space $\mathbf{MT}$ is
a non-singular quadric in $\mathbf{CP}^{3}$, and its canonical orientation
is induced by the volume form $D\lambda\wedge D\mu$, with $D\lambda\coloneqq\varepsilon_{AB}\lambda^{A}d\lambda^{B}$
and $D\mathcal{\mu}\coloneqq\varepsilon^{\dot{A}\dot{B}}\mu_{\dot{A}}d\mu_{\dot{B}}$.

We now invoke the celestial RSVW identity derived in Eq. (\ref{eq:-49}),
which allows us to rewrite the expression for the ``reduced'' leaf
amplitude in terms of integrals over minitwistor space:
\begin{align}
\text{m}_{n}\left(z_{i},\bar{z}_{i},\Delta_{i}\right) & \,=\,\prod_{r=1}^{n}\,\,\,\underset{\mathbf{MT}\times\mathbf{CP}^{0|8}}{\int}\,\,\,D\lambda_{r}\wedge D\mu_{r}\wedge d^{0|8}\zeta_{r}\,\,\,\boldsymbol{\psi}_{2h_{r}}\left(\zeta_{r}^{\alpha}\right)\,\widetilde{\mathcal{F}}_{2h_{r}}\left(\lambda_{r}^{A},\mu_{r\dot{A}};z_{r}^{A},\bar{z}_{r\dot{A}}\right)\label{eq:-55}\\
 & \underset{\mathbf{RP}^{3|16}}{\int}\,D^{3|16}\mathbb{X}\,\,\,\prod_{i=1}^{n}\,\,\,\overline{\delta}_{2h_{i}}\left(\mu_{i\dot{A}},\lambda_{i}^{A}\frac{X_{A\dot{A}}}{\left|X\right|}\right)\,\delta^{\left(8\right)}\left(\zeta_{i}^{\alpha}-\lambda_{i}^{A}\theta_{A}^{\alpha}\right)\,\,\,\frac{1}{\lambda_{i}\cdot\lambda_{i+1}}.
\end{align}

The subsequent step in our derivation requires a specification of
the domain over which the ``reduced'' gravitational leaf amplitude
is defined. Observe that the first integrals appearing in the above
expression are taken over the product space $\mathbf{MT}\times\mathbf{CP}^{0|8}$.
As will be discussed in detail in Section \ref{sec:Minitwistor-String-Theories},
where our sigma model is introduced, the minitwistor superspace\footnote{For a rigorous mathematical discussion, see \citet{samann2009mini}.}
is identified with the trivial superbundle $\mathbf{MT}^{2|8}\simeq\mathbf{MT}\times\mathbf{CP}^{0|8}$.
Here, the base manifold is the nonsingular quadric, while the typical
fibre isomorphic to $\mathbf{CP}^{0|8}$ follows from the projectivisation
of the $\mathbf{Z}_{2}$-graded vector space spanned by the Grassmann-valued
homogeneous coordinates $\zeta^{\alpha}$, with $\alpha=1,...,8$.

The orientation on the total space of this superbundle is defined
by the Berezin-de Witt volume superform:
\begin{equation}
D^{2|8}\mathsf{Z}\,\coloneqq\,D\lambda\wedge D\mu\wedge d^{0|8}\zeta.
\end{equation}

Following Witten's original construction\footnote{See \citet{witten2004perturbative,edward2004parity}.},
the \emph{minitwistor wavefunction} associated with a graviton of
\emph{celestial conformal weight} $\Delta$ is identified from Eq.
(\ref{eq:-55}) as:
\begin{equation}
\Psi_{\Delta}\left(\mathsf{Z}^{I};z^{A},\bar{z}_{\dot{A}}\right)\,\coloneqq\,\boldsymbol{\varphi}_{\Delta}\left(\zeta^{\alpha}\right)\,\widetilde{\mathcal{F}}_{\Delta}\left(\lambda^{A},\mu_{\dot{A}};z^{A},\bar{z}_{\dot{A}}\right).
\end{equation}
Consequently, the ``reduced'' gravitational leaf amplitude can be
expressed as a Fourier transform over minitwistor superspace:
\begin{equation}
\text{m}_{n}\left(z_{i},\bar{z}_{i},\Delta_{i}\right)\,=\,\underset{\left(\mathbf{MT}^{2|8}\right)^{\times n}}{\int}\,\bigwedge_{i=1}^{n}\,D^{2|8}\mathsf{Z}_{i}\,\,\,\Phi_{2h_{i}}\left(\mathsf{Z}_{i};z_{i},\bar{z}_{i}\right)\,\widetilde{\text{m}}_{n}\left(\mathsf{Z}_{i}\right),
\end{equation}
where the Fourier-transformed amplitude is given by:
\begin{equation}
\widetilde{\text{m}}_{n}\left(\mathsf{Z}_{i}\right)\,=\,\underset{\mathbf{RP}^{3|16}}{\int}D^{3|16}\mathbb{X}\,\,\,\prod_{i=1}^{n}\,\,\,\overline{\delta}_{2h_{i}}\left(\mu_{i\dot{A}},\lambda_{i}^{A}\frac{X_{A\dot{A}}}{\left|X\right|}\right)\,\delta^{\left(8\right)}\left(\zeta_{i}^{\alpha}-\lambda_{i}^{A}\theta_{A}^{\alpha}\right)\,\,\,\frac{1}{\lambda_{i}\cdot\lambda_{i+1}}.
\end{equation}

Thus, the Fourier-transformed leaf amplitudes for gravitons in $\mathcal{N}=8$
Supergravity\footnote{At tree-level for configurations characterised by maximal helicity
violation.} is expressed as an integral over the moduli space of minitwistor
lines $\mathcal{L}\left(X\right)$ embedded in the non-singular quadric
$\mathbf{MT}$. This interpretation arises from the fact that the
integration with respect to the measure $D^{3}X$ is localised by
the weighted distributional forms $\overline{\delta}_{2h_{i}}$, which
impose the incidence relations characterising the curves $\mathcal{L}\left(X\right)$.
Therefore, the support of the resulting integral is restricted to
configurations where the graviton insertion points lie along such
minitwistor lines.

It must be emphasised that this integral does not represent a conventional
volume over the moduli space of conic curves. The distributional forms
$\overline{\delta}_{2h_{i}}$ possess specific degrees of homogeneity
that encode the celestial scaling dimensions $\Delta_{i}$ of the
gravitons involved in the scattering process. It would be interesting
to reformulate the expression for $\widetilde{\text{m}}_{n}\left(\mathsf{Z}_{i}\right)$
using the formalism proposed by \citet{movshev2006berezinian} and
\citet{adamo2013moduli}.

\subsection{Generating Functional\label{subsec:Generating-Functional-1}}

We now proceed to construct a generating functional for gravitational
leaf superamplitudes by employing the geometric interpretation derived
above. To incorporate the weight-shifting operator $\fintop\,dh_{1}...dh_{n}$,
it is convenient to reformulate the amplitude $M_{n}\left(z_{i},\bar{z}_{i},\Delta_{i}\right)$
as follows.

\subsubsection{Preliminaries: A Celestial Correlator}

Let $\boldsymbol{O}\coloneqq\boldsymbol{O}[\omega^{A},\mathsf{P}_{i},\Psi_{\Delta_{i}}]$
be an operator depending on the auxiliary spinor $\omega^{A}$, the
weight-shifting operators $\mathsf{P}_{i}$, and the minitwistor graviton
wavefunctions $\Psi_{\Delta_{i}}\left(\mathsf{Z}_{i};z_{i},\bar{z}_{i}\right)$.
Define the expectation value of $\boldsymbol{O}$ over the auxiliary
spinor and the fermionic doublet $(\boldsymbol{\chi},\boldsymbol{\chi}^{\dagger})$
as:
\begin{equation}
\left\langle \boldsymbol{O}\right\rangle _{\omega,n}\,=\,\underset{C_{n}}{\oint}\,\left\langle \omega d\omega\right\rangle \,\,\,\langle\omega\big|\,\boldsymbol{O}[\omega^{A},\mathsf{P}_{i},\Psi_{\Delta_{i}}]\,\big|\omega\rangle,\label{eq:}
\end{equation}
where $C$$_{n}$ denotes a small contour centred at $z_{n}$, the
insertion point of the $n$-th graviton on the celestial sphere.

Next, introduce a sequence of Grassmann-valued scalars $\mathsf{c}_{i}$
for $i=1,...,n$, satisfying the normalisation condition:
\begin{equation}
\underset{\mathbf{CP}^{0|1}}{\int}d^{0|1}\mathsf{c}_{i}\,\left(\alpha_{i}+\mathsf{c}_{i}\beta_{i}\right)=\beta_{i},
\end{equation}
for all $\alpha_{i},\beta_{i}\in\mathbf{C}$. Denote by $\Lambda\coloneqq\Lambda[\mathsf{c}_{1},...,c_{n}]$
the Grassmann algebra generated by the variables $\mathsf{c}_{i}$,
which may be interpreted as \emph{ghost-like fields} corresponding
to anti-commuting spinless coordinates.

Let $\mathcal{D}\left(\mathscr{H}_{c}\right)$ be the ring of bounded
linear operators on the Hilbert space $\mathscr{H}_{c}$ of celestial
wavefunctions, and extend this structure to the $\mathbf{Z}_{2}$-graded
ring:
\begin{equation}
\mathcal{D}_{\Lambda}(\mathscr{H}_{c})\,\coloneqq\,\Lambda[\mathsf{c}]\otimes\mathcal{D}\left(\mathscr{H}_{c}\right).
\end{equation}
Using this formalism, the graviton vertex operators $\mathcal{U}_{i}$
and $\mathcal{V}_{i}$ can be unified into a single element of $\mathcal{D}_{\Lambda}\left(\mathscr{H}_{c}\right)$
of the form:
\begin{equation}
\mathcal{U}_{i}+\mathsf{c}_{i}\mathcal{V}_{i}\in\mathcal{D}_{\Lambda}\left(\mathscr{H}_{c}\right).
\end{equation}

With these constructions at hand, the gravitational leaf amplitude
$M_{n}\left(z_{i},\bar{z}_{i},\Delta_{i}\right)$ can be expressed
as the correlation function of the following operator:
\begin{equation}
\boldsymbol{M}_{n}\left(z_{i},\bar{z}_{i},\Delta_{i}\right)\,=\,\underset{\mathbf{RP}^{3|16}}{\int}D^{3|16}\mathbb{X}\,\,\,\prod_{i=1}^{n}\,\,\,\left(\mathcal{U}_{i}+\mathsf{c}_{i}\mathcal{V}_{i}\right)\,\boldsymbol{\psi}_{2h_{i}}\left(z_{i}\cdot\theta^{\alpha}\right)\,K_{2h_{i}}\left(X;z_{i},\bar{z}_{i}\right)\,\,\,\frac{1}{z_{i}\cdot z_{i+1}}.
\end{equation}
In terms of the correlator $\left\langle \boldsymbol{O}\right\rangle _{\omega,n}$
previously introduced in Eq. (\ref{eq:}), the graviton leaf amplitude
admits the following representation:
\begin{equation}
M_{n}\left(z_{i},\bar{z}_{i},\Delta_{i}\right)\,=\,\left\langle \int d\mathsf{c}_{2}...d\mathsf{c}_{n-2}\,\,\,\boldsymbol{M}_{n}\left(z_{i},\bar{z}_{i},\Delta_{i}\right)\right\rangle _{\omega,n}.
\end{equation}

\subsubsection{Minitwistor Gravitational Background}

We now proceed to construct the generating functional. The first step
is to introduce the \emph{minitwistor gravitational background potential}
as a representative of a cohomology class:
\begin{equation}
\boldsymbol{\omega}_{i}\in\mathcal{D}_{\Lambda}[\mathscr{H}_{c}]\otimes\Omega^{0,1}\big(\mathbf{MT}^{2|8},\mathcal{O}(-2,-\Delta)\big).
\end{equation}
Recall that the minitwistor superline $\mathcal{L}\left(X,\theta\right)\subset\mathbf{MT}^{2|8}$,
associated with the point $\mathbb{X}^{\hat{I}}=\left(X_{A\dot{A}},\theta_{A}^{\alpha}\right)$
in projective superspace $\mathbf{RP}^{3|16}$, is define by the locus:
\begin{equation}
\mathcal{L}\left(X,\theta\right)\,\coloneqq\,\left\{ \,\big(\lambda^{A},\mu_{\dot{A}},\xi^{\alpha}\big)\in\mathbf{MT}^{2|4}\,\Big|\,\mu_{\dot{A}}=\lambda^{A}\frac{X_{A\dot{A}}}{\left|X\right|},\,\xi^{\alpha}=\lambda^{A}\theta_{A}^{\alpha}\,\right\} .
\end{equation}
In addition, let $\overline{\partial}\,\big|_{\mathcal{L}\left(X,\theta\right)}$
denote the restriction of the Dolbeault operator to the supercurve
$\mathcal{L}\left(X,\theta\right)$. We then define the \emph{generating
functional }as:
\begin{equation}
\digamma[\boldsymbol{\omega}_{i}]\,\coloneqq\,\underset{\mathbf{RP}^{3|16}}{\int}D^{3|16}\mathbb{X}\,\,\,\log\det\left(\overline{\partial}\,+\boldsymbol{\omega}_{i}\right)\big|_{\mathcal{L}\left(X,\theta\right)}.
\end{equation}
We claim that $\digamma[\boldsymbol{\omega}_{i}]$ generates $\boldsymbol{M}_{n}\left(z_{i},\bar{z}_{i},\Delta_{i}\right)$
upon evaluation at the gravitational background potential:
\begin{equation}
\boldsymbol{\omega}_{i}\,=\,\underset{\mathcal{P}\times\mathbf{C}^{2}}{\int}\boldsymbol{\beta}_{i}\,\,\,\left(\mathcal{U}_{i}+\mathsf{c}_{i}\mathcal{V}_{i}\right)\,\boldsymbol{\psi}_{\Delta_{i}}\left(z_{i}\cdot\theta^{\alpha}\right)\,\mathcal{F}_{\Delta_{i}}\left(\lambda_{i},\mu_{i};z_{i},\bar{z}_{i}\right).\label{eq:-17}
\end{equation}
Here, $\mathcal{P}\coloneqq1+i\mathbf{R}$ denotes the continuous
principal series to which the celestial conformal weight $\Delta_{i}$
belongs, and $\mathcal{F}_{\Delta_{i}}\left(\lambda_{i},\mu_{i};z_{i},\bar{z}_{i}\right)$
is the minitwistor representative defined in Eq. (\ref{eq:-22}).
The volume form on $\mathcal{P}\times\mathbf{C}^{2}$ is given by:
\begin{equation}
\boldsymbol{\beta}_{i}\,\coloneqq\,d\Delta_{i}\wedge dz_{i}\wedge d\bar{z}_{i}.
\end{equation}

To establish this claim, we expand the Quillen determinant as:
\begin{equation}
\digamma[\boldsymbol{\omega}_{i}]\,=\,\sum_{m\geq2}\frac{\left(-1\right)^{m+1}}{m}\underset{\mathbf{RP}^{3|16}}{\int}D^{3|16}\mathbb{X}\,\,\,\underset{\left(\mathcal{L}\left(X,\theta\right)\right)^{\times m}}{\int}\,\,\,\bigwedge_{i=1}^{m}\,D\lambda_{i}\,\,\,\boldsymbol{\omega}_{i}\,\big|_{\mathcal{L}\left(X,\theta\right)}\,\,\,\frac{1}{\lambda_{i}\cdot\lambda_{i+1}}.
\end{equation}
Substituting the expression for $\boldsymbol{\omega}_{i}$ from Eq.
(\ref{eq:-17}) into the expansion yields:
\begin{align}
\digamma[\boldsymbol{\omega}_{i}]\,\,\, & =\,\,\,\sum_{m\geq2}\,\,\,\frac{\left(-1\right)^{m+1}}{m}\,\,\,\underset{\mathbf{RP}^{3|16}}{\int}\,\,\,D^{3|16}\mathbb{X}\,\,\,\underset{\left(\mathcal{P}\times\mathbf{C}^{2}\right)^{\times m}}{\int}\,\,\,\bigwedge_{i=1}^{m}\,\,\,\boldsymbol{\beta}_{i}\\
 & \,\left(\mathcal{U}_{i}+\mathsf{c}_{i}\mathcal{V}_{i}\right)\,\,\,\boldsymbol{\psi}_{2h_{i}}\left(z_{i}\cdot\theta^{\alpha}\right)\,K_{2h_{i}}\left(X;z_{i},\bar{z}_{i}\right)\,\,\,\frac{1}{z_{i}\cdot z_{i+1}}.
\end{align}
Finally, differentiating $\digamma[\boldsymbol{\omega}_{i}]$ with
respect to the classical expectation values $\mathsf{h}_{2h_{i}}^{\ell_{i}}\left(z_{i},\bar{z}_{i}\right)$
associated with the annihilation operator for a graviton with helicity
$\ell_{i}$ and celestial scaling dimension $2h_{i}$, we obtain:
\begin{equation}
\frac{\delta}{\delta\mathsf{h}_{2h_{1}}^{\ell_{1}}\left(z_{1},\bar{z}_{1}\right)}...\frac{\delta}{\delta\mathsf{h}_{2h_{n}}^{\ell_{n}}\left(z_{n},\bar{z}_{n}\right)}\,\digamma[\boldsymbol{\omega}_{i}]\,\Bigg|_{\boldsymbol{\psi}=0}=\boldsymbol{M}_{n}\left(z_{i},\bar{z}_{i},\Delta_{i}\right),
\end{equation}
thus completing the proof.

\section{Minitwistor Celestial CFT\label{sec:Minitwistor-String-Theories}}
\begin{notation*}
The integral operator $\int_{\mathbf{CP}^{1}}\,$ is understood to
act on every term appearing on the right-hand side of an expression,
irrespective of whether such terms are enclosed within brackets. For
example, if $a$ and $b$ are $\left(0,1\right)$-forms on $\mathbf{CP}^{1}$,
then $\int_{\mathbf{CP}^{1}}\,a+b\coloneqq\int_{\mathbf{CP}^{1}}\left(a+b\right).$ 

Local coordinates on minitwistor space $\mathbf{MT}$ are denoted
by \textbf{$\mathsf{Z}^{I}\coloneqq\big(\lambda^{A},\mu_{\dot{A}}$$\big)$},
and to avoid unnecessary repetition, spinor indices in the arguments
of functions (or functionals) will be shown only at the point of their
initial definition. 
\end{notation*}

\subsection{Physical Motivation\label{subsec:Physical-Motivation}}

Twistor string theory\footnote{See \citet{abou2008einstein,adamo2016twistor,geyer2014ambitwistor,geyer2015ambitwistor,geyer2022sagex,roiban2004googly,witten2004perturbative,Wolf:2006me}.}
may be formulated as a theory of maps $\big(Z^{I},\widetilde{Z}^{I}\big):\mathbf{CP}^{1}\longrightarrow\mathbf{PT}^{s}\times\mathbf{PT}^{s}$,
where $Z^{I}$ and $\widetilde{Z}^{I}$ denote holomorphic and anti-holomorphic
fields, respectively, encoding the embedding of the Riemann sphere
into supersymmetric projective twistor space $\mathbf{PT}^{s}$. The
action functional is obtained by gauging the action of a holomorphic
$\beta\gamma$-system, where the gauging procedure implements the
local rescaling symmetry of the \emph{projective} geometry of twistor
space.

Explicitly, the action may be expressed as:
\begin{equation}
\underset{\mathbf{CP}^{1}}{\int}d\sigma\,\,\,Y_{I}\overline{\nabla}Z^{I}+\widetilde{Y}_{I}\nabla\widetilde{Z}^{I}+...,
\end{equation}
where $Y_{I}$ and $\widetilde{Y}_{I}$ are the canonical conjugate
fields to $Z^{I}$ and $\widetilde{Z}^{I}$, respectively. The ellipsis
$\left(...\right)$ represents contributions from an auxiliary conformal
matter system, which gives rises to worldsheet WZNW currents necessary
for anomaly cancellation. The covariant derivatives $\nabla=\partial+\boldsymbol{a}$
and $\overline{\nabla}=\overline{\partial}+\overline{\boldsymbol{a}}$
are Dolbeault operators twisted by gauge fields $\boldsymbol{a}$
and $\overline{\boldsymbol{a}}$ corresponding to the gauging of the
rescaling symmetry of projective twistor space, under which the fields
transform as:
\begin{equation}
Z^{I}\,\mapsto\,\alpha Z^{I},\,\,\,Y_{I}\,\mapsto\,\alpha^{-1}Y_{I},\,\,\,\overline{\boldsymbol{a}}\,\mapsto\,\overline{\boldsymbol{a}}\,-\,\overline{\partial}\,\log\alpha,
\end{equation}
where $\alpha:\mathscr{U}\subseteq\mathbf{CP}^{1}\longrightarrow\mathbf{C}^{*}$
is a nowhere vanishing holomorphic function, and $\mathscr{U}$ the
domain of a local trivialisation. Analogous transformations hold in
the anti-holomorphic sector for $\widetilde{Z}^{I}$, $\widetilde{Y}_{I}$,
and the gauge field $\boldsymbol{a}$.

We propose to follow a similar construction in our theory, with the
goal of deriving equations of motion that describe the embedding of
the celestial sphere into minitwistor space $\mathbf{MT}$ as a minitwistor
line. Unlike the anomaly-free twistor string theory described above,
our theory will be defined only at the semiclassical level, due to
the presence of quantum anomalies that preclude a full quantum formulation.
In the semiclassical limit, we will demonstrate that the on-shell
effective action reproduces the celestial leaf amplitudes associated
with tree-level scattering processes for gluons in $\mathcal{N}=4$
SYM theory and for gravitons in $\mathcal{N}=8$, in configurations
characterised by maximal helicity violation.

\subsection{Action Functional}

\subsubsection*{First Step}

Recall that the celestial sphere is modelled as the complex projective
line. In addition, the minitwistor space $\mathbf{MT}$ admits a holomorphic
embedding of the Riemann sphere into minitwistor lines defined by
the incidence relation. This embedding is formalised through the following
sequence of constructions.

Let $F_{\dot{A}}\left(X_{A\dot{A}},\lambda^{A}\right)$ be a holomorphic
map\footnote{Rigorously, $F_{\dot{A}}$ is a section of the projective spinor bundle.},
possessing homogeneity of degree one with respect to the spinor $\lambda^{A}$
and invariant under re-scalings of the ``spacetime'' coordinate $X_{A\dot{A}}\in H_{3}^{+}$,
such that:
\begin{equation}
F_{\dot{A}}\left(X_{A\dot{A}},\lambda^{A}\right)\,\coloneqq\,\lambda^{A}\,\frac{X_{A\dot{A}}}{\left|X\right|}.
\end{equation}
The minitwistor line $\mathcal{L}\left(X\right)$ in $\mathbf{MT}$
associated with $X_{A\dot{A}}$ is defined as the locus of points
satisfying the incidence relation:
\begin{equation}
\mathcal{L}\left(X\right)\,\coloneqq\,\big\{\,\left(\lambda^{A},\mu_{\dot{A}}\right)\in\mathbf{MT}\,\,\,\big|\,\,\,\mu_{\dot{A}}=F_{\dot{A}}\left(X,\lambda\right)\,\big\}.
\end{equation}
Following the sheaf-theoretic conventions employed by \citet{forster1981compact},
the restriction homomorphism $\rho_{X}\coloneqq\rho_{\mathcal{L}\left(X\right)}$
associated with the curve $\mathcal{L}\left(X\right)$ acts on sections
of the holomorphic vector bundle $\mathcal{O}\left(p,q\right)\longrightarrow\text{\textbf{MT}}$
by mapping a representative $g$ of a cohomology class as follows:
\begin{equation}
\rho_{X}\left(g\right)\left(\lambda^{A}\right)\,\coloneqq\,g\left(\lambda^{A},F_{\dot{A}}\left(X,\lambda\right)\right).
\end{equation}

Next, consider the canonical surjection $\pi_{X}:\mathcal{L}\left(X\right)\longrightarrow\mathbf{CP}^{1}$,
which projects the curve $\mathcal{L}\left(X\right)$ onto the Riemann
sphere. The natural embedding of the celestial sphere into the quadric
$\mathbf{MT}$ (in the form of a minitwistor line) can be understood
as a section of this fibration:
\begin{equation}
\sigma_{X}:\mathbf{CP}^{1}\longrightarrow\mathcal{L}\left(X\right),\,\,\,\sigma_{X}\left(\lambda^{A}\right)\,\coloneqq\,\left(\lambda^{A},F_{\dot{A}}\left(X,\lambda\right)\right),
\end{equation}
where, by construction, the composition satisfies $\pi_{X}\circ\sigma_{X}=id_{\mathbf{CP}^{1}}$.

On the other hand, the holomorphic function $F_{\dot{A}}\left(X,\lambda\right)$,
which \emph{defines} the embedding of the Riemann sphere into the
minitwistor space through the incidence relation, is \emph{entirely}
determined by its homogeneity properties and holomorphicity. Explicitly,
it satisfies:
\begin{equation}
F_{\dot{A}}\left(X,\cdot\right)\in\Omega^{0,1}\left(X,\mathcal{O}\left(1\right)\right),\,\,\,\overline{\partial}\,\big|_{X}\,F_{\dot{A}}\left(X,\lambda\right)=0.
\end{equation}

Therefore, by analogy with twistor string theory, one might be tempted
to introduce an action integral over the Riemann sphere $\mathbf{CP}^{1}$,
in terms of the embedding function $F_{\dot{A}}$, as follows:
\begin{equation}
\underset{\mathbf{CP}^{1}}{\int}D\lambda\,\,\,F_{\dot{A}}\,\overline{\partial}\,\big|_{X}\,F^{\dot{A}}.
\end{equation}
The corresponding equations of motion would yield the constraint $\overline{\partial}\,\big|_{X}F_{\dot{A}}\left(X,\lambda\right)=0$,
which, as previously noted, completely determines the embedding map
$\sigma_{X}:\mathbf{CP}^{1}\longrightarrow\mathcal{L}\left(X\right)$. 

However, the action integral written above is mathematically ill-defined
due to a lack of projective invariance. Consider the integration measure
$D\lambda\coloneqq\varepsilon_{AB}\lambda^{A}d\lambda^{B}$, which
transforms under a rescaling $\lambda^{A}\mapsto\alpha\lambda^{A}$
as $D\lambda\mapsto\alpha^{2}D\lambda$. Simultaneously, the ``kinetic
term'' transforms as:
\begin{equation}
F_{\dot{A}}\,\overline{\partial}\,\big|_{X}\,F^{\dot{A}}\mapsto\alpha^{2}F_{\dot{A}}\,\overline{\partial}\,\big|_{X}\,F^{\dot{A}}.
\end{equation}
Thus, the integrand $D\lambda\,\,\,F_{\dot{A}}\,\overline{\partial}\,\big|_{X}\,F^{\dot{A}}$
fails to exhibit the required projective invariance under the transformation
$\lambda^{A}\mapsto\alpha\lambda^{A}$, precluding its interpretation
as a meaningful action integral.

\subsubsection*{Second Step}

Our objective is to construct an action functional whose associated
Euler-Lagrange equations precisely reproduce the embedding of the
celestial sphere as a minitwistor line in  $\mathbf{MT}$. A formulation
of such an action can be introduced as follows\footnote{A similar solution was formulated by \citet{adamo2021twistor}, \citet{sharma2022twistor}
and \citet{chiou2005massless}. For a mathematically rigorous discussion,
see \citet{samann2006aspects,samann2009mini}, \citet{dunajski2009twistor}
and \citet{adamo2018minitwistors}.}.

Let $(\kappa_{1}^{A},\kappa_{2}^{A})$ denote a normalised spinor
basis satisfying the condition $\varepsilon_{AB}\kappa_{1}^{A}\kappa_{2}^{B}=1$.
The conic curve $\mathcal{L}\left(X\right)$ can be parametrised by
trivialising the fibration $\pi_{X}:\mathcal{L}\left(X\right)\longrightarrow\mathbf{CP}^{1}$
through the introduction of local coordinates $\omega^{A}$, defined
by:
\begin{equation}
\omega^{A}\,\coloneqq\,\omega_{1}\kappa_{1}^{A}\,+\,\omega_{2}\,\kappa_{2}^{A}.
\end{equation}
These coordinates $\omega^{A}$ are postulated to be projectively
related to the minitwistor coordinates $\lambda^{A}$ through the
relation:
\begin{equation}
\lambda^{A}\,\equiv\,\frac{\omega^{A}}{\omega_{1}\omega_{2}}\,=\,\frac{\kappa_{1}^{A}}{\omega_{2}}\,+\,\frac{\kappa_{2}^{A}}{\omega_{1}}.\label{eq:-24}
\end{equation}
Within this trivialisation, the holomorphic function $F_{\dot{A}}\left(X,\lambda\right)$,
which determines the embedding map $\sigma_{X}:\mathbf{CP}^{1}\longrightarrow\mathcal{L}\left(X\right)$,
yields a representative function:
\begin{equation}
M_{\dot{A}}\left(X_{A\dot{A}},\omega^{A}\right)\,\,\,\coloneqq\,\,\,F_{\dot{A}}\left(X_{A\dot{A}},\lambda^{A}\left(\omega\right)\right)\,\,\,=\,\,\,\frac{\kappa_{1}^{A}}{\omega_{2}}\,\frac{X_{A\dot{A}}}{\left|X\right|}\,+\,\frac{\kappa_{2}^{A}}{\omega_{1}}\,\frac{X_{A\dot{A}}}{\left|X\right|}.
\end{equation}

Two observations are now in order. First, the projective relation
between $\lambda^{A}$ and $\omega^{A}$ guarantees that the expression
for $M_{\dot{A}}$ can be inverted to recover $F_{\dot{A}}$. Hence,
specifying $M_{\dot{A}}$ fully determines the embedding $\sigma_{X}:\mathbf{CP}^{1}\longrightarrow\mathcal{L}\left(X\right)$.
Second, the function $M_{\dot{A}}$ is uniquely characterised as \emph{the}
solution to the following first-order partial differential equation\footnote{Recall that $\overline{\delta}\left(z\right)\,\coloneqq\,\left(2\pi i\right)^{-1}\,\overline{\partial}\,z^{-1}$
for all $z\in\mathbf{C}^{*}$.}:
\begin{equation}
\overline{\partial}\,\big|_{X}\,M_{\dot{A}}\left(X,\omega\right)\,=\,2\pi i\,\overline{\delta}\left(\omega_{2}\right)\,w_{1\dot{A}}+2\pi i\,\overline{\delta}\left(\omega_{1}\right)\,w_{2\dot{A}},\label{eq:-23}
\end{equation}
where the ``boundary conditions'' are specified by the residues at
$\omega_{2}$ and $\omega_{1}$, respectively, as:
\begin{equation}
w_{1\dot{A}}\,=\,\kappa_{1}^{A}\,\frac{X_{A\dot{A}}}{\left|X\right|},\,\,\,w_{2\dot{A}}\,=\,\kappa_{2}^{A}\,\frac{X_{A\dot{A}}}{\left|X\right|}.
\end{equation}

The above discussion implies that, in seeking an action functional
for a \emph{minitwistor celestial CFT}, the function $M_{\dot{A}}$
may be regarded as a dynamical variable whose Euler-Lagrange equations
yield precisely the differential equation (\ref{eq:-23}), while the
boundary conditions are enforced through external source terms.

Further, the function $M_{\dot{A}}$, as opposed to its counterpart
$F_{\dot{A}}$, possesses an additional advantage due to its homogeneity
properties in the trivialisation $\omega^{A}$. Specifically, $M_{\dot{A}}$
is homogeneous of degree $-1$ in $\omega^{A}$, which renders the
kinetic term:
\begin{equation}
M_{\dot{A}}\,\overline{\partial}\,\big|_{X}\,M^{\dot{A}},
\end{equation}
suitably weighted for integration against the measure $D\omega\coloneqq\varepsilon_{AB}\omega^{A}d\omega^{B}$.
Indeed, under a rescaling $\omega^{A}\mapsto\alpha\omega^{A}$, the
measure transforms as $D\omega\mapsto\alpha^{2}D\omega$, while the
kinetic term transforms as:
\begin{equation}
M_{\dot{A}}\,\overline{\partial}\,\big|_{X}\,M^{\dot{A}}\,\mapsto\,\alpha^{-2}M_{\dot{A}}\overline{\partial}\,\big|_{X}\,M^{\dot{A}},
\end{equation}
ensuring that the integrand:
\begin{equation}
D\omega\,\,\,M_{\dot{A}}\overline{\partial}\,\big|_{X}\,M^{\dot{A}},
\end{equation}
is weightless under such re-scalings. 

Consequently, we are naturally lead to consider the action functional:
\begin{equation}
\mathcal{S}_{0}\big[M_{\dot{A}}(\omega^{A})\big]\,=\,\frac{1}{b}\underset{\mathcal{L}\left(X\right)}{\int}\,D\omega\,\,\,M_{\dot{A}}\,\overline{\partial}\,\big|_{X}\,M^{\dot{A}}\,+4\pi i\overline{\delta}\left(\omega_{2}\right)\left[w_{1}M\right]+4\pi i\overline{\delta}\left(\omega_{1}\right)\left[w_{2}M\right],
\end{equation}
where we have employed the spinor-helicity bracket notation to express
the contraction of $M^{\dot{A}}$ with the external sources as $\left[w_{i}M\right]\coloneqq w_{i\dot{A}}M^{\dot{A}}$
for $i=1,2$. Although the bracket notation is employed for the source
terms, we refrain from using it in the kinetic term to avoid notational
clutter.

Finally, a more suggestive rewriting of the action is possible by
interpreting the sources $w_{i\dot{A}}$ ($i=1,2$) as originating
from an external current\footnote{We follow the terminology of \citet{schwinger1966particles}.}
by defining:
\begin{equation}
J_{\dot{A}}\,\,\,\coloneqq\,\,\,4\pi i\,\overline{\delta}\left(\omega_{2}\right)\,\kappa_{1}^{A}\,\frac{X_{A\dot{A}}}{\left|X\right|}\,+\,4\pi i\,\overline{\delta}\left(\omega_{1}\right)\,\kappa_{2}^{A}\,\frac{X_{A\dot{A}}}{\left|X\right|}.\label{eq:-37}
\end{equation}
The action functional then takes the form:
\begin{equation}
\mathcal{S}_{0}=\,\frac{1}{b}\underset{\mathcal{L}\left(X\right)}{\int}\,D\omega\,\,\,M_{\dot{A}}\,\overline{\partial}\,\big|_{X}\,M^{\dot{A}}\,+\,\left[JM\right],
\end{equation}
where $[JM]$ denotes the contraction of $J_{\dot{A}}$ with $M^{\dot{A}}$.

\subsection{Supersymmetry\label{subsec:Supersymmetry}}

Our extension of the RSVW formalism to celestial leaf amplitudes in
$\mathcal{N}=4$ SYM theory and $\mathcal{N}=8$ Supergravity shows
that the corresponding superamplitudes can be written as a Fourier
transform on minitwistor superspace. This implies that any attempt
to reproduce these amplitudes using a sigma model on the celestial
sphere must include a supersymmetric version of $\mathbf{MT}$ as
the target space. 

In the subsequent subsections, we shall present an informal discussion
of the minitwistor superspace, referring the reader to \citet{rogers2007supermanifolds}
for a rigorous mathematical treatment\footnote{See also \citet{dewitt1992supermanifolds,leites1980introduction,manin1997introduction}.}.
Thereafter, we proceed to extend our construction of the sigma model
action functional by formulating a generalisation in which the domain
is the celestial \emph{supersphere} and the corresponding target space
is taken to be the minitwistor \emph{superspace}.

\subsubsection{Projective Superspace}

We begin by observing that the minitwistor space $\mathbf{MT}$ is
the space of oriented geodesics on the hyperboloid $H_{3}^{+}$. A
model for the hyperbolic geometry of $H_{3}^{+}$ can be derived from
the projective geometry of the three-dimensional real projective space
$\mathbf{RP}^{3}$, as reviewed in the Appendix. Accordingly, our
discussion of the minitwistor \emph{superspace} $\mathbf{MT}^{2|\mathcal{N}}$
starts with a preliminary definition of supermanifold\emph{ $\mathbf{RP}^{3|2\mathcal{N}}$.} 

We identify the projective superspace $\mathbf{RP}^{3|2\mathcal{N}}$
with the trivial superbundle $\mathbf{RP}^{3}\times\mathbf{CP}^{0|2\mathcal{N}}$,
in which the three-dimensional real projective space $\mathbf{RP}^{3}$
serves as the base manifold and the fibre is parametrised by ``fermionic
degrees of freedom.'' To be precise, the typical fibre is charted
by the introduction of Grassmann-valued \emph{homogeneous} spinorial
coordinates $\theta_{A}^{\alpha}$, subject to an equivalence relation
under projective rescaling by a nonzero complex scalar $\alpha$,
such that $\theta_{A}^{\alpha}\sim\alpha\theta_{A}^{\alpha}$. The
normalisation of the fermionic ``dimensions'' is imposed through a
Berezin integral over the fibre, given by:
\begin{equation}
\int d^{0|2}\theta\,\,\,\theta_{A}^{\alpha}\theta_{B}^{\alpha}\,=\,\varepsilon_{AB}.
\end{equation}

The natural orientation of the total space of the superbundle $\mathbf{RP}^{3|2\mathcal{N}}$
is induced by the Berezin-de Witt volume superform, defined as:
\begin{equation}
D^{3|2\mathcal{N}}\mathbb{X}\,\coloneqq\,\frac{D^{3}X}{\left|X\right|^{4}}\,\wedge\,d^{0|2\mathcal{N}}\theta.
\end{equation}

\subsubsection{Minitwistor Superspace}

The minitwistor superspace $\mathbf{MT}^{2|\mathcal{N}}$ (associated
with the hyperbolic geometry modelled on $\mathbf{RP}^{3|2\mathcal{N}}$)
is constructed by extending the bosonic minitwistor space $\mathbf{MT}$
through the inclusion of Grassmann-odd coordinates that encode the
``fermionic dimensions.'' More precisely, we define $\mathbf{MT}^{2|\mathcal{N}}$
as the trivial superbundle:
\begin{equation}
\mathbf{MT}^{2|\mathcal{N}}\,\coloneqq\,\mathbf{MT}\times\mathbf{CP}^{0|\mathcal{N}},
\end{equation}
where the base manifold is the bosonic minitwistor space $\mathbf{MT}$,
and the typical fibre is charted by the Grassmann-odd \emph{homogeneous}
scalar coordinates $\zeta^{\alpha}$, with $\alpha\in\{1,...,\mathcal{N}\}$.

A bundle superchart on $\mathbf{MT}^{2|\mathcal{N}}$ is given by
a local trivialisation $(\mathscr{U},\mathsf{Z}^{I})$, where $\mathscr{U}\subseteq\mathbf{MT}^{2|\mathcal{N}}$
denotes an open neighbourhood and $\mathsf{Z}^{I}:\mathscr{U}\longrightarrow U\times\mathbf{CP}^{0|\mathcal{N}}$
is a local coordinate map. The image of $\mathsf{Z}^{I}$ is contained
in $U\subset\mathbf{CP}^{1|0}\times\mathbf{CP}^{1|0}$, and the coordinate
functions $\mathsf{Z}^{I}$ decompose as:
\begin{equation}
p\in\mathscr{U}\mapsto\mathsf{Z}^{I}\left(p\right)\coloneqq(\lambda^{A}\left(p\right),\mu_{\dot{A}}\left(p\right),\zeta^{\alpha}\left(p\right)).
\end{equation}

The canonical orientation of the minitwistor superspace $\mathbf{MT}^{2|\mathcal{N}}$
is specified by the Berezin-de Witt volume superform on the total
space of the superbundle, and is defined by:
\begin{equation}
D^{2|\mathcal{N}}\mathsf{Z}\,\coloneqq\,D\lambda\,\wedge\,D\mu\,\wedge\,d^{0|\mathcal{N}}\zeta.
\end{equation}

\subsubsection{Minitwistor Superlines}

In minitwistor superspace, the minitwistor lines are generalised to
\emph{superlines} $\mathcal{L}\left(X,\theta\right)$, each associated
with a ``spacetime'' point $(X_{A\dot{A}},\theta_{A}^{\alpha})\in\mathbf{RP}^{3|2\mathcal{N}}$.
The incidence relation defining these supercurves is determined by
sections of the \emph{projective spinor superbundle}, $\mathbf{PS}_{3|2\mathcal{N}}\coloneqq\mathbf{RP}^{3|2\mathcal{N}}\times\mathbf{CP}^{1|0}$,
which are given by:
\begin{equation}
F_{\dot{A}}(X_{A\dot{A}},\theta_{A}^{\alpha};\lambda^{A})\,\coloneqq\,\lambda^{A}\,\frac{X_{A\dot{A}}}{\left|X\right|},\,\,\,G^{\alpha}(X_{A\dot{A}},\theta_{A}^{\alpha};\lambda^{A})\,\coloneqq\,\lambda^{A}\theta_{A}^{\alpha}.
\end{equation}
The supercurve $\mathcal{L}\left(X,\theta\right)$, which we shall
refer to as the \emph{minitwistor superline }based at $\left(X,\theta\right)$,
is then defined as the locus of points in minitwistor superspace satisfying
the incidence relation:
\begin{equation}
\mathcal{L}\left(X,\theta\right)\,\coloneqq\,\big\{\,\mathsf{Z}^{I}\,\in\,\mathbf{MT}^{2|\mathcal{N}}\,\big|\,\mu_{\dot{A}}=F_{\dot{A}}\left(X,\theta;\lambda\right),\,\zeta^{\alpha}=G^{\alpha}\left(X,\theta;\lambda\right)\,\big\},
\end{equation}
where $\mathsf{Z}^{I}=(\lambda^{A},\mu_{\dot{A}},\zeta^{\alpha})$
denotes local coordinates on the minitwistor superspace.

To formalise the restriction of holomorphic sections to a curve $\mathcal{L}\left(X,\theta\right)$,
let us consider a representative $g$ of a cohomology class associated
to the vector bundle $\mathcal{O}\left(p,q\right)\longrightarrow\mathbf{MT}$.
The restriction homomorphism $\rho_{\mathcal{L}\left(X,\theta\right)}$
maps this section to a section of the fibration $\mathcal{L}\left(X,\theta\right)\longrightarrow\mathbf{CP}^{1}$
by evaluating it along the incidence relations defining $\mathcal{L}\left(X,\theta\right)$:
\begin{equation}
g\,\big|_{\mathcal{L}\left(X,\theta\right)}\left(\lambda^{A}\right)\,\coloneqq\,\rho_{\mathcal{L}\left(X,\theta\right)}\left(g\right)\left(\lambda^{A}\right)\,\coloneqq g\left(\lambda^{A},F_{\dot{A}}\left(X,\theta;\lambda\right),G^{\alpha}\left(X,\theta;\lambda\right)\right).
\end{equation}

\paragraph*{Embedding as a Section.}

We now consider the canonical projection $\pi_{\left(X,\theta\right)}:\mathcal{L}\left(X,\theta\right)\longrightarrow\mathbf{CP}^{1}$
which maps each point on the supercurve to its projective coordinate
$[\lambda^{A}]\in\mathbf{CP}^{1}$. A holomorphic embedding of the
celestial sphere onto the minitwistor superline $\mathcal{L}\left(X,\theta\right)$
can thus be identified with a section of this fibration. We denote
this section by:
\begin{equation}
\sigma_{\left(X,\theta\right)}:\mathbf{CP}^{1|0}\longrightarrow\mathcal{L}\left(X,\theta\right),\,\,\,\sigma_{\left(X,\theta\right)}\left(\lambda^{A}\right)\coloneqq\left(\lambda^{A},F_{\dot{A}}\left(X,\theta;\lambda\right),G^{\alpha}\left(X,\theta;\lambda\right)\right).
\end{equation}
Hence, $\pi_{\left(X,\theta\right)}\circ\sigma_{\left(X,\theta\right)}=id_{\mathbf{CP}^{1}}$
by construction.

\subsubsection{Construction of the Action}

\paragraph*{First Step: Dynamical Variables.}

To formulate a theory whose solutions to the equations of motion correspond
to embeddings of the celestial sphere into minitwistor superspace
$\mathbf{MT}^{2|\mathcal{N}}$, it suffices to specify the holomorphic
sections $F_{\dot{A}}\left(X,\theta;\lambda\right)$ and $G^{\alpha}\left(X,\theta;\lambda\right)$
introduced previously as dynamical variables. However, as observed,
both $F_{\dot{A}}\left(X,\theta;\lambda\right)$ and $G^{\alpha}\left(X,\theta;\lambda\right)$
exhibit homogeneity of degree one in the spinor variable $\lambda^{A}$.
Consequently, any attempt to construct a weightless kinetic term using
these functions directly would be inconsistent with the homogeneity
degree of the measure $D\lambda=\varepsilon_{AB}\lambda^{A}d\lambda^{B}$.

To resolve this issue, we employ the same procedure used in the bosonic
case: we introduce a normalised spinor basis $(\kappa_{1}^{A},\kappa_{2}^{A})$
satisfying the normalisation condition $\varepsilon_{AB}\kappa_{1}^{A}\kappa_{2}^{B}=1$,
and we define a trivialisation of the fibration $\pi_{\left(X,\theta\right)}:\mathcal{L}\left(X,\theta\right)\longrightarrow\mathbf{CP}^{1}$
via a new spinor $\omega^{A}$ expressed as a linear combination of
the basis spinors, $\omega^{A}\coloneqq\omega_{1}\kappa_{1}^{A}+\omega_{2}\kappa_{2}^{A}$.
We also postulate that the spinor $\omega^{A}$ is projectively related
to $\lambda^{A}$ by Eq. (\ref{eq:-24}).

The next step is to reformulate the section $F_{\dot{A}}\left(X,\theta;\lambda\right)$
in terms of this new spinor basis. Specifically, we define the function:
\begin{equation}
M_{\dot{A}}\left(X_{A\dot{A}},\theta_{A}^{\alpha};\omega^{A}\right)\,\coloneqq\,F_{\dot{A}}\left(X_{A\dot{A}},\theta_{A}^{\alpha};\omega^{A}\left(\lambda\right)\right),
\end{equation}
which contains the same information as $F_{\dot{A}}\left(X,\theta;\lambda\right)$
in terms of the new spinor $\omega^{A}$. The properties of this function
have already been analysed in our prior discussion on the bosonic
model.

We now focus our attention on $G^{\alpha}\left(X,\theta;\lambda\right)$,
which we reformulate analogously as a new function:
\begin{equation}
N^{\alpha}\left(X_{A\dot{A}},\theta_{A}^{\alpha};\omega^{A}\right)\,\coloneqq\,G^{\alpha}\left(X_{A\dot{A}},\theta_{A}^{\alpha};\omega^{A}\left(\lambda\right)\right),
\end{equation}
which can be rewritten as:
\begin{equation}
N^{\alpha}\left(X,\theta;\omega\right)\,=\,\frac{\widetilde{w}_{1}^{\alpha}}{\omega_{2}}\,+\,\frac{\widetilde{w}_{2}^{\alpha}}{\omega_{1}};\,\,\,\text{where}\,\,\,\widetilde{w}_{i}^{\alpha}\,\coloneqq\,\kappa_{i}^{A}\theta_{A}^{\alpha}\,\left(i=1,2\right).
\end{equation}
Since $\lambda^{A}$ and $\omega^{A}$ are projectively related, the
above expression for $N^{\alpha}\left(X,\theta;\omega\right)$ can
be inverted to recover $G^{\alpha}\left(X,\theta;\lambda\right)$
if the function $N^{\alpha}\left(X,\theta;\omega\right)$ is given.
Therefore, specifying the embedding $\sigma_{\left(X,\theta\right)}:\mathbf{CP}^{1}\longrightarrow\mathcal{L}\left(X,\theta\right)$
into a minitwistor superline is equivalent to specifying the pair
of functions $M_{\dot{A}}\left(X,\theta;\omega\right)$ and $N^{\alpha}\left(X,\theta;\omega\right)$.

We thus arrive at the following conclusion: the set of functions $\{M_{\dot{A}},N^{\alpha}\}$
can be taken as the dynamical fields in the supersymmetric generalisation
of the minitwistor celestial CFT.

An additional remark is in order about the function $N^{\alpha}\left(X,\theta;\omega\right)$.
It can be seem that this function is uniquely determined as \emph{the}
solution to the following partial differential equation:
\begin{equation}
\overline{\partial}\,\big|_{\mathcal{L}\left(X,\theta\right)}\,N^{\alpha}\left(X,\theta;\lambda\right)\,=\,2\pi i\,\overline{\delta}\left(\omega_{2}\right)\,\widetilde{w}_{1}^{\alpha}\,+\,2\pi i\,\overline{\delta}\left(\omega_{1}\right)\,\widetilde{w}_{2}^{\alpha},\label{eq:-38}
\end{equation}
where the boundary conditions are specified by the residues $\widetilde{w}_{1}^{\alpha}$
and $\widetilde{w}_{2}$ at the poles $\omega_{2}$ and $\omega_{1}$,
respectively. 

\paragraph*{Second Step: Action Functional }

We now arrive at the construction of an action that gives rise to
a well-posed variational principle, whose associated Euler-Lagrange
equations describe the embedding of the celestial sphere into minitwistor
superspace as minitwistor supercurves. The proposed action is given
by:
\begin{equation}
\mathcal{S}\,[M_{\dot{A}},N^{\alpha},e_{\alpha}]\,=\,\underset{\mathcal{L}\left(X,\theta\right)}{\int}\,D\omega\,\,\,M_{\dot{A}}\,\overline{\partial}\,\big|_{\mathcal{L}\left(X,\theta\right)}\,M^{\dot{A}}\,+\,e_{\alpha}\,\overline{\partial}\,\big|_{\mathcal{L}\left(X,\theta\right)}\,N^{\alpha}\,+\,J_{\dot{A}}M^{\dot{A}}-e_{\alpha}K^{\alpha}.
\end{equation}
The external bosonic current $J_{\dot{A}}$ appearing in the action
is identical to the one introduced in Eq. (\ref{eq:-37}), while the
newly introduced fermionic current $K^{\alpha}$ is defined as:
\begin{equation}
K^{\alpha}\,\coloneqq\,2\pi i\,\overline{\delta}\left(\omega_{2}\right)\,\kappa_{1}^{A}\theta_{A}^{\alpha}\,+\,2\pi i\,\overline{\delta}\left(\omega_{1}\right)\,\kappa_{2}^{A}\theta_{A}^{\alpha}.
\end{equation}
Note that $e_{\alpha}$ is a Lagrange multiplier, and that the equations
of motion are given by Eqs. (\ref{eq:-23}) and (\ref{eq:-38}).

\paragraph*{Celestial Supersphere.}

The final level of abstraction is achieved through the extension of
the celestial sphere to its $\mathcal{N}=2$ supersymmetric generalisation,
the \emph{celestial supersphere}. This extension is obtained by adjoining
to the homogeneous coordinates $[\omega^{A}]\in\mathbf{CP}^{1}$ two
Grassmann-odd variables, $\eta$ and $\bar{\eta}$, which satisfy
the Berezin normalisation condition $\int d^{0|2}\eta\,\,\,\eta\bar{\eta}=1$.
The celestial supersphere is equipped with a natural orientation given
by the Berezin-de Witt superform:
\begin{equation}
d^{1|2}z\,\coloneqq\,D\omega\,\wedge\,d^{0|2}\eta.
\end{equation}
Hereafter, we denote by $\widetilde{\mathcal{L}}\left(X,\theta\right)$
the minitwistor line obtained from the embedding $\mathbf{CP}^{1|2}\longrightarrow\widetilde{\mathcal{L}}\left(X,\theta\right)$. 

To describe the local geometry of the celestial supersphere, we introduce
a \emph{super-vielbein $E_{\dot{A}}^{\,\,\,\alpha}$ }that satisfies
the orthonormality condition $E_{\dot{A}}^{\,\,\,\alpha}E_{\,\,\,\beta}^{\dot{A}}=\delta_{\,\,\,\beta}^{\alpha}$.
Now, note the following \emph{algebraic }identities:
\begin{align}
 & M_{\dot{A}}\,\overline{\partial}\,\big|_{\mathcal{L}\left(X,\theta\right)}\,M^{\dot{A}}+\,e_{\alpha}\,\overline{\partial}\,\big|_{\mathcal{L}\left(X,\theta\right)}\,N^{\alpha}\\
 & =\int d^{0|2}\eta\,\,\,\big(\,\eta M_{\dot{A}}+\bar{\eta}E_{\dot{A}}^{\,\,\,\alpha}e_{\alpha}\,\big)\,\overline{\partial}\,\big|_{\mathcal{L}\left(X,\theta\right)}\,\big(\,\bar{\eta}M^{\dot{A}}-\eta E_{\,\,\,\beta}^{\dot{A}}N^{\beta}\,\big),
\end{align}
and:
\begin{align}
 & J_{\dot{A}}M^{\dot{A}}-e_{\alpha}K^{\alpha}\\
 & =\int d^{0|2}\eta\,\,\,\big(\,\eta J_{\dot{A}}+\bar{\eta}E_{\dot{A}}^{\,\,\,\alpha}e_{\alpha}\,\big)\,\overline{\partial}\,\big|_{\mathcal{L}\left(X,\theta\right)}\,\big(\,\bar{\eta}M^{\dot{A}}+\eta E_{\,\,\,\beta}^{\dot{A}}K^{\beta}\,\big).
\end{align}

To facilitate a more compact notation, we define the \emph{superfields:}
\begin{equation}
P_{\dot{A}}\left(\omega,\eta,\bar{\eta}\right)\,\coloneqq\,\eta M_{\dot{A}}+\bar{\eta}E_{\dot{A}}^{\,\,\,\alpha}e_{\alpha},\,\,\,Q^{\dot{A}}\left(\omega,\eta,\bar{\eta}\right)\,\coloneqq\,\bar{\eta}M^{\dot{A}}-\eta E_{\,\,\,\beta}^{\dot{A}}N^{\beta},
\end{equation}
and introduce the \emph{super-currents}: 
\begin{equation}
j_{\dot{A}}\left(\omega,\eta,\bar{\eta}\right)\,\coloneqq\,\eta J_{\dot{A}}+\bar{\eta}E_{\dot{A}}^{\,\,\,\alpha}e_{\alpha},\,\,\,k^{\dot{A}}\left(\omega,\eta,\bar{\eta}\right)\,\coloneqq\,\bar{\eta}M^{\dot{A}}+\eta E_{\,\,\,\beta}^{\dot{A}}K^{\beta}.
\end{equation}
Finally, the action for the supersymmetric minitwistor celestial CFT
can be expressed in terms of these superfields and super-currents
as:
\begin{equation}
\mathcal{S}[P,Q,j,k]\,=\,\frac{1}{b}\underset{\widetilde{\mathcal{L}}\left(X,\theta\right)}{\int}d^{1|2}z\,\,\,P_{\dot{A}}\,\overline{\partial}\,\big|_{\mathcal{L}\left(X,\theta\right)}\,Q^{\dot{A}}\,+\,[jk],
\end{equation}
where the bracket $[jk]$ denotes the spinor contraction of the super-currents
$j_{\dot{A}}$ and $k^{\dot{A}}$.

\subsection{Phenomenology}

In conventional superstring theory, phenomenological considerations
arise from the study of Calabi-Yau compactifications or D-brane configurations.
In twistor and ambitwistor string theories, such considerations are
incorporated within an auxiliary conformal matter system, which we
briefly discussed in Subsection \ref{subsec:Physical-Motivation}.
The minitwistor celestial CFTs constructed in this work follow an
analogous procedure: the relevant physical features are encoded in
an auxiliary matter model, which we shall now proceed to define.

\subsubsection{Fermionic System}

We begin with the following physical motivations. First, the generating
functional $\digamma[\boldsymbol{\omega}]$, derived in Subsection
\ref{subsec:Generating-Functional} for $\mathcal{N}=4$ SYM theory
and in Subsection \ref{subsec:Generating-Functional-1} for $\mathcal{N}=8$
Supergravity, is expressed as an integral over projective superspace
$\mathbf{RP}^{3|2\mathcal{N}}$ of the Quillen determinant.

On the other hand, let $\mathbf{G}$ be a gauge Lie group with Lie
algebra $\mathfrak{g}\simeq\left(T_{e}\left(\mathbf{G}\right),[\cdot,\cdot]\right)$,
Denote by $\{\mathsf{T}^{a}\}$ a basis of generators of $\mathfrak{g}$
in the fundamental representation $\pi$, satisfying the normalisation
condition $\mathsf{Tr}\left(\mathsf{T}^{a}\mathsf{T}^{b}\right)=\frac{1}{2}\boldsymbol{k}^{ab}$,
where $\boldsymbol{k}^{ab}$ denotes the Cartan-Killing form of $\mathbf{G}$.
The generators obey the commutation relations $[\mathsf{T}^{a},\mathsf{T}^{b}]=if^{abc}\mathsf{T}^{c}$,
where $f^{abc}$ are the structure constants of $\mathfrak{g}$.

Consider a $\mathfrak{g}$-valued connection one-form $\boldsymbol{\omega}=\boldsymbol{\omega}^{a}\mathsf{T}^{a}$
on a principal $\mathbf{G}$-bundle over the Riemann sphere $\mathbf{CP}^{1}$.
In addition, let $\left(q,\bar{q}\right)$ denote a fermionic system
defined on the associated vector bundle, whose typical fibre is isomorphic
to the representation space of $\pi$. The dynamics of the fermionic
system $\left(q,\bar{q}\right)$ is governed by the action functional:
\begin{equation}
\mathcal{S}_{(q,\bar{q})}\,\coloneqq\,\underset{\mathbf{CP}^{1}}{\int}d\sigma\,\,\,\bar{q}^{\text{i}}\left(\overline{\partial}+\boldsymbol{\omega}^{a}\mathsf{T}_{\text{ij}}^{a}\right)q^{\text{j}},
\end{equation}
where $d\sigma$ denotes the ``volume'' form on $\mathbf{CP}^{1}$
and $\overline{\partial}$ is the Dolbeault operator.

Now, it is well-known\footnote{See \citet{nair2005chern}.} that the
corresponding quantum effective action for this system is given by
the chiral determinant of the associated Dirac operator (or twisted
Dolbeault operator):
\begin{equation}
\mathcal{W}_{(q,\bar{q})}\,\propto\,\mathsf{Tr}\log\left(\overline{\partial}+\boldsymbol{\omega}\right).\label{eq:-57}
\end{equation}

Keeping these remarks in mind, we proceed by considering the principal
$\mathbf{G}$-bundle over a minitwistor supercurve $\mathcal{L}\left(X,\theta\right)$,
equipped with a $\mathfrak{g}$-valued connection one-form $\boldsymbol{\omega}=\boldsymbol{\omega}^{a}\mathsf{T}^{a}$.
Introduce the fermionic system $\left(q,\bar{q}\right)$ defined on
the associated vector bundle over $\mathcal{L}\left(X,\theta\right)$,
and define the superfields:
\begin{equation}
\psi\left(\omega,\eta,\bar{\eta}\right)\,\coloneqq\,\bar{\eta}\,q,\,\,\,\bar{\psi}\left(\omega,\eta,\bar{\eta}\right)\,\coloneqq\,\eta\,\bar{q}.
\end{equation}
We propose that the \emph{interaction term} in the action functional
is given by:
\begin{equation}
\mathcal{S}_{int}\,\coloneqq\,\underset{\mathcal{L}\left(X,\theta\right)}{\int}d^{1|2}z\,\,\,\bar{\psi}\left(\overline{\partial}+\boldsymbol{\omega}\right)\,\big|_{\mathcal{L}\left(X,\theta\right)}\,\psi.
\end{equation}
Therefore, the complete action describing the minitwistor celestial
CFT is:
\begin{equation}
\mathcal{I}\,=\,\,\frac{1}{b}\underset{\widetilde{\mathcal{L}}\left(X,\theta\right)}{\int}d^{1|2}z\,\,\,P_{\dot{A}}\,\overline{\partial}\,\big|_{\mathcal{L}\left(X,\theta\right)}\,Q^{\dot{A}}\,+\,[jk]\,+\,b\,\bar{\psi}\left(\overline{\partial}+\boldsymbol{\omega}\right)\,\big|_{\mathcal{L}\left(X,\theta\right)}\,\psi,\label{eq:-56}
\end{equation}
where the parameter $b$ controls the semiclassical approximation,
and is analogous to the Liouville coupling constant\footnote{Cf. \citet{ribault2005h3+}.}.

\subsubsection{Semiclassical Analysis}

We turn now to the task of demonstrating that the semiclassical effective
action arising from the theory described by Eq. (\ref{eq:-56}) yields
the generating functional $\digamma[\boldsymbol{\omega}]$.

Prior to presenting the path-integral formulation of the effective
action, we recall the key observation made in the concluding remarks
of Section \ref{sec:-Supersymmetric-Yang-Mills}. The dynamical fields
in the present theory correspond to the embedding maps of the celestial
sphere into minitwistor superspace $\mathbf{MT}^{2|\mathcal{N}}$,
realised as minitwistor superlines $\mathcal{L}\left(X,\theta\right)$.
Formally, these embeddings are represented by sections:
\begin{equation}
\sigma_{\left(X,\theta\right)}:\mathbf{CP}^{1}\longrightarrow\mathcal{L}\left(X,\theta\right),
\end{equation}
of the canonical fibration:
\begin{equation}
\pi_{\left(X,\theta\right)}:\mathcal{L}\left(X,\theta\right)\longrightarrow\mathbf{CP}^{1}.
\end{equation}
Thus, an integral such as:
\begin{equation}
\underset{\mathbf{RP}^{3|2\mathcal{N}}}{\int}\,D^{3|2\mathcal{N}}\mathbb{X}\,\,\,\underset{\mathcal{L}\left(X,\theta\right)}{\int}\,\,\,\left(...\right)\,\,\,,
\end{equation}
must be interpreted as an integral over the moduli space of \emph{embeddings
}of the celestial sphere into $\mathbf{MT}^{2|\mathcal{N}}$, rather
than an integral over multiple distinct celestial spheres parametrised
by coordinates $\mathbb{X}^{\hat{I}}=\left(X_{A\dot{A}},\theta_{A}^{\alpha}\right)$
in projective superspace $\mathbf{RP}^{3|2\mathcal{N}}$. This is
physically important, because semiclassically the path integral is
not summing over disconnected celestial spheres but rather over all
possible embeddings of a single celestial sphere as a minitwistor
superline.

Consequently, the semiclassical effective action $\mathcal{W}$ must
account for contributions from all possible embeddings of the celestial
sphere into $\mathbf{MT}^{2|\mathcal{N}}$, and is given by:
\begin{equation}
\mathcal{W}\,=\,-\lim_{b\rightarrow0^{+}}b\,\,\,\log\underset{\mathbf{RP}^{3|2\mathcal{N}}}{\int}\,D^{3|2\mathcal{N}}\mathbb{X}\,\,\,\int\left[dMdNd\psi d\bar{\psi}\right]\,\,\,\exp\left(-\mathcal{I}\right).
\end{equation}

In the semiclassical limit $b\rightarrow0^{+}$, the dominant contribution
to the path integral over the embedding fields $M_{\dot{A}}$ and
$N^{\alpha}$ arises from the saddle-point approximation, which enforces
the classical equations of motion, given by Eqs. (\ref{eq:-23}) and
(\ref{eq:-38}), and are precisely those that define the embedding
maps $\mathbf{CP}^{1}\longrightarrow\mathcal{L}\left(X,\theta\right)$.
Accordingly, the path integral over $M_{\dot{A}}$ and $N^{\alpha}$
in the limit $b\rightarrow0^{+}$ reduces to an evaluation of the
remaining terms in the action at the locus defined by the incidence
relation $\mathcal{L}\left(X,\theta\right)$, which is achieved by
the application of the restriction homomorphism $\rho_{\mathcal{L}\left(X,\theta\right)}$.

Having restricted the bosonic path integral to the moduli space of
embeddings of the celestial sphere, we now consider the fermionic
contributions. The path integral over the fermionic system $(\psi,\bar{\psi})$
results in the chiral determinant, which, as shown in Eq. (\ref{eq:-38}),
is the supersymmetric extension of the Quillen determinant. Consequently,
the semiclassical effective action $\mathcal{W}$ becomes:
\begin{equation}
\mathcal{W}\,=\,\underset{\mathbf{RP}^{3|2\mathcal{N}}}{\int}\,D^{3|2\mathcal{N}}\mathbb{X}\,\,\,\log\det\left(\overline{\partial}+\boldsymbol{\omega}\right)\,\big|_{\mathcal{L}\left(X,\theta\right)}.
\end{equation}

The above expression recovers the minitwistor generating functional
$\digamma[\boldsymbol{\omega}]$ for the background potential $\boldsymbol{\omega}$
appropriate to the spacetime theory under consideration. This demonstrates
that the minitwistor sigma model reproduces semiclassically the tree-level
MHV celestial leaf superamplitudes for gluons in $\mathcal{N}=4$
SYM theory and for gravitons in $\mathcal{N}=8$ Supergravity.

\section{Discussion\label{sec:Discussion}}

The supersymmetric minitwistor celestial CFTs developed in the previous
section reproduce the tree-level celestial leaf superamplitudes in
$\mathcal{N}=4$ SYM theory and $\mathcal{N}=8$ Supergravity for
MHV configurations. \citet{banerjee2021mhv} proposed that the MHV
sector of celestial CFT might serve as a minimal model for celestial
holography. The minitwistor celestial CFTs introduced here could contribute
to addressing this conjecture for the following reason.

The formalism presented in this paper relies on minitwistor wavefunctions,
defined as cohomology classes in minitwistor space. This approach
extends the RSVW prescription to celestial amplitudes and offers a
new perspective on the celestial leaf amplitudes. In this setup, tree-level
MHV leaf amplitudes are expressed as integrals over the moduli space
of minitwistor lines. It is therefore reasonable to suggest that celestial
amplitudes corresponding to next-to-MHV (NMHV) configurations could
be described by integrals over the moduli space of higher-degree curves
in minitwistor space.

We propose that future work should focus on extending the minitwistor
generating functional developed here to systematically incorporate
NMHV celestial amplitudes. Additionally, it would be worthwhile to
study a celestial version of the Cachazo-Svrcek-Witten (CSW) expansion
for leaf amplitudes. We expect that such an extension could offer
a concrete path toward proving the Banerjee-Gosh conjecture.

The integral over the moduli space of special curves in minitwistor
superspace that arises from our extension of the RSVW formalism to
celestial amplitudes is not a standard volume integral. This is because
the delta functions $\overline{\delta}_{\Delta_{i}}$, which localise
the superspace integral to the locus of incidence in minitwistor space,
have non-trivial homogeneity properties given by the celestial conformal
weights $\Delta_{i}$. It would be interesting to reformulate these
integrals using the formalism introduced by \citet{movshev2006berezinian}
and further elaborated by \citet{adamo2013moduli}.

In the semiclassical limit $b\rightarrow0^{+}$ of our minitwistor
celestial CFTs, where the parameter $b$ plays a role similar to the
Liouville coupling constant, we have been able to reproduce celestial
amplitudes at tree level. An important direction for future research
is to explore whether extending beyond the semiclassical limit could
produce loop corrections, similar to those studied in the context
of celestial Liouville theory in \citet{mol2024partial}. Another
promising approach would be to consider a complex scaling reduction
of Berkovits' original twistor string theory to minitwistor space.
Investigating whether the resulting minitwistor string theory can
generate NMHV celestial leaf amplitudes as integrals over the moduli
space of higher-degree algebraic curves in minitwistor superspace
$\mathbf{MT}^{2|\mathcal{N}}$ would be an important step forward.

\appendix

\section{Minitwistor Geometry\label{sec:Minitwistor-Geometry}}

The non-singular quadric $\mathbf{MT}\subset\mathbf{CP}^{3}$ upon
which our celestial CFTs will be defined corresponds to the minitwistor
space associated with the hyperboloid $\mathbf{H}_{3}$. Consequently,
we shall succinctly review how the hyperbolic geometry of $\mathbf{H}_{3}$
arises from the projective geometry of $\mathbf{CP}^{3}$. Then, we
recapitulate the Hitchin construction\footnote{As developed in \citet{hitchin1982twistor}.}
of minitwistor space, and discuss the mapping from representatives
of cohomology classes in $\mathbf{MT}$ to conformal primaries of
the $H_{3}^{+}$-WZNW model. 

Here, we follow the notation of \citet{kobayashi1996foundations}.

\textcompwordmark{}

\paragraph*{Note.}

The present subsection adopts a more mathematical style compared to
the remainder of this manuscript, and it is intended to acquaint the
reader with basic notions of minitwistor geometry.

\subsection{Hyperbolic Space from Projective Geometry\label{subsec:Hyperbolic-Space-from}}

Let the four-vector $X^{A\dot{A}}$ be homogeneous coordinates on
$\mathbf{CP}^{3}$, subject to the equivalence relation $X^{A\dot{A}}\sim a\cdot X^{A\dot{A}}$,
where $a$ is any non-zero complex scalar. A necessary and sufficient
condition for a set of components of a holomorphic metric (given in
the chart $X^{A\dot{A}}$) to be well-defined on $\mathbf{CP}^{3}$,
is that such components must be invariant under the ``gauge transformations''
$X^{A\dot{A}}\mapsto a\cdot X^{A\dot{A}}$ $\left(a\in\mathbf{C}^{*}\right)$,
and they must not possess any components along the scaling dimension
defined by this equivalence relation. 

A first fundamental form satisfying these conditions is given by:
\begin{equation}
ds^{2}=-\frac{dX^{2}}{X^{2}}+\left(\frac{X\cdot dX}{X^{2}}\right)^{2}.\label{eq:-21}
\end{equation}
The invariance of $ds^{2}$ under rescalings is manifest. Introducing
the metric tensor $\boldsymbol{g}_{A\dot{A}B\dot{B}}$ associated
with $ds^{2}$, for which $ds^{2}=\boldsymbol{g}_{A\dot{A}B\dot{B}}dX^{A\dot{A}}\otimes dX^{B\dot{B}}$,
we observe that:
\begin{equation}
\text{£}_{\Upsilon}\boldsymbol{g}=0,
\end{equation}
where $\Upsilon\coloneqq X^{A\dot{A}}\frac{\partial}{\partial X^{A\dot{A}}}\in\mathscr{X}\left(\mathbf{CP}^{3}\right)$
is the Euler vector field, and $\text{£}_{\Upsilon}$ is the Lie derivative
along the flow of $\Upsilon$. 

Consequently, $ds^{2}$ is devoid of components along the scaling
dimension, and defines a first fundamental form on the open submanifold
$\mathbf{CP}^{3}-\mathcal{B}$. Here, $\mathcal{B}$ denotes the closed
submanifold on which $\boldsymbol{g}_{A\dot{A}B\dot{B}}$ becomes
singular, and is defined as:
\begin{equation}
\mathcal{B}\coloneqq\left\{ \left(X^{A\dot{A}}\right)\in\mathbf{CP}^{3}\,\big|\,\left\langle X,X\right\rangle =0\right\} .
\end{equation}
Rewriting $ds^{2}$ in terms of normalised coordinates, $X^{A\dot{A}}/\left|X\right|$,
we find:
\begin{equation}
ds^{2}=-\varepsilon_{A\dot{A}}\varepsilon_{B\dot{B}}d\left(\frac{X^{A\dot{A}}}{\left|X\right|}\right)d\left(\frac{X^{B\dot{B}}}{\left|X\right|}\right),\,\,\,\left|X\right|^{2}\coloneqq\left\langle X,X\right\rangle .
\end{equation}
It follows that the submanifold $\mathbf{CP}^{3}-\mathcal{B}$, endowed
with the geometry induced by $\boldsymbol{g}_{A\dot{A}B\dot{B}}$,
is isometric to complexified $AdS_{3}$, with $\mathcal{B}\simeq\partial AdS_{3}$
corresponding to its conformal boundary.

To obtain either the Kleinian hyperboloid $\mathbf{H}_{3}$ or the
Lorentzian $AdS_{3}$, we restrict the metric tensor $\boldsymbol{g}_{AB\dot{A}\dot{B}}$
to an appropriate slice of $\mathbf{CP}^{3}$. This is achieved by
imposing suitable reality conditions on the components of $X^{A\dot{A}}$.

\subsection{Minitwistor Space}

A minitwistor space\footnote{See \citet{hitchin1982monopoles} and \citet{jones1985minitwistor}.}
$\mathcal{M}$ is defined as any two-dimensional complex manifold
containing a rational curve $\mathscr{C}$ (a holomorphic embedding
of $\mathbf{CP}^{1}$ into $\mathcal{M}$) with a normal bundle isomorphic
to $\mathcal{O}\left(2\right)$. Any such curve $\mathscr{C}$ is
referred to as a \emph{minitwistor line }or a \emph{Hitchin special
curve}. By virtue of a theorem due to Kodaira\footnote{Cf. \citet{kodaira1963structure} or \citet{kobayashi1996foundations}},
the manifold $\mathcal{M}$ admits a three-parameter family $\mathscr{F}$
of such special curves. Denoting the corresponding parameter space
by $\mathbf{W}\left(\mathscr{F}\right)$, a corollary of Kodaira's
theorem establishes the existence of an isomorphism between the tangent
space $T_{x}\big(\mathbf{W}\left(\mathscr{F}\right)\big)$, at any
point $x\in\mathbf{W}\left(\mathscr{F}\right)$, and the space of
globally defined smooth sections of the normal bundle $\mathcal{N}_{x}$
of the Hitchin special curve $\mathscr{C}_{x}\subset\mathcal{M}$
corresponding to $x$, given by $\Gamma\left(\mathscr{C}_{x},\mathcal{N}_{x}\right)$.
Furthermore, it has been demonstrated by Hitchin (1982) that $\mathbf{W}\coloneqq\mathbf{W}\left(\mathscr{F}\right)$,
constructed in this manner, can be endowed with the structure of a
Weyl manifold.

For our purposes, we specialise to the non-singular quadric $\mathbf{MT}$
$\coloneqq\mathbf{CP}^{1}\times\mathbf{CP}^{1}$. Let $\left(\lambda^{A},\mu_{\dot{A}}\right)$
denote homogeneous coordinates on $\mathbf{CP}^{1}\times\mathbf{CP}^{1}$.
The quadric $\mathbf{MT}$ is realised as an embedding into $\mathbf{CP}^{3}$
through the mapping:
\begin{equation}
\left(\lambda^{A},\mu_{\dot{A}}\right)\in\text{\textbf{MT}}\mapsto[\lambda^{A}\mu_{\dot{A}}]\in\mathbf{CP}^{3}.
\end{equation}
For any point $X^{A\dot{A}}\in\mathbf{CP}^{3}$, the locus of incidence:
\begin{equation}
\mathcal{L}\left(X\right)\coloneqq\left\{ \left(\lambda^{A},\mu_{\dot{A}}\right)\in\mathbf{MT}\,\big|\,\mu_{\dot{A}}=\lambda^{A}X_{A\dot{A}}\right\} ,
\end{equation}
defines a conic contained within $\mathbf{MT}$. This conic satisfies
the following property: there exists a hyperplane $\mathcal{H}\subset\mathbf{CP}^{3}$
such that $\mathcal{L}\left(X\right)=\mathbf{MT}\cap\mathcal{H}$.
Any such conics intersect in precisely two points. As a consequence,
these conics have a normal bundle isomorphic to $\mathcal{O}\left(2\right)$
and thus constitute special curves in the sense of Hitchin. Moreover,
the condition $\det\big(X^{A\dot{A}}\big)=0$ is both necessary and
sufficient for a plane section to be tangent to $\mathbf{MT}$. Any
such plane section determines $X^{A\dot{A}}$ up to a proportionality
factor. Accordingly, the space $\mathbf{W}$ of plane sections that
are not tangent to $\mathbf{MT}$ may be identified with the set of
non-null rays emanating from the origin in Minkowski space $\mathbf{R}^{\left(1,3\right)}$,
thereby realising $\mathbf{W}$ as a hyperboloid embedded into $\mathbf{R}^{\left(1,3\right)}$.

Now consider two points $X^{A\dot{A}},Y^{A\dot{A}}\in\mathbf{W}$
that are null-separated in the Weyl conformal structure. In this case,
either the corresponding conics in $\mathbf{MT}$ intersect, or the
associated planes in $\mathbf{CP}^{3}$ intersect along a line $\mathcal{L}$
that is tangent to $\mathbf{MT}$. Consequently, there exists a unique
plane passing through $\mathcal{L}$ and tangent to $\mathbf{MT}$.
This implies that the algebraic equation:
\begin{equation}
\det\big(X^{A\dot{A}}+tY^{A\dot{A}}\big)=0,
\end{equation}
admits solutions with multiplicity greater than one. By restricting
$X^{A\dot{A}}$ and $Y^{A\dot{A}}$ to be unit vectors in Minkowski
space and expanding the preceding equation in $t$, it follows that
$X^{A\dot{A}}$ and $Y^{A\dot{A}}$ are null-separated in $\mathbf{W}$
if and only if:
\begin{equation}
X^{A\dot{A}}Y_{A\dot{A}}=\varepsilon_{A\dot{A}}\varepsilon_{B\dot{B}}X^{A\dot{A}}Y^{B\dot{B}}=1.
\end{equation}
However, it is well known that the distance $\delta$ between the
points $X^{A\dot{A}}$ and $Y^{A\dot{A}}$ in the first fundamental
form induced on the hyperboloid\textbf{ }is given by\textbf{ $\delta=\cosh^{-1}\big(X^{A\dot{A}}Y_{A\dot{A}}\big)$.
}Thus, the conformal metric associated with Hitchin's construction
is equivalent to the class of metrics conformally related to the standard
hyperbolic metric on \textbf{$\mathbf{H}_{3}$.}

We now establish that the Weyl conformal structure determined via
the Hitchin correspondence from $\mathbf{MT}$ is equivalent to the
conformal structure obtained from the hyperboloid $\mathbf{H}_{3}$,
modulo conformal rescalings of the metric; consequently, $\mathbf{W}$
and $\mathbf{H}_{3}$ are conformally related. In fact, let $X^{A\dot{A}},Y^{A\dot{A}}\in\mathbf{W}$
be two points, and let $\Pi_{X},\Pi_{Y}\subset\mathbf{CP}^{3}$ denote
the planes corresponding to these points. Consider the line of intersection
$\mathcal{L}\coloneqq\Pi_{X}\cap\Pi_{Y}$. Let $\mathscr{F}$ denote
the one-parameter family of planes passing through $\mathcal{L}$.
The family $\mathscr{F}$ defines a set of conics contained within
$\mathbf{MT}$ that intersect at $\mathcal{L}$ and defines a Weyl
geodesic $\gamma$ in $\mathbf{W}$. However, $\gamma$ is precisely
the intersection of the hyperboloid $\mathbf{H}_{3}$ with the two-dimensional
subspace spanned by $X^{A\dot{A}}$ and $Y^{A\dot{A}}$ in Minkowski
space. Since $\mathbf{H}_{3}$ is endowed with a hyperbolic metric,
the curve $\gamma$ is a geodesic of the metric induced on $\mathbf{H}_{3}$.
This establishes the equivalence of the conformal structure on $\mathbf{W}$
with that on $\mathbf{H}_{3}$, thereby completing the proof that
the non-singular quadric $\mathbf{MT}$ is the minitwistor space of
$\mathbf{H}_{3}$ via the Hitchin correspondence.

\subsection{The Holomorphic Vector Bundle $\mathcal{O}\left(p,q\right)\protect\longrightarrow\mathbf{MT}$\label{subsec:The-Holomorphic-Vector}}

In this subsection, we shall define the holomorphic vector bundle
$\mathcal{O}\left(p,q\right)\longrightarrow\mathbf{MT}$ which serves
as the domain upon which the minitwistor Penrose transform is defined.
We establish that this bundle can be identified with the infinite-dimensional
function space $\mathscr{C}_{p,q}^{\infty}\left(\mathbf{MT}\right)$,
consisting of smooth complex-valued functions:
\begin{equation}
h:\left(\mathbf{C}^{*}\right)^{2}\times\left(\mathbf{C}^{*}\right)^{2}\longrightarrow\mathbf{C},
\end{equation}
satisfying the homogeneity property:
\begin{equation}
h\left(a\cdot\lambda^{A},b\cdot\mu_{\dot{A}}\right)=a^{p}b^{q}\,\,\,h\left(\lambda^{A},\mu_{\dot{A}}\right),
\end{equation}
for every pair of nonzero complex scalars $a$ and $b$.

\textcompwordmark{}

\paragraph*{Definition. }

Recall that the space of dotted two-component spinors forms a holomorphic
vector bundle $\mathcal{O}\left(1\right)\oplus\mathcal{O}\left(1\right)$
fibered over $\mathbf{CP}^{1}$. Let $\mu_{\mathscr{U}}\coloneqq\left(\mathscr{U},\vec{\mu}\right)$
denote a local trivialisation of this bundle, where $\mathscr{U}\subset\mathbf{CP}^{1}$
is an open neighbourhood and $\vec{\mu}:\mathscr{U}\longrightarrow\mathbf{C}^{2}$
is a coordinate map, which associates each dotted spinor $\mu_{\dot{A}}$
with its coordinate representation:
\begin{equation}
\left(\mu\right)_{\mathscr{U}}\coloneqq\left(\mu_{\dot{1}},\mu_{\dot{2}}\right).
\end{equation}
Similarly, the space of undotted two-component spinors can be described
by a corresponding holomorphic vector bundle. Let $\lambda_{\mathscr{V}}\coloneqq(\mathscr{V},\vec{\lambda})$
be an analogous local trivialisation, where $\mathscr{V}$ is an open
neighbourhood, and the map $\vec{\lambda}:\mathscr{V}\longrightarrow\mathbf{C}^{2}$
assigns each undotted spinor $\lambda^{A}$ to its coordinate representation:
\begin{equation}
\lambda_{\mathscr{V}}\coloneqq(\lambda^{1},\lambda^{2}).
\end{equation}
Next, define the set:
\begin{equation}
\mathcal{P}\,\coloneqq\,\big\{\,\big(\vec{\lambda},\vec{\mu},s\big)\,\big|\,\vec{\lambda}\in\left(\mathbf{C}^{*}\right)^{2},\,\vec{\mu}\in\left(\mathbf{C}^{*}\right),\,s\in\mathbf{C}\,\big\},
\end{equation}
and introduce an equivalence relation $\boldsymbol{\simeq}$ on $\mathcal{P}$
as follows. For any pair of nonzero complex scalars $a,b\in\mathbf{C}^{*}$,
we impose the condition:
\begin{equation}
\big(\vec{\lambda},\vec{\mu},s\big)\,\boldsymbol{\simeq}\,\big(a\cdot\vec{\lambda},\,b\cdot\vec{\mu},\,\,\,a^{p}b^{q}\,\cdot\,s\big).
\end{equation}
Taking the quotient of $\mathcal{P}$ by $\boldsymbol{\simeq}$ yields
a new manifold,
\begin{equation}
\mathcal{O}\left(p,q\right)\coloneqq\mathcal{P}/\boldsymbol{\simeq},
\end{equation}
which is endowed with the quotient topology. 

To establish the vector bundle structure of $\mathcal{O}\left(p,q\right)\longrightarrow\mathbf{MT}$,
consider the following diagram:\begin{equation} \label{triangle-diagram} \begin{tikzcd}[row sep=large, column sep=large] \mathbf{C}^{2} \times \mathbf{C}^{2} \arrow[dr, "\pi"', font=\Large] & & \mathcal{O}(p,q) \arrow[dl, "\mathcal{Q}", font=\Large] \\ & \mathbf{MT} & \end{tikzcd} \end{equation}with
the mappings in the diagram defined as follows:
\begin{equation}
\pi(\vec{\lambda},\vec{\mu})\coloneqq([\lambda^{A}],[\mu_{\dot{A}}]),\,\,\,\mathcal{Q}(\lambda^{A},\mu_{\dot{A}})\coloneqq\pi(\vec{\lambda},\vec{\mu}).
\end{equation}
Consequently, $\mathcal{Q}$ is a surjection with fibres isomorphic
to $\mathbf{C}$, and the transition functions between local trivialisations
satisfy the holomorphic cocycle condition, thereby showing that $\mathcal{Q}$
defines a holomorphic bundle over $\mathbf{MT}$.

\textcompwordmark{}

\paragraph*{Proposition.}

Let $\mathscr{C}_{p,q}^{\infty}\left(\mathbf{MT}\right)$ denote the
space of $\mathscr{C}^{\infty}$ complex-valued functions $h$ defined
on $\left(\mathbf{C}^{*}\right)^{2}\times\left(\mathbf{C}^{*}\right)^{2}$
that satisfy the homogeneity property:
\begin{equation}
h\left(a\cdot\lambda^{A},b\cdot\mu_{\dot{A}}\right)=a^{p}b^{q}\,\,\,h\left(\lambda^{A},\mu_{\dot{A}}\right),
\end{equation}
for every pair of nonzero complex scalars $a$ and $b$. The function
space $\mathscr{C}_{p,q}^{\infty}\left(\mathbf{MT}\right)$ can be
canonically identified with the module $\Gamma^{\infty}(\mathcal{O}(p,q))$
of smooth sections on the holomorphic vector bundle $\mathcal{O}\left(p,q\right)\longrightarrow\mathbf{MT}$.

\paragraph*{Proof.}

First, consider $h\in\mathscr{C}_{p,q}^{\infty}\left(\mathbf{MT}\right)$.
For each minitwistor $\widetilde{\mathsf{Z}}^{I}\coloneqq(\widetilde{\lambda}^{A},\widetilde{\mu}_{\dot{A}})\in\mathbf{MT}$,
define the set:
\begin{equation}
\digamma\coloneqq\left\{ \left(Z,h\left(Z\right)\right)\,\big|\,Z=(\vec{\lambda},\vec{\mu})\in\mathbf{C}^{2}\times\mathbf{C}^{2},\,\mathcal{\pi}\left(Z\right)=\widetilde{\mathsf{Z}}^{I}\right\} .
\end{equation}
By the homogeneity condition satisfied by $h$, the set $\digamma$
consists of equivalence classes under $\boldsymbol{\simeq}$. Consequently,
this set determines a unique point $\sigma(\widetilde{\mathsf{Z}}^{I})$
in the fibre of $\mathcal{O}\left(p,q\right)$ over $\widetilde{\mathsf{Z}}^{I}$.
Thus, the function $h$ defines a section $\widetilde{\mathsf{Z}}^{I}\mapsto\sigma(\widetilde{\mathsf{Z}}^{I})$
contained in $\Gamma^{\infty}(\mathcal{O}(p,q))$.

Conversely, suppose $\sigma\in\Gamma^{\infty}(\mathcal{O}(p,q))$
is a smooth section. For each minitwistor $\widetilde{\mathsf{Z}}^{I}\in\mathbf{MT}$,
there exists a class $\mathsf{C}=\{(Z,s)\}$ of equivalent pairs such
that $\pi(Z)=\widetilde{\mathsf{Z}}^{I}$. Choosing a representative
$Z=(\vec{\lambda},\vec{\mu})\in\mathbf{C}^{2}\times\mathbf{C}^{2}$,
we find a unique pair $(Z,s)\in\mathsf{C}$ in the class. The section
$\sigma$ therefore determines a function $f\in\mathscr{C}_{p,q}^{\infty}\left(\mathbf{MT}\right)$
defined by $Z\mapsto f\left(Z\right)\coloneqq s$, which satisfies
the homogeneity property by construction. Hence, the spaces $\mathscr{C}_{p,q}^{\infty}\left(\mathbf{MT}\right)$
and $\Gamma^{\infty}(\mathcal{O}(p,q))$ are isomorphic, as claimed.

\section{Leaf Amplitudes Review\label{sec:Leaf-Amplitudes-Review}}

\subsection{Klein Space}

The \emph{leaf representation} of celestial amplitudes is our main
motivation for introducing multi-gluon and multi-graviton wavefunctions
using minitwistor variables. This formalism requires the analytic
continuation of Minkowski spacetime $\mathbf{R}^{\left(1,3\right)}$
from a Lorentzian $\left(-+++\right)$ to a Kleinian $\left(--++\right)$
signature. For this purpose, we briefly review the relevant aspects
of Kleinian geometry, which have been investigated in detail by \citet{barrett1994kleinian},
\citet{bhattacharjee2022celestial}, \citet{crawley2022black}, \citet{cheung2003families}
and \citet{duary2024spectral}.

We begin by considering Cartesian coordinates $X^{\mu}$ $(0\le\mu,\nu,...\leq3)$
on $\mathbf{R}^{4}$, and define the Kleinian metric tensor $h_{\mu\nu}$
as:
\begin{equation}
h_{\mu\nu}\coloneqq\text{diag}\left(-1,-1,+1,+1\right).
\end{equation}
The resulting vector space $\mathbf{R}^{\left(2,2\right)}\coloneqq\left(\mathbf{R}^{4},\left\langle \cdot,\cdot\right\rangle \right)$,
equipped with the inner product:
\begin{equation}
\left\langle X,Y\right\rangle \coloneqq h_{\mu\nu}X^{\mu}Y^{\nu},\,\,\,\text{for}\,X^{\mu},Y^{\mu}\in\mathbf{R}^{\left(2,2\right)},
\end{equation}
is known as the \emph{four-dimensional Klein space. }

We identify three distinguished submanifolds contained in $\mathbf{R}^{\left(2,2\right)}$:
the \emph{null cone} $\Lambda$, the \emph{time-like wedge} $\mathbf{W}^{-}$,
and the \emph{space-like wedge $\mathbf{W}^{+}$}, defined (respectively)
as: 
\begin{align}
 & \Lambda\coloneqq\{\left(X^{\mu}\right)\in\mathbf{R}^{\left(2,2\right)}\,\big|\,\left\langle X,X\right\rangle =0\},\\
 & \mathbf{W}^{-}\coloneqq\{\left(X^{\mu}\right)\in\mathbf{R}^{\left(2,2\right)}\,\big|\,\left\langle X,X\right\rangle <0\},\\
 & \mathbf{W}^{+}\coloneqq\{\left(X^{\mu}\right)\in\mathbf{R}^{\left(2,2\right)}\,\big|\,\left\langle X,X\right\rangle >0\}.
\end{align}

\paragraph*{Time-like Wedge.}

To chart the time-like wedge $\mathbf{W}^{-}$, we employ the coordinate
system $X_{-}^{\mu}:\mathbf{W}^{-}\longrightarrow\mathbf{R}^{4}$,
defined as:
\begin{align}
 & X_{-}^{0}\coloneqq\tau\cos\left(\psi\right)\cosh\left(\rho\right),\,\,\,X_{-}^{1}\coloneqq\tau\sin\left(\psi\right)\cosh\left(\rho\right),\\
 & X_{-}^{2}\coloneqq\tau\cos\left(\varphi\right)\sinh\left(\rho\right),\,\,\,X_{-}^{3}\coloneqq\tau\sin\left(\varphi\right)\sinh\left(\rho\right),
\end{align}
where $\tau,\rho\in\left(0,\infty\right)$ and $\left(\psi,\varphi\right)\in S^{1}\times S^{1}$.

With respect to this parametrisation, the induced first fundamental
form on $\mathbf{W}^{-}$ takes the form:
\begin{equation}
\left(ds^{2}\right)_{\mathbf{W}^{-}}=-d\tau^{2}+\tau^{2}\left(d\rho^{2}-\cosh^{2}\left(\rho\right)d\psi^{2}+\sinh^{2}\left(\rho\right)d\phi^{2}\right).\label{eq:-26}
\end{equation}
It follows that the hypersurfaces of constant $\tau$ contained in
$\mathbf{W}^{-}$ are diffeomorphic to the three-dimensional Lorentzian
anti-de Sitter space with periodic time, denoted $AdS_{3}/\mathbf{Z}$.
Furthermore, the integration measure on $\mathbf{W}^{-}$ in the coordinate
chart $X_{-}^{\mu}$ is given by the volume form:
\begin{equation}
\left(d^{4}X\right){}_{\mathbf{W}^{-}}\coloneqq\frac{1}{2}\tau^{3}\sinh\left(2\rho\right)d\tau d\rho d\psi d\varphi,\label{eq:-27}
\end{equation}
where the juxtaposition of differentials is understood as the exterior
product of differential forms, $d\tau d\rho d\psi d\varphi=d\tau\wedge d\rho\wedge d\psi\wedge d\varphi$. 

A closed submanifold of $\mathbf{W}^{-}$, which plays a central role
in the subsequent discussion, is the \emph{standard Kleinian hyperboloid
$\mathbf{H}_{3}$}, defined as:
\begin{equation}
\mathbf{H}_{3}\coloneqq\{\left(X^{\mu}\right)\in\mathbf{R}^{\left(2,2\right)}\,\big|\,\left\langle X,X\right\rangle =-1\}.
\end{equation}
We chart $\mathbf{H}_{3}$ using the coordinate system $x^{\mu}:\mathbf{H}_{3}\longrightarrow\mathbf{R}^{4}$,
given by the functions:
\begin{align}
 & x^{0}\coloneqq\cos\left(\psi\right)\cosh\left(\rho\right),\,\,\,x^{1}\coloneqq\sin\left(\psi\right)\cosh\left(\rho\right),\\
 & x^{2}\coloneqq\cos\left(\varphi\right)\sinh\left(\rho\right),\,\,\,x^{3}\coloneqq\sin\left(\varphi\right)\sinh\left(\rho\right).
\end{align}

The restriction of the first fundamental form of $\mathbf{W}^{-}$
to $\mathbf{H}_{3}$, obtained via the pull-back $\iota_{-}^{*}$
of the inclusion map $\iota_{-}:\mathbf{H}_{3}\longrightarrow\mathbf{W}^{-}$,
endows $\mathbf{H}_{3}$ with a Lorentzian metric. This structure
implies the existence of an isometry from $\mathbf{H}_{3}$ onto $AdS_{3}/\mathbf{Z}$.
The corresponding integration measure on $\mathbf{H}_{3}$ is given
by the volume form:
\begin{equation}
d^{3}x\coloneqq\frac{1}{2}\sinh\left(2\rho\right)d\rho d\psi d\varphi.\label{eq:-28}
\end{equation}
Combining Eqs. (\ref{eq:-27}) and (\ref{eq:-28}) for the volume
forms on $\mathbf{W}^{-}$ and $\mathbf{H}_{3}$, respectively, we
deduce:
\begin{equation}
\left(d^{4}X\right){}_{\mathbf{W}^{-}}=\tau^{3}d\tau d^{3}x.\label{eq:-33}
\end{equation}

\paragraph*{Space-like Wedge.}

The space-like wedge $\mathbf{W}^{+}$ is parametrised by the coordinate
system $X_{+}^{\mu}:\mathbf{W}^{+}\longrightarrow\mathbf{R}^{4}$,
defined by:
\begin{equation}
X_{+}^{0}\coloneqq\tau\cos\left(\psi\right)\sinh\left(\rho\right),\,\,\,X_{+}^{1}\coloneqq\tau\sin\left(\psi\right)\sinh\left(\rho\right),
\end{equation}
\begin{equation}
X_{+}^{2}\coloneqq\tau\cos\left(\varphi\right)\cosh\left(\rho\right),\,\,\,X_{+}^{3}\coloneqq\tau\sin\left(\varphi\right)\cosh\left(\rho\right).
\end{equation}
The first fundamental form induced on $\mathbf{W}^{+}$ from the Kleinian
metric $h_{\mu\nu}$ is given by:
\begin{equation}
\left(ds^{2}\right)_{\mathbf{W}^{+}}=d\tau^{2}-\tau^{2}\left(d\rho^{2}+\sinh^{2}\left(\rho\right)d\psi^{2}-\cosh^{2}\left(\rho\right)d\varphi^{2}\right).
\end{equation}
The integration measure on $\mathbf{W}^{+}$ is represented by the
volume form:
\begin{equation}
\left(d^{4}X\right)_{\mathbf{W}^{+}}=-\frac{1}{2}\tau^{3}\sinh\left(2\rho\right)d\tau d\rho d\psi d\varphi.\label{eq:-29}
\end{equation}

Contained in $\mathbf{W}^{+}$ is the \emph{unit hyperboloid }$\mathbf{H}_{3}^{+}\coloneqq\{X_{+}^{\mu}\in\mathbf{R}^{(2,2)}\,\big|\,\left\langle X_{+},X_{+}\right\rangle =1\}$,
which is charted by the coordinate system $y^{\mu}:\mathbf{H}_{3}^{+}\longrightarrow\mathbf{R}^{4}$
defined by the functions:
\begin{equation}
y^{0}\coloneqq\cos\left(\psi\right)\sinh\left(\rho\right),\,\,\,y^{1}=\sin\left(\psi\right)\sinh\left(\rho\right),
\end{equation}
\begin{equation}
y^{2}\coloneqq\cos\left(\varphi\right)\cosh\left(\rho\right),\,\,\,y^{3}\coloneqq\sin\left(\varphi\right)\cosh\left(\rho\right).
\end{equation}

The first fundamental form on $\mathbf{H}_{3}^{+}$, induced by the
pull-back of the inclusion map $\iota_{+}:\mathbf{H}_{3}^{+}\longrightarrow\mathbf{W}^{+}$,
is given in these coordinates by:
\begin{equation}
\left(ds^{2}\right)_{\mathbf{H}_{3}^{+}}=d\rho^{2}+\sinh^{2}\left(\rho\right)d\psi^{2}-\cosh^{2}\left(\rho\right)d\varphi^{2}.
\end{equation}

The geometric distinction between $\mathbf{H}_{3}$ (from the time-like
wedge) and $\mathbf{H}_{3}^{+}$ lies in the interchange of the time-like
and space-like orientations. Despite this difference, $\mathbf{H}_{3}^{+}$
remains diffeomorphic to $AdS_{3}/\mathbf{Z}$. The volume form on
$\mathbf{H}_{3}^{+}$ is expressed as:
\begin{equation}
d^{3}y=-\frac{1}{2}\sinh\left(2\rho\right)d\rho d\psi d\varphi.\label{eq:-30}
\end{equation}
Finally, from Eqs. (\ref{eq:-29}) and (\ref{eq:-30}), we deduce
the following decomposition of the integration measure:
\begin{equation}
\left(d^{4}X\right)_{\mathbf{W}^{+}}=\tau^{3}d^{3}y.\label{eq:-34}
\end{equation}

\subsection{Spinor Algebra}

Employing the Van der Waerden formalism, as reviewed by \citet{veblen1933geometry}
and \citet{penrose1984spinors}, the inner product of two undotted
two-component spinors, denoted $\mu^{A}$ and $\nu^{A}$, is defined
as:
\begin{equation}
\left\langle \mu\nu\right\rangle \coloneqq\mu\cdot\nu\coloneqq\varepsilon_{AB}\mu^{A}\nu^{B}=\mu^{A}\nu_{A},
\end{equation}
where $\varepsilon_{AB}$ is the Levi-Civita symbol, antisymmetric
under index exchange, and normalised according to $\varepsilon_{12}=-\varepsilon_{21}=1$.
The lowering and raising of spinor indices adhere to the standard
convention, according to which $\mu_{A}\coloneqq\varepsilon_{AB}\mu^{B}$
and $\nu^{A}=\varepsilon^{AB}\nu_{B}$ with $\varepsilon^{AB}$ satisfying
$\varepsilon^{AC}\varepsilon_{CB}=\delta_{\,\,\,B}^{A}$.

Similarly, the inner product of two dotted spinors, $\bar{\mu}_{\dot{A}}$
and $\bar{\nu}_{\dot{A}}$, is defined as:
\begin{equation}
[\bar{\mu}\bar{\nu}]\coloneqq\varepsilon^{\dot{A}\dot{B}}\bar{\mu}_{\dot{A}}\bar{\nu}_{\dot{B}},
\end{equation}
where $\varepsilon^{\dot{A}\dot{B}}$ is the antisymmetric Levi-Civita
symbol for the dotted spinor space. Index manipulation for dotted
spinors follows analogous rules, employing the conventions $\bar{\mu}^{\dot{A}}\coloneqq\varepsilon^{\dot{A}\dot{B}}\bar{\mu}_{\dot{B}}$
and $\bar{\nu}_{\dot{A}}\coloneqq\varepsilon_{\dot{A}\dot{B}}\bar{\nu}^{\dot{B}}$,
where $\varepsilon_{\dot{A}\dot{B}}$ satisfies $\varepsilon_{\dot{A}\dot{C}}\varepsilon^{\dot{C}\dot{B}}=\delta_{\dot{A}}^{\,\,\,\dot{B}}$.

The \emph{Kleinian Pauli matrices}, which establish a correspondence
between vector and spinor indices in Kleinian signature, are defined
as follows:
\begin{equation}
\boldsymbol{\sigma}^{0}=\sigma^{3},\,\,\,\boldsymbol{\sigma}^{1}=\sigma^{1},\,\,\,\boldsymbol{\sigma}^{2}=\mathbb{I},\,\,\,\boldsymbol{\sigma}^{3}=i\sigma^{2},
\end{equation}
where $(\sigma^{\mu})_{A\dot{A}}$ denotes the Lorentzian Pauli matrices,
defined in the $(-+++)$ signature.

The reason for this choice becomes apparent through the following
construction. Consider the map from $\mathbf{R}^{\left(2,2\right)}$
into the space of real $2\times2$ matrices, given by:
\begin{equation}
X^{\mu}\mapsto(\underline{X})_{A\dot{A}}\coloneqq\sum_{\mu=0}^{3}X^{\mu}(\boldsymbol{\sigma}^{\mu})_{A\dot{A}}=\begin{pmatrix}X^{0}+X^{2} & X^{1}+X^{3}\\
X^{1}-X^{3} & X^{2}-X^{0}
\end{pmatrix}.\label{eq:-31}
\end{equation}
In this expression, the Einstein summation convention (according to
which indices are raised and lowered using the Kleinian metric $h_{\mu\nu}$)
is temporarily suspended, and ordinary summation is employed instead. 

The matrix $\underline{X}$ satisfies two important properties:
\begin{equation}
\underline{X}^{*}=\underline{X},\,\,\,\det\underline{X}=\left\langle X,X\right\rangle .\label{eq:-9}
\end{equation}
The first condition imposes a reality constraint, while the second
relates the determinant of $\underline{X}$ to the Kleinian inner
product of the four-vector $X^{\mu}$. 

For a transformation of the form:
\begin{equation}
\underline{X}\mapsto U\underline{X}\widetilde{U}^{-1},
\end{equation}
to preserve the conditions given by Eq. (\ref{eq:-9}), it is both
necessary and sufficient that $U$ and $\widetilde{U}$ take the forms:
\begin{equation}
U=\exp\left(\sum_{i=1}^{3}\lambda_{i}\boldsymbol{\sigma}^{i}\right),\,\,\,\widetilde{U}=\exp\left(\sum_{i=1}^{3}\widetilde{\lambda}_{i}\boldsymbol{\sigma}^{i}\right),
\end{equation}
with $\lambda_{i},\widetilde{\lambda}_{i}\in\mathbf{R}^{3}$. 

This construction realises the $\left(2,2\right)$ representation
of $SL\left(2,\mathbf{R}\right)\times\widetilde{SL\left(2,\mathbf{R}\right)}$
within the vector representation of $O\left(2,2\right)$, as anticipated
by the well-known isomorphism:
\begin{equation}
Spin\left(2,2\right)\simeq SL\left(2,\mathbf{R}\right)\times\widetilde{SL\left(2,\mathbf{R}\right)}.
\end{equation}
Thus, the mapping defined by Eq. (\ref{eq:-31}) allows the conversion
between vector and spinor indices in Kleinian signature, as we claimed.

\paragraph*{Standard Null-vector.}

To facilitate the parametrisation of the celestial sphere at null
infinity, we introduce the pair of two-component spinors $\eta^{A}$
and $\bar{\eta}_{\dot{A}}$. These spinors are defined in terms of
the angular coordinates $\bar{\zeta},\zeta\in\mathbf{CP}^{1}$, arising
from the stereographic projection of the celestial sphere onto the
complex plane, and are given by:
\begin{equation}
\eta^{A}\coloneqq\begin{pmatrix}\zeta\\
1
\end{pmatrix},\,\,\,\bar{\eta}_{\dot{A}}\coloneqq\begin{pmatrix}1 & -\bar{\zeta}\end{pmatrix}.
\end{equation}
The \emph{standard null four-vector} $q^{\mu}=q^{\mu}(\zeta,\bar{\zeta})$
is defined as:
\begin{equation}
q^{\mu}(\zeta,\bar{\zeta})\coloneqq\eta^{A}(\boldsymbol{\sigma}^{\mu})_{A\dot{A}}\bar{\eta}^{\dot{A}}=\left(\zeta\bar{\zeta}-1,\zeta+\bar{\zeta},1+\zeta\bar{\zeta},\zeta-\bar{\zeta}\right).\label{eq:-35}
\end{equation}

\subsection{Celestial Wavefunctions and the Leaf Amplitude Representation\label{subsec:Celestial-Wavefunctions-and}}

The leaf representation of celestial amplitudes, introduced by \citet{melton2023celestial},
arises from the consideration of the following integral over spacetime:
\begin{equation}
\mathcal{I}\left(\pi_{i},\bar{\pi}_{i}\right)\coloneqq\int_{\mathbf{R}^{\left(2,2\right)}}d^{4}X\,\,\,\prod_{i=1}^{n}\phi_{2h_{i}}\left(X\big|\pi_{i},\bar{\pi}_{i}\right).\label{eq:-32}
\end{equation}
Here, $\phi_{\Delta}\left(X\big|\pi_{i},\bar{\pi}_{i}\right)$ denotes
the \emph{celestial conformal primary wavefunction} for massless scalars
with conformal weight $\Delta$, defined by:
\begin{equation}
\phi_{\Delta}\left(X\big|\pi_{i},\bar{\pi}_{i}\right)\coloneqq\frac{\mathcal{C}\left(\Delta\right)}{\left(i\varepsilon+\langle\pi_{i}\big|X\big|\bar{\pi}_{i}]\right)^{\Delta}},\,\,\,\mathcal{C}\left(\Delta\right)\coloneqq i^{-\Delta}\Gamma\left(\Delta\right).\label{eq:-3}
\end{equation}

The spacetime integral in Eq. (\ref{eq:-32}) contains the key features
of the leaf representation, which we subsequently employ to reduce
from twistor to minitwistor variables. Thus, we briefly review its
computation.

Since the null cone $\Lambda$ in \textbf{$\mathbf{R}^{\left(2,2\right)}$
}has zero measure, the spacetime integral can be decomposed as the
sum of contributions from the time-like and space-like wedges:
\begin{equation}
\mathcal{I}=\int_{\mathbf{W}^{-}}d^{4}X\,\prod_{i=1}^{n}\phi_{2h_{i}}\left(X\big|\pi_{i},\bar{\pi}_{i}\right)+\int_{\mathbf{W}^{+}}d^{4}X\,\prod_{i=1}^{n}\phi_{2h_{i}}\left(X\big|\pi_{i},\bar{\pi}_{i}\right).
\end{equation}

Employing the decomposition of the measures for $\mathbf{W}^{-}$
and $\mathbf{W}^{+}$, given (respectively) by Eqs. (\ref{eq:-33})
and (\ref{eq:-34}), the above expression can be reformulated as:
\begin{equation}
\mathcal{I}=\int_{\left(0,\infty\right)}d\tau\,\tau^{3}\int_{\mathbf{H}_{3}}d^{3}x\,\prod_{i=1}^{n}\phi_{2h_{i}}\left(\tau x^{\mu}\big|\pi_{i},\bar{\pi}_{i}\right)+\int_{\left(0,\infty\right)}d\tau\,\tau^{3}\int_{\mathbf{H}_{3}^{+}}d^{3}x\,\prod_{i=1}^{n}\phi_{2h_{i}}\left(\tau y^{\mu}\big|\pi_{i},\bar{\pi}_{i}\right).
\end{equation}

An important property of these integrals follows from the parametrisation
of the null four-vector introduced in Eq. (\ref{eq:-35}). In fact,
it can be shown that: 
\begin{equation}
\int_{\mathbf{H}_{3}^{+}}d^{3}x\,\prod_{i=1}^{n}\phi_{2h_{i}}\left(\tau y^{\mu}\big|\pi_{i},\bar{\pi}_{i}\right)=\int_{\mathbf{H}_{3}}d^{3}x\,\prod_{i=1}^{n}\phi_{2h_{i}}\left(\tau x^{\mu}\big|\pi_{i},-\bar{\pi}_{i}\right).
\end{equation}

This result allows one to replace the integration over the unit hyperboloid
in the space-like wedge, $\mathbf{H}_{+}^{3}$, with an integration
over the standard Kleinian hyperboloid in the time-like wedge, $\mathbf{H}_{3}$,
provided the substitution $\bar{\pi}_{i\dot{A}}\mapsto-\bar{\pi}_{i\dot{A}}$
is made for all $1\leq i\leq n$.

Accordingly, the integral $\mathcal{I}$ can be expressed as:
\begin{equation}
\mathcal{I}=\int_{\left(0,\infty\right)}d\tau\,\tau^{3}\int_{\mathbf{H}_{3}}d^{3}x\,\prod_{i=1}^{n}\phi_{2h_{i}}\left(\tau x^{\mu}\big|\pi_{i},\bar{\pi}_{i}\right)+(\bar{\pi}_{i\dot{A}}\rightarrow-\bar{\pi}_{i\dot{A}}),\label{eq:-36}
\end{equation}
where $(\bar{\pi}_{i\dot{A}}\rightarrow-\bar{\pi}_{i\dot{A}})$ signifies
the repetition of the first term with the indicated substitution.

Factorising the $\tau$-dependence, we rewrite the integrand as:
\begin{equation}
\prod_{i=1}^{n}\phi_{2h_{i}}\left(\tau x^{\mu}\big|\pi_{i},\bar{\pi}_{i}\right)=\tau^{-2\sum_{i=1}^{n}h_{i}}\prod_{i=1}^{n}\frac{\mathcal{C}\left(2h_{i}\right)}{\left(i\varepsilon'+\langle\pi_{i}\big|x\big|\bar{\pi}_{i}]\right)^{2h_{i}}},
\end{equation}
where $\varepsilon'$ is a redefined infinitesimal regulator. 

Substituting this expression into Eq. (\ref{eq:-36}), we obtain:
\begin{equation}
\mathcal{I}=\int_{\left(0,\infty\right)}d\tau\,\tau^{3-2\sum_{i=1}^{n}h_{i}}\int_{\mathbf{H}_{3}}d^{3}x\,\prod_{i=1}^{n}\frac{\mathcal{C}\left(2h_{i}\right)}{\left(i\varepsilon'+\langle\pi_{i}\big|x\big|\bar{\pi}_{i}]\right)^{2h_{i}}}+(\bar{\pi}_{i\dot{A}}\rightarrow-\bar{\pi}_{i\dot{A}}).
\end{equation}

Finally, using the generalised Dirac delta function, analytically
continued to the complex domain as explained in \citet{donnay2020asymptotic},
we express this integral as:
\begin{equation}
\mathcal{I}=2\pi\delta\left(\beta\right)\int_{\mathbf{H}_{3}}d^{3}x\,\prod_{i=1}^{n}G_{2h_{i}}\left(x\big|\pi_{i},\bar{\pi}_{i}\right)+(\bar{\pi}_{i\dot{A}}\rightarrow-\bar{\pi}_{i\dot{A}}),\label{eq:-4}
\end{equation}
where $\beta\coloneqq4-2\sum_{i=1}^{n}h_{i},$ and $G_{\Delta}$ is
the bulk-to-boundary Green's function\footnote{The Green's function $G_{\Delta}$ may be obtained by the analytic
continuation of the corresponding $AdS_{3}$ propagator, as studied
by \citet{costa2014spinning} and reviewed by \citet{penedones2017tasi}.
Alternatively, $G_{\Delta}$ may also be derived by an analogous procedure
starting from the $H_{3}^{+}$-WZNW conformal primaries, discussed
by \citet{teschner1999mini}, employing the techniques developed by
\citet{gelfandgeneralized}.} for the covariant Laplacian on $\mathbf{H}_{3}$, given by:
\begin{equation}
G_{\Delta}\left(x\big|\pi,\bar{\pi}\right)\coloneqq\frac{\mathcal{C}\left(\Delta\right)}{\left(i\varepsilon+\langle\pi\big|x\big|\bar{\pi}]\right)^{\Delta}}.\label{eq:-11}
\end{equation}

\bibliographystyle{../TTCFT/revtex-tds/bibtex/bst/revtex/aipnum4-2}
\bibliography{CCFT2}

\end{document}